\documentclass[fleqn,usenatbib]{mnras}

\usepackage{newtxtext,newtxmath}
\usepackage{pdflscape}

\usepackage[T1]{fontenc}

\DeclareRobustCommand{\VAN}[3]{#2}
\let\VANthebibliography\thebibliography
\def\thebibliography{\DeclareRobustCommand{\VAN}[3]{##3}\VANthebibliography}

\usepackage{graphicx}	
\usepackage{amsmath}	
\usepackage{tabularx}
\usepackage{subfig}

\title{CO spectra of the ISM in the Host Galaxies of the Most Luminous WISE-Selected AGNs}

\author[Lee R. Martin et al.]{
Lee R. Martin,$^{1}$\thanks{E-mail: lm493@leicester.ac.uk (LRM);}
Andrew W. Blain,$^{1}$\thanks{ab520@leicester.ac.uk (AWB)}
Tanio Díaz-Santos,$^{2,3}$
Roberto J. Assef,$^{4}$ \newauthor
Chao-Wei Tsai,$^{5,6,7}$ 
Hyunsung D. Jun,$^{8,9}$
Peter R.M. Eisenhardt,$^{10}$
Jingwen Wu,$^{5,6}$ \newauthor
Andrey Vayner,$^{11}$ 
and Román Fernández Aranda$^{2,12}$
\\
$^{1}$University of Leicester, Physics and Astronomy, University Road, Leicester LE1 7RH, UK\\
$^{2}$Institute of Astrophysics, Foundation for Research and Technology-Hellas (FORTH), Heraklion, 70013, Greece\\
$^{3}$School of Sciences, European University Cyprus, Diogenes street, Engomi, 1516 Nicosia, Cyprus \\
$^{4}$Instituto de Estudios Astrof\'isicos, Facultad de Ingenier\'ia y Ciencias, Universidad Diego Portales, Av. Ej\'ercito Libertador 441, Santiago, Chile \\
$^{5}$University of Chinese Academy of Sciences, Beijing 100049, China \\
$^{6}$National Astronomical Observatories, Chinese Academy of Sciences, 20A Datun Road, Beijing 100101, China \\
$^{7}$Institute for Frontiers in Astronomy and Astrophysics, Beijing Normal University, Beijing 102206, China \\
$^{8}$Department of Physics, Northwestern College, 101 7th St SW, Orange City, IA 51041, USA \\
$^{9}$School of Physics, Korea Institute for Advanced Study, 85 Hoegiro, Dongdaemun-gu, Seoul 02455, Republic of Korea \\
$^{10}$Jet Propulsion Laboratory, California Institute of Technology, 4800 Oak Grove Drive, Pasadena, CA 91109, USA \\
$^{11}$Caltech/IPAC, 1200 E. California Blvd. Pasadena, CA 91125, USA\\
$^{12}$Department of Physics, University of Crete, 70013, Heraklion, Greece
}

\date{Accepted XXX. Received YYY; in original form ZZZ}

\pubyear{2024}

\begin{document}
\label{firstpage}
\pagerange{\pageref{firstpage}--\pageref{lastpage}}
\maketitle

\begin{abstract}
We present observations of mid-\textit{J} (\textit{J}~=~4--3 or \textit{J}~=~5--4) carbon monoxide (CO) emission lines and continuum emission from a sample of ten of the most luminous (\textit{L}\textsubscript{bol}~$\geq$~10$^{14}$\,L\textsubscript{$\rm\odot$}) Hot Dust-Obscured Galaxies (Hot DOGs) discovered by the Wide-field Infrared Survey Explorer (WISE) with redshifts up to 4.6. We uncover broad spectral lines (FWHM~$\geq$~400\,km\,s$^{-1}$) in these objects, suggesting a turbulent molecular interstellar medium (ISM) may be ubiquitous in Hot DOGs. A halo of molecular gas, extending out to a radius of 5\,kpc is observed in W2305--0039, likely supplied by 940\,km\,s$^{-1}$ molecular outflows. W0831+0140 is plausibly the host of a merger between at least two galaxies, consistent with observations made using ionized gas. These CO(4--3) observations contrast with previous CO(1--0) studies of the same sources: the CO(4--3) to CO(1--0) luminosity ratios exceed 300 in each source, suggesting that the lowest excited states of CO are underluminous. These findings show that the molecular gas in Hot DOGs is consistently turbulent, plausibly a consequence of AGN feedback, triggered by galactic mergers.
\end{abstract}




\section{Introduction} \label{sec:intro}
Feedback from active galactic nuclei (AGN) can have a profound effect on their host galaxy; stifling or enhancing star formation, heating the interstellar medium (ISM), and expelling large quantities of gas in outflows \citep{davies2022galaxy}. The supermassive black hole (SMBH) at the core of an AGN is fuelled by accreting infalling material; a process plausibly triggered by galactic mergers \citep{gao2020mergers}. The co-evolution of massive galaxies and their central SMBH is therefore important in understanding how galaxies evolve through cosmic time \citep{hopkins2008cosmological}. The precise interactions between obscured AGN and the ISM are debated, partly on account of the difficulty in detecting the AGN \citep{stern2012mid}. Observations of the ISM of a host galaxy, which could be strongly affected by an AGN and supply the gas to fuel it, provide a potent approach to probe these interactions. Furthermore, the best opportunities to determine how SMBHs affect their gas will come from the most luminous and extreme AGN, where the processes should be the most pronounced. \\
\indent Hot Dust-Obscured Galaxies (Hot DOGs) are hyper-luminous infrared galaxies (HyLIRGs). Approximately 10 per cent of the Wide-field Infrared Survey Explorer \citep[WISE;][]{wright2010wide} colour-selected Hot DOG population are extremely luminous \citep[\textit{L}\textsubscript{IR}~$\geq$~10$^{14}$\,L\textsubscript{$\rm\odot$};][]{tsai2015most} and are amongst the most luminous galaxies in the universe showing no signs of gravitational lensing. Hot DOGs are a subset of `W1W2 dropouts' on account of their weak detection in the WISE bands W1 (3.4\,$\mu$m) and W2 (4.6\,$\mu$m), but strong detections in the W3 (12\,$\mu$m) and W4 (22\,$\mu$m) bands \citep{eisenhardt2012first}. Further observations revealed high minimum dust temperatures \citep[\textit{T}\textsubscript{dust}~$\geq$~60\,K;][]{wu2012submillimeter}, with their spectral energy distributions (SEDs) showing significant contributions from dust at temperatures as hot as \textit{T}\textsubscript{dust}~$\approx$~450\,K \citep{tsai2015most}. This is hotter than dust detected in dust-obscured galaxies (DOGs), where the average dust temperature of even the most luminous targets is \textit{T}\textsubscript{dust}~$\approx$~45\,K \citep{blain2003submillimetre,bussmann2009infrared,casey2014dusty}. Investigations of Hot DOG SEDs are consistent with obscured, highly active AGN \citep{eisenhardt2012first,wu2012submillimeter,assef2020hot} and they have been found to exist in overdense environments \citep{jones2014submillimetre,assef2015half,diaz2018multiple,ginolfi2022detection,luo2022overdensity,zewdie2023overdensity}, particularly with neighbouring submillimetre galaxies \citep[SMGs;][]{blain2002submillimeter, casey2014dusty} and Lyman-break galaxies \citep{ginolfi2022detection,zewdie2023overdensity}, suggesting that galactic mergers may be a common and effective way of funnelling large quantities of gas into their  core, powering the AGN. The AGNs in these distant quasars are so extreme that their luminosity is close to (or potentially above) their Eddington limit \citep{wu2018eddington,tsai2018super,finnerty2020fast}, though some Eddington ratios may be overestimated due to an underestimated SMBH mass, or overestimated bolometric luminosity \citep{jun2020spectral}. Ionized gas outflows have been observed in Hot DOGs \citep{jun2020spectral,finnerty2020fast}, suggesting that they are in the process of expelling a portion of their gas, likely prompted by feedback from the AGN. Since they are blowing away their enshrouding material, Hot DOGs are thought to be a transitional phase in galaxy evolution and will subsequently evolve into optically unobscured quasars \citep{wu2012submillimeter,ricci2017growing,jun2021dust,assef2022imaging}. However, there is a recent suggestion that the phase of dust obscuration may be a recurrent event \citep[][]{diaz2021kinematics}, supported by simulations which recreate the formation of dusty galaxies \citep{narayanan2015formation}. With extreme luminosities and high dust temperatures, Hot DOGs should be excellent laboratories for studying how ISM gas is processed and excited by AGN. \\
\indent Carbon monoxide (CO) is the second most abundant molecule in the universe \citep{omont2007molecules} and is more easily excited and observed than H\textsubscript{2}. CO observations can reveal information about the bulk of molecular gas in a system, assuming that the abundance of H\textsubscript{2} and CO are linked \citep{solomon2005molecular}. Evidence of galactic mergers and outflows have been identified in Hot DOGs using fine-structure atomic lines \citep{diaz2016strikingly,diaz2018multiple,jun2020spectral,finnerty2020fast,diaz2021kinematics}, and so we can expect to probe the same properties when observing molecular gas and in some cases potentially spatially resolve the kinematics in this component, as has been achieved with high-resolution [CII] observations \citep{diaz2021kinematics}.\\
\indent CO observations in Hot DOGs are currently limited in number. So far, CO(1--0) in Hot DOGs has a total of five confirmed detections \citep{penney2020cold}. These low-\textit{J} observations provided information about some of the coldest and most extended gas. The Karl G. Jansky Very Large Array \citep[JVLA;][]{perley2011expanded} detected the molecular gas, but the observations lacked spatially-resolved velocity information. A handful of other low-\textit{J} observations are published \citep[CO(2--1), CO(3--2);][]{diaz2018multiple,fan2019alma}. Mid-\textit{J} CO observations probe more-excited, hotter (\textit{T}\textsubscript{ex}~$\geq$~55\,K), high-pressure molecular gas and have around ten confirmed detections \citep[CO(4--3), CO(6--5);][]{fan2018alma, ginolfi2022detection,sun2024physical}. Observations of high-\textit{J} transitions therefore reveal information about more energetic gas. There are currently few high-\textit{J} CO detections \citep[CO(7--6), CO(9--8);][]{ginolfi2022detection,aranda2024benchmark}. More detections are needed to place limits on the size, morphology and dynamics of the molecular gas, as well as provide information about the abundance of fuel which supplies the AGN.\\
\indent The Atacama Large Millimetre Array \citep[ALMA;][]{wootten2009atacama} has observed the atomic and molecular gas of Hot DOGs in striking resolved detail \citep[e.g.][]{diaz2016strikingly,fan2018alma,diaz2021kinematics}. With a wavelength range of 0.3--8.6\,mm, ALMA can probe millimetre and submillimetre emission at any redshift, and provide unparalleled resolution compared to rest-frame ultraviolet (UV) observations \citep{tsai2015most} from the Multi Unit Spectroscopic Explorer \citep[MUSE;][]{bacon2010muse}. \\
\indent In this study, we observe mid-\textit{J} CO and continuum emission from a sample of ten Hot DOGs, selected on account of being among the 10 per cent most luminous known to date \citep[\textit{L}\textsubscript{bol}~$\geq$~10$^{14}$\,L\textsubscript{$\rm\odot$};][]{tsai2015most}. We examine CO(4--3) in nine sources and CO(5--4) in another, to obtain properties of warm gas in a sizeable sample and provide spatially-resolved detail unseen in several targets with low-\textit{J} observations \citep{penney2020cold}. Due to the luminosity of these targets, we likely investigate during an era when SMBH growth was at its most extreme \citep{ricci2017growing}.\\
\indent The structure of this paper is as follows: a summary of the sources and data reduction are presented in Section \ref{sec:observations}. Section \ref{sec:results} presents assessments of the dust continuum and mid-\textit{J} CO line emission at a wavelength around 3\,mm; and of CO emission line redshifts, velocity dispersions, luminosities, and the dynamical and molecular gas mass. We discuss the implications of these results in Section \ref{sec:discussion} and summarise in Section \ref{sec:summary}. For brevity, we consistently state the truncated source names in place of their full WISE designation (e.g. WISEJ011601.41--050504.0 is quoted as W0116--0505). Throughout this paper, we refer to CO transitions as either low-\textit{J} [CO(1--0), CO(2--1), CO(3--2)], mid-\textit{J} [CO(4--3), CO(5--4), CO(6--5)] or high-\textit{J} [CO(7--6) and above]. We assume cosmological parameters of H\textsubscript{0}~=~70\,km\,s$^{-1}$\,Mpc$^{-1}$, $\Omega$\textsubscript{m}~=~0.7 and $\Omega$\textsubscript{$\Lambda$}~=~0.3.
\section{Observations \& Data Reduction}\label{sec:observations}
We obtained images of ten of the most luminous Hot DOGs \citep[\textit{L}\textsubscript{bol}~$\geq$~10$^{14}$\,L\textsubscript{$\rm\odot$};][]{tsai2015most} with ALMA (PID 2017.1.00358.S, PI: R.~J.~Assef) during December 2017 (Table~\ref{table:sources}). The on-source observing time was 60--90 minutes for each target using ALMA's 12-m array, with W0134--2922, W1322--0328, and W2305--0039 being observed twice. Observations of nine sources examined the CO(4--3) emission line, while CO(5--4) was observed in W2246--0526 due to its higher redshift. ALMA's band 3 (84--116\,GHz) was used for the observations that comprised four spectral windows (spw), each with around 2\,GHz bandwidth and 70 channels. The four windows are in two adjacent groups, with a separation of 10\,GHz between the groups. The reference (line) spw targeted the CO line, at a frequency determined using rest-frame UV redshifts \citep{wu2012submillimeter, tsai2015most}. The three remaining spws included line-free dust continuum. Six out of ten line observation recover a mean signal-to-noise S/N~$\geq$~3, while W0134--2922, W0615--5716, W1248--2154, and W2042--3245 have S/N~$<$~3. A summary of the observations is presented in Table~\ref{table:sources}.\\
\begin{table*}
    \centering
    \caption{Properties of the Hot DOG sample observed using ALMA. (1) Right ascension; (2) Declination; (3-4) Size of the major and minor axis of the beam; (5) Position angle of the beam; (6) On-source observing time; (7) Root mean square (RMS) noise per 75\,km\,s$^{-1}$ channel of the continuum-subtracted line spw; (8) Mean signal-to-noise ratio of the CO(4--3) line, except W2246--0526 which shows the CO(5--4) line.}
    \label{table:sources}
    \setlength\extrarowheight{2pt}
    \begin{tabular}{c c c c c c c c c}
        \hline 
         Source & R.A. (J2000)  & Dec (J2000) & \multicolumn{2}{c}{Beam Size} & Beam PA & t\textsubscript{obs} & RMS & S/N \\
        & hhmmss.ss & ddmmss.s & [arcsec] & [kpc] & [deg] & [s] & [mJy~beam$^{-1}$] &  \\
        & (1) & (2) & (3) & (4) & (5) & (6) & (7) & (8) \\
        \hline
          W0116--0505 & 01:16:01.42 & -05:05:04.2 & 0.32$\times$0.23 & 2.5$\times$1.8 & -66 & 4465 & 0.24 & 3.2\\
          W0134--2922 & 01:34:35.71 & -29:22:45.4 & 0.39$\times$0.28 & 3.1$\times$2.2 & -64 & 6588 & 0.21 & $<$~3 \\
          W0615--5716 & 06:15:11.07 & -57:16:14.6 & 0.33$\times$0.28 & 2.5$\times$2.1 & -45 & 3639 & 0.25 & $<$~3 \\
          W0831+0140 & 08:31:53.26 & +01:40:10.8 & 0.38$\times$0.30 & 2.7$\times$2.1 & 70 & 3152 & 0.39 & 5.5 \\
          W1248--2154 & 12:48:15.21 & -21:54:20.4 & 0.54$\times$0.48 & 4.1$\times$3.7 & 28 & 3667 & 0.19 & $<$~3 \\
          W1322--0328 & 13:22:32.57 & -03:28:42.2 & 0.52$\times$0.45 & 4.1$\times$3.5 & 21 & 6578 & 0.30 & 6.1 \\
          W2042--3245 & 20:42:49.28 & -32:45:17.9 & 0.55$\times$0.39 & 3.9$\times$2.8 & -76 & 3039 & 0.20 & $<$~3 \\
          W2246--0526 & 22:46:07.57 & -05:26:35.0 & 0.37$\times$0.24 & 2.5$\times$1.6 & -62 & 3426 & 0.29 & 4.2 \\
          W2246--7143 & 22:46:12.07 & -71:44:01.3 & 0.40$\times$0.30 & 3.0$\times$2.2 & 13 & 3946 & 0.32 & 5.3 \\
          W2305--0039 & 23:05:25.88 & -00:39:25.7 & 0.39$\times$0.34 & 3.0$\times$2.6 & 58 & 5238 & 0.36 & 5.6 \\
         \hline
    \end{tabular}
\end{table*}
\indent These sources, originally identified by WISE, are the most luminous Herschel/submillimetre-identified Hot DOGs with well-constrained submillimetre SEDs and confident redshifts \citep{tsai2015most}. The observations were limited only by ALMA's accessible sky declination. Five sources in the sample are observed in [CII] \citep{diaz2018multiple}, and four sources have CO(1--0) observations \citep{penney2020cold}. Examining these mid-\textit{J} CO lines in the most luminous cases will allow us to be as complete as possible in observing the conditions of the molecular ISM, where the processes at play should be most pronounced.\\ 
\indent The mid-to-far-IR luminosities of these sources is significantly larger than other high-redshift ULIRGs: attributed to their substantial AGN luminosities. SED modelling of Hot DOGs indicates up to 99 per cent of the bolometric luminosity can be attributed to an AGN \citep{eisenhardt2012first}. An extreme, compact starburst could reproduce the luminosity of Hot DOGs, but would require star formation rates (SFR) an order of magnitude higher than currently observed in even the most extreme systems \citep{tsai2015most}. AGNs are, therefore, the most plausible engines of Hot DOGs. \\
\indent The data were reduced using the Common Astronomy Software Applications package \citep[CASA;][]{mcmullin2007casa} and calibrated using the CASA pipeline. For the three sources which were observed twice, the measurement sets were combined using CASA's \texttt{concat} task, to within a frequency accuracy of 1\,MHz ($\Delta$\textit{v}~$\approx$~3\,km\,s$^{-1}$). All image cubes were generated using a 75\,km\,s$^{-1}$ ($\approx$~25\,MHz) channel width. A uniform pixel size of 0.05$^{\prime\prime}$ was used to sample the beams (resolution 0.2--0.5$^{\prime\prime}$) to examine up to 10 pixels across the main lobe of the beam. Initial images of the line emission were produced using CASA's \texttt{tclean} task. We experimented with the \texttt{robust} parameter within \texttt{briggs} weighting, and compared cubes close to uniform weighting, between uniform and natural weighting, and close to natural weighting. We investigated whether \texttt{uvtaper} would yield a greater S/N. In the final cubes, to obtain the highest S/N, \texttt{robust} was set to a value of 2.0; close to natural weighting without \texttt{uvtaper}. The continuum was subtracted from the line spw using \texttt{uvcontsub}, with the closest neighbouring line-free spw used to estimate the continuum, as using the other spws would sample dust continuum 10\,GHz away from the frequency of the CO line.\\
\indent The resulting data were exported and analysed using a variety of Python packages: \texttt{spectral-cube} \citep{robitaille2016spectral} was used to generate moment maps, while 1-D spectra were extracted using a circular aperture, centred on the CO line emission, using \texttt{photutils} \citep{bradley2016photutils}. The optimal aperture position and radius were found through an iterative process; the aperture was centred on each pixel in the central 10$\times$10 pixels of each cube, then the spectrum was extracted using circular apertures of radius 6--13 pixels. Thus, up to 800 apertures were examined per cube. The chosen apertures are those that recovered the largest S/N. Gaussian functions were fit to the spectra using \texttt{LMFIT} \citep{newville2016lmfit}.
\section{Observational Results} \label{sec:results}
\subsection{Dust Continuum}\label{sec:continuum}
\begin{figure*}
 \begin{tabular}{ccc}
 \noindent\subfloat{\includegraphics[width=6cm]{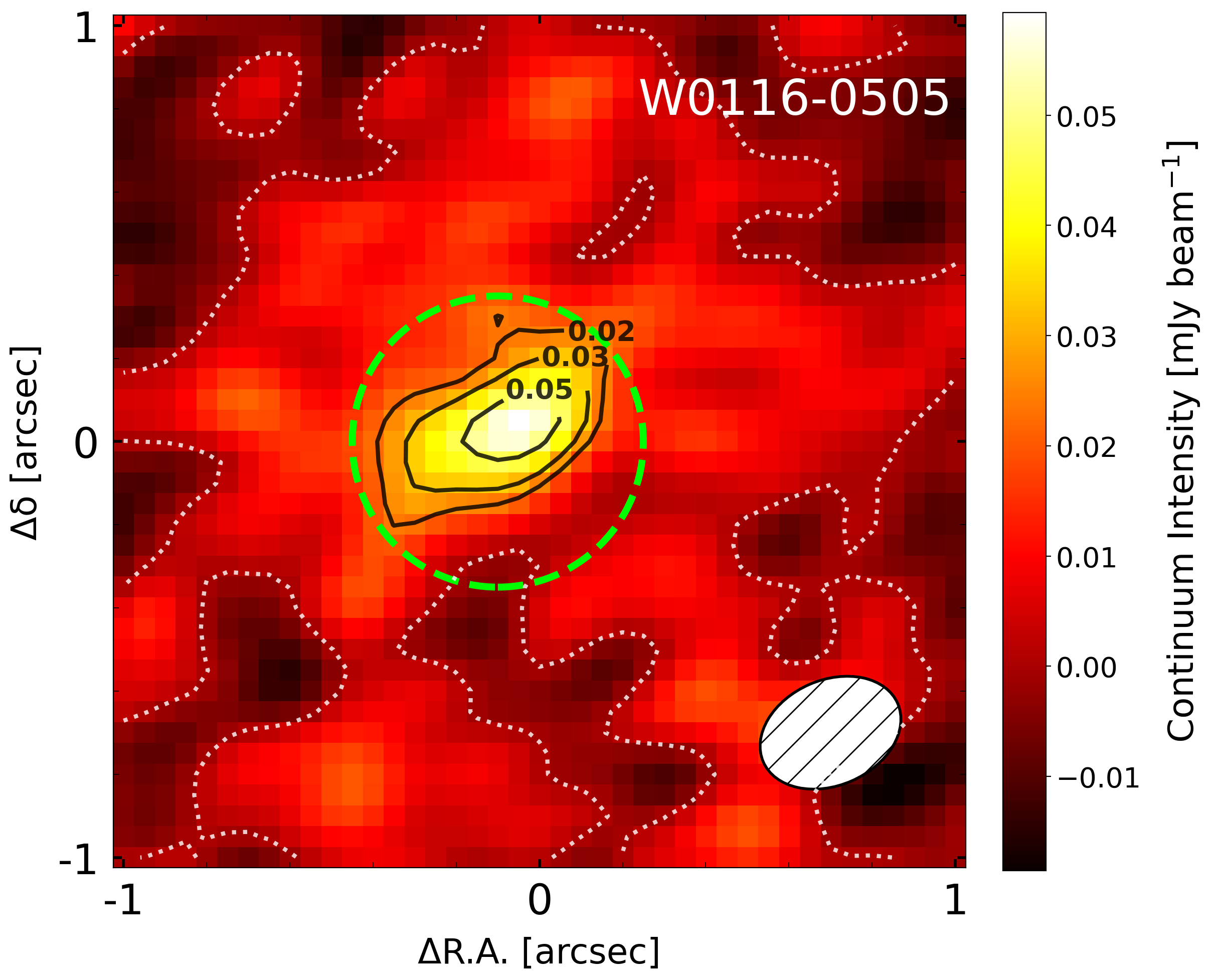}} \hspace{0.2cm}
 \subfloat{\includegraphics[width=6cm]{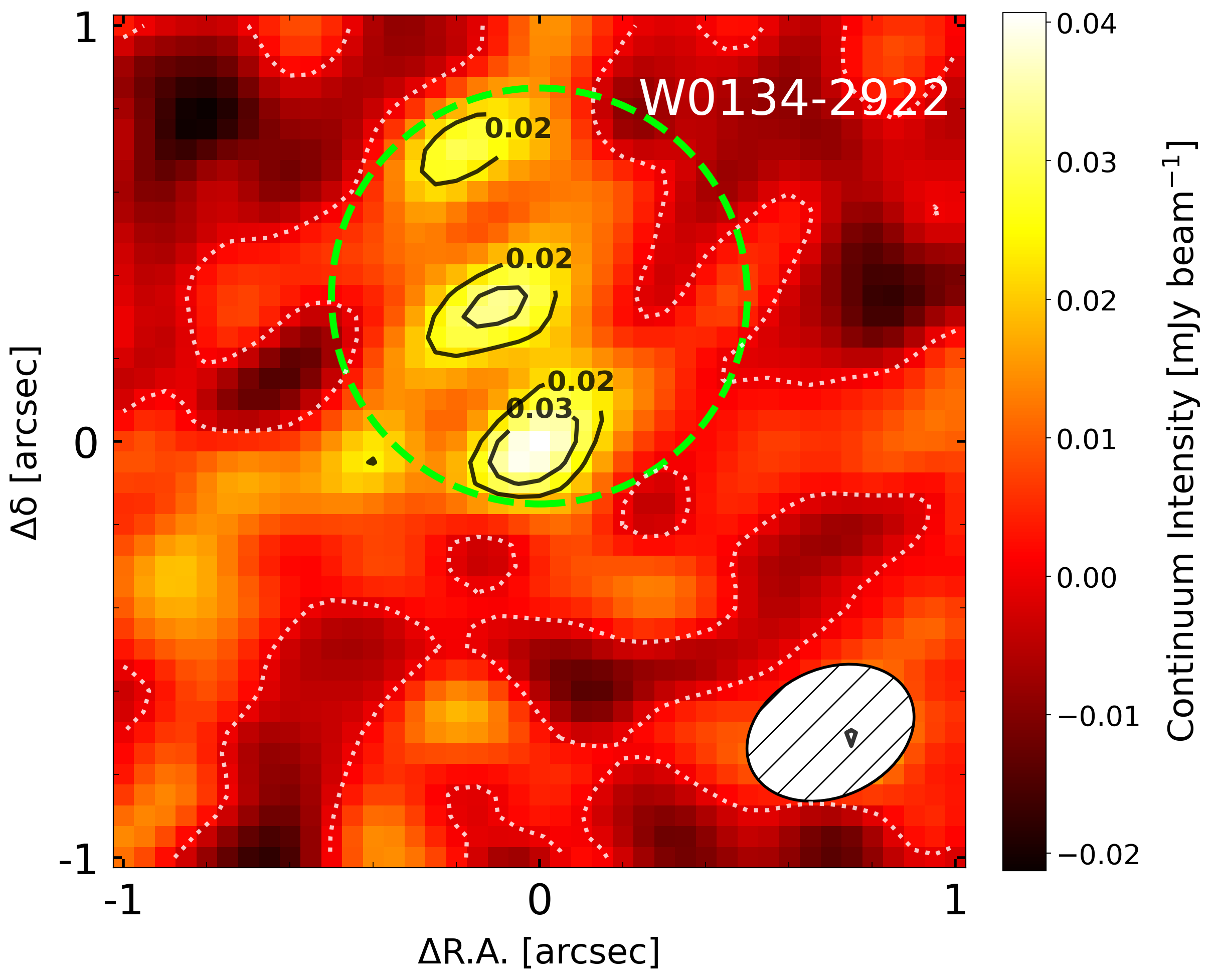}} \hspace{0.2cm}
 \subfloat{\includegraphics[width=6cm]{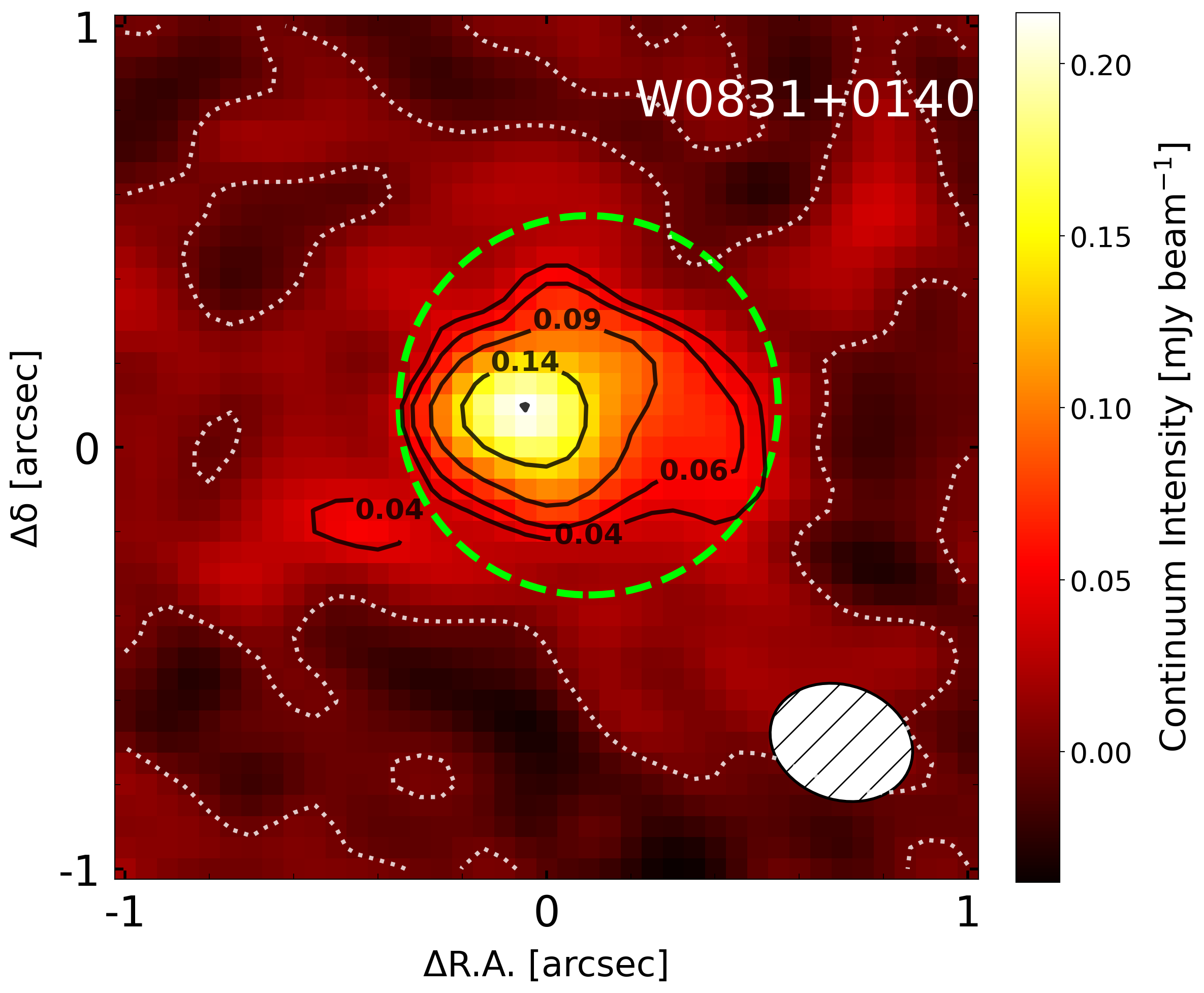}} \\
 \subfloat{\includegraphics[width=6cm]{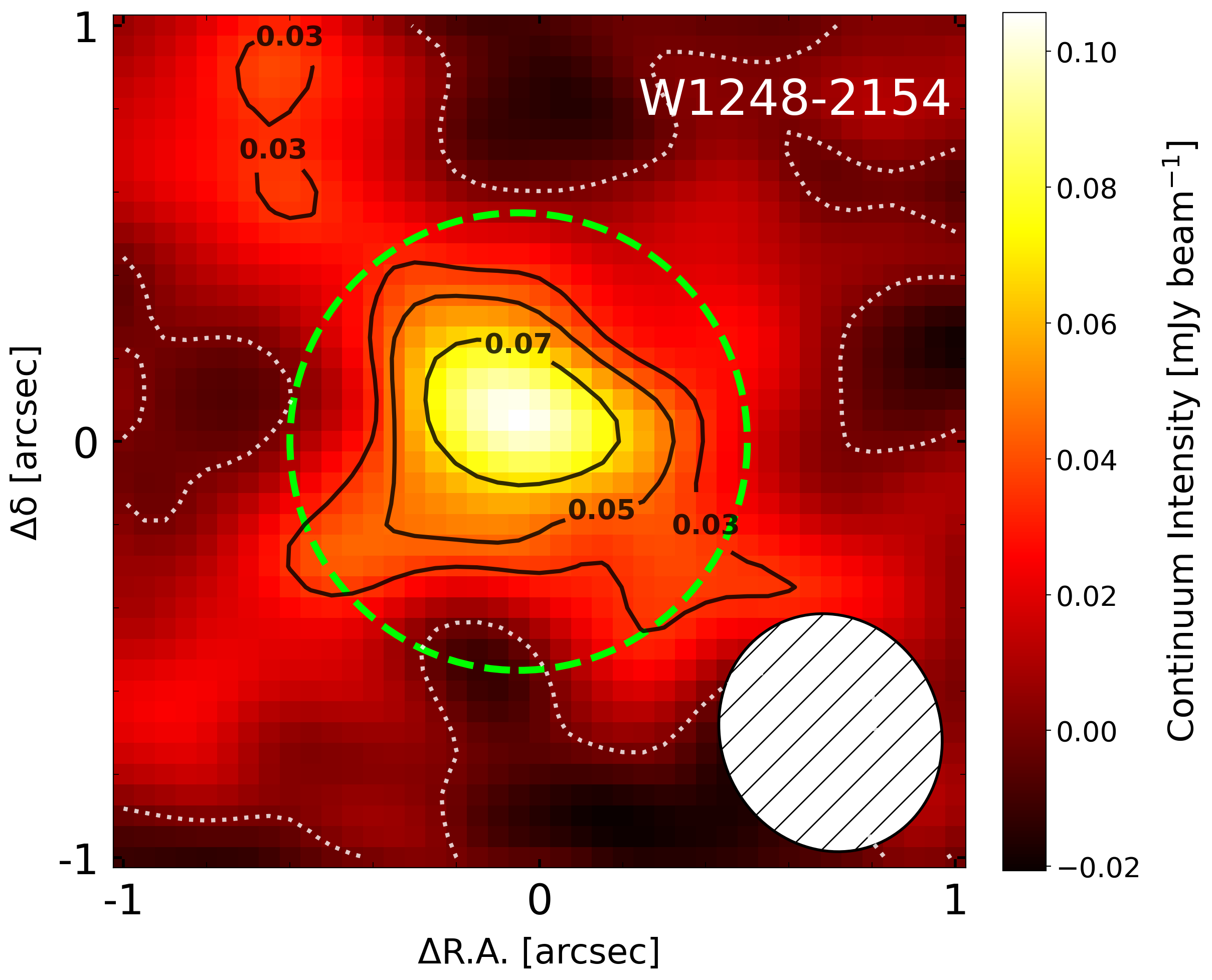}} \hspace{0.2cm}
 \subfloat{\includegraphics[width=6cm]{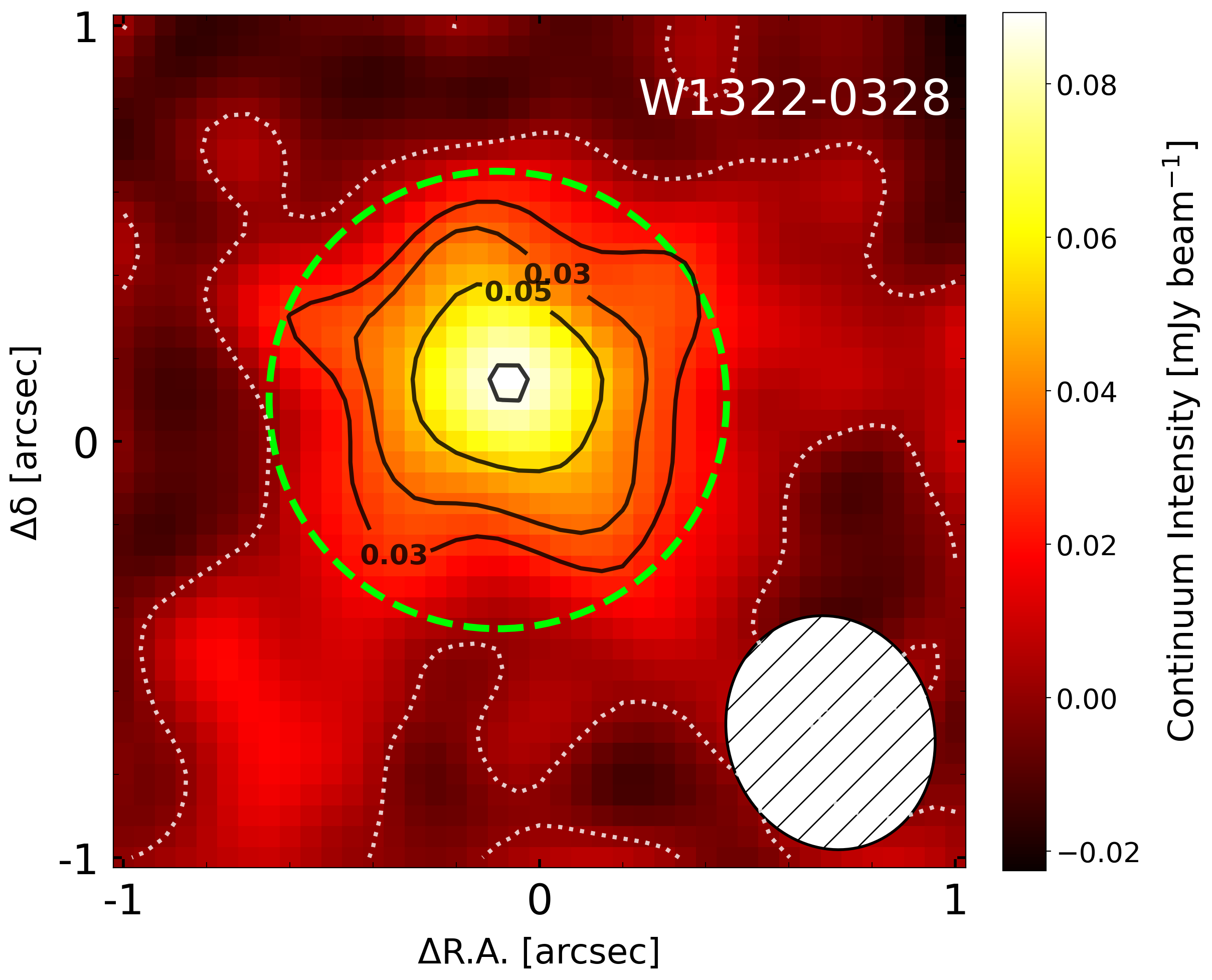}} \hspace{0.2cm}
 \subfloat{\includegraphics[width=6cm]{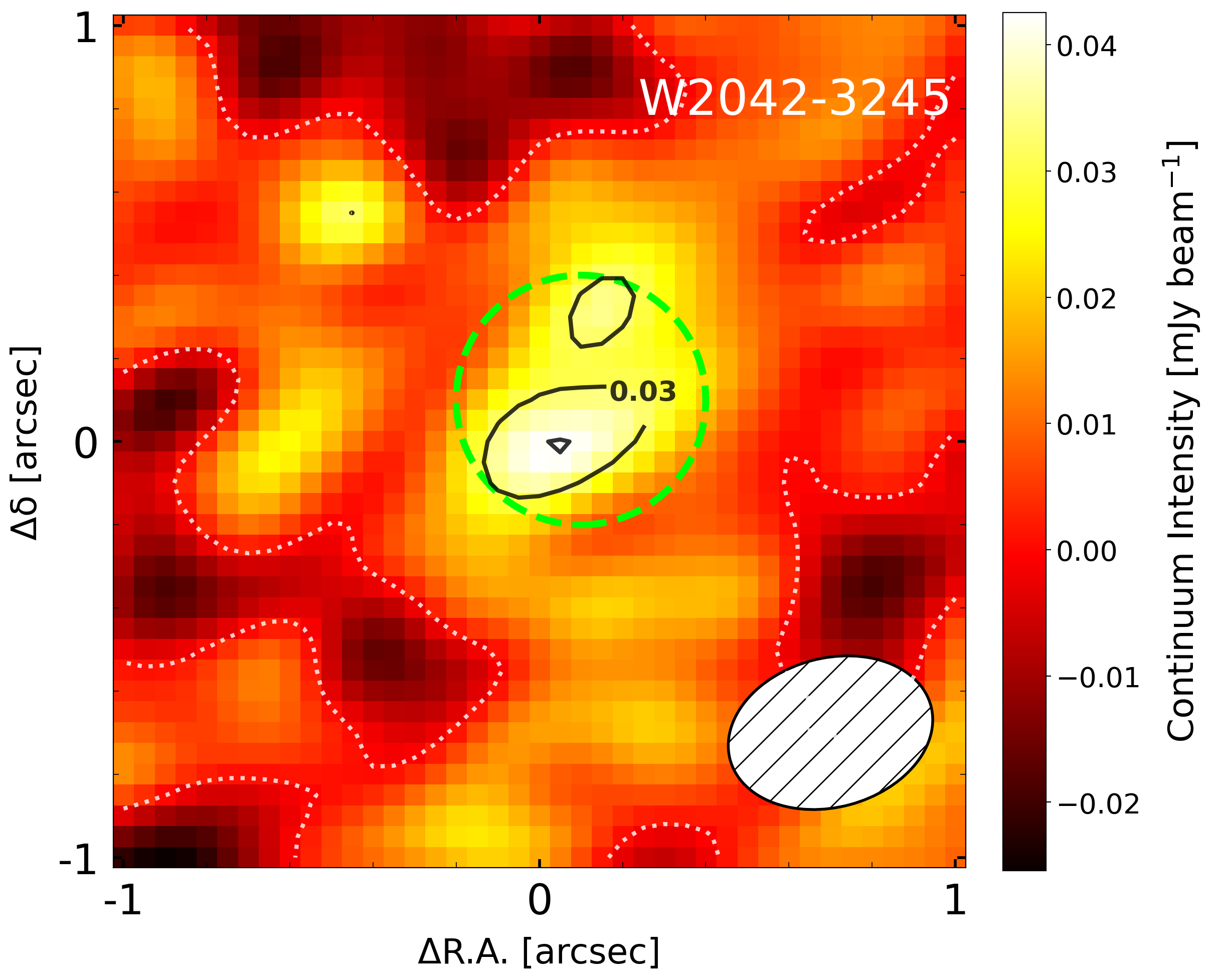}}\\
 \subfloat{\includegraphics[width=6cm]{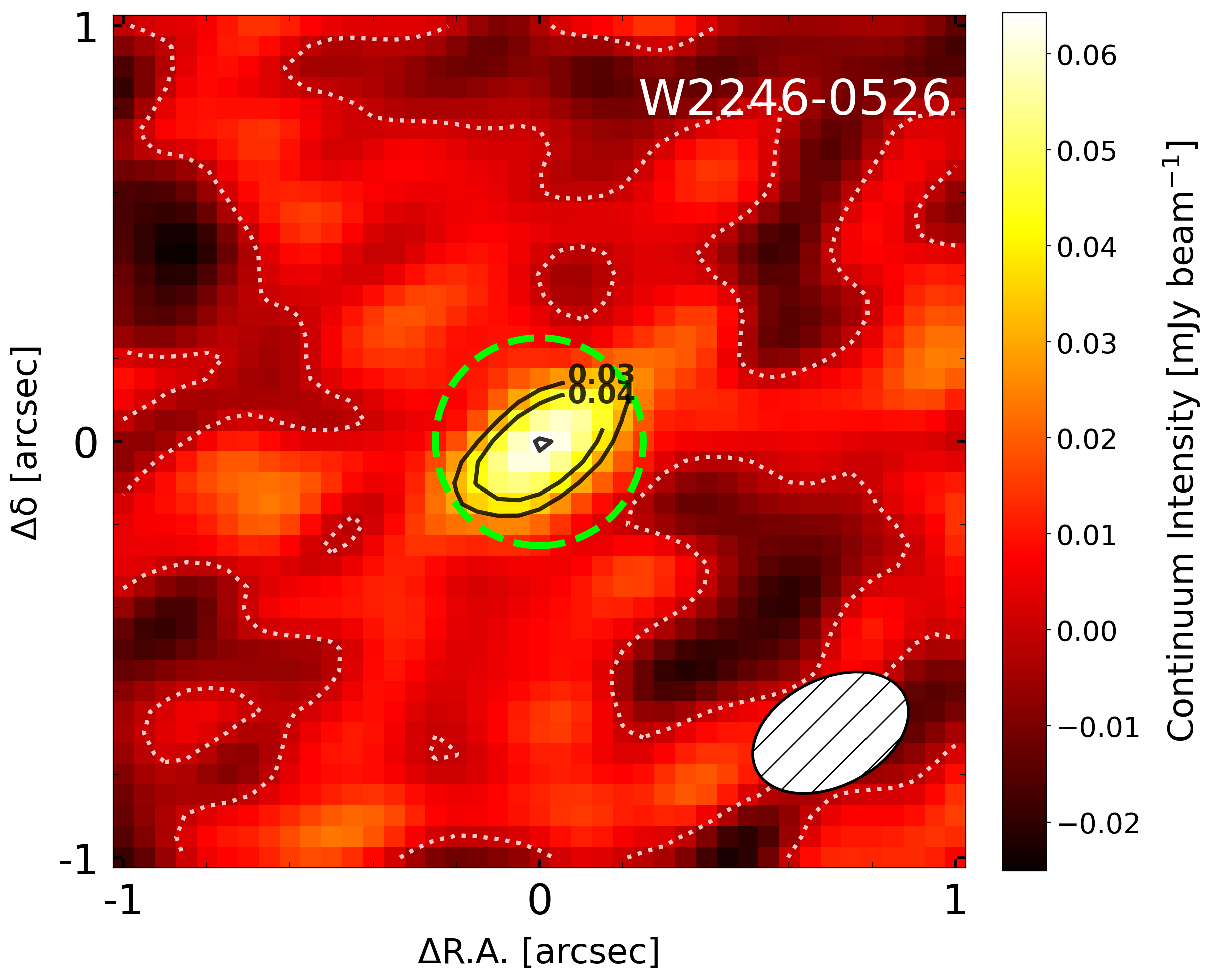}} \hspace{0.2cm}
 \subfloat{\includegraphics[width=6cm]{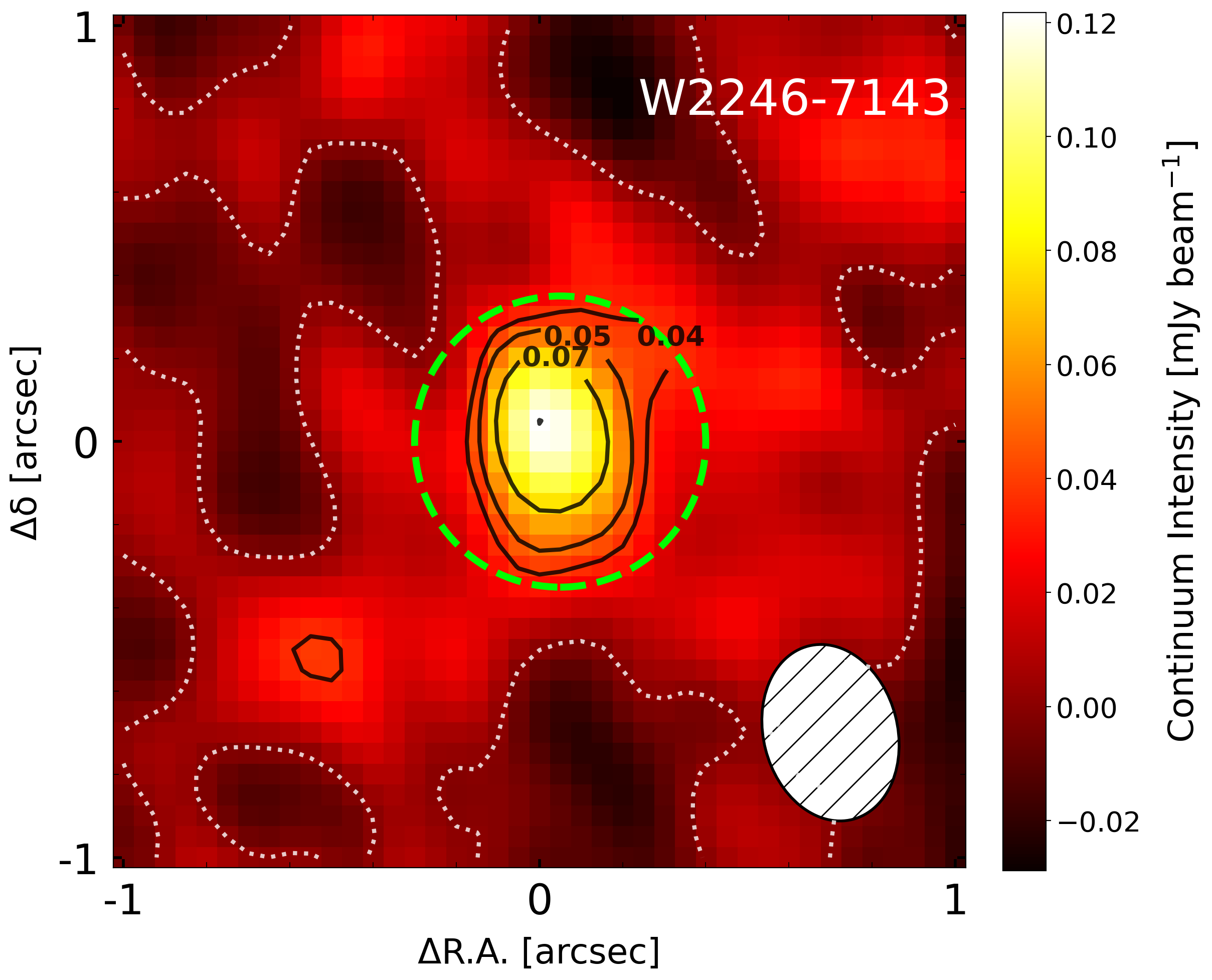}} \hspace{0.2cm}
 \subfloat{\includegraphics[width=6cm]{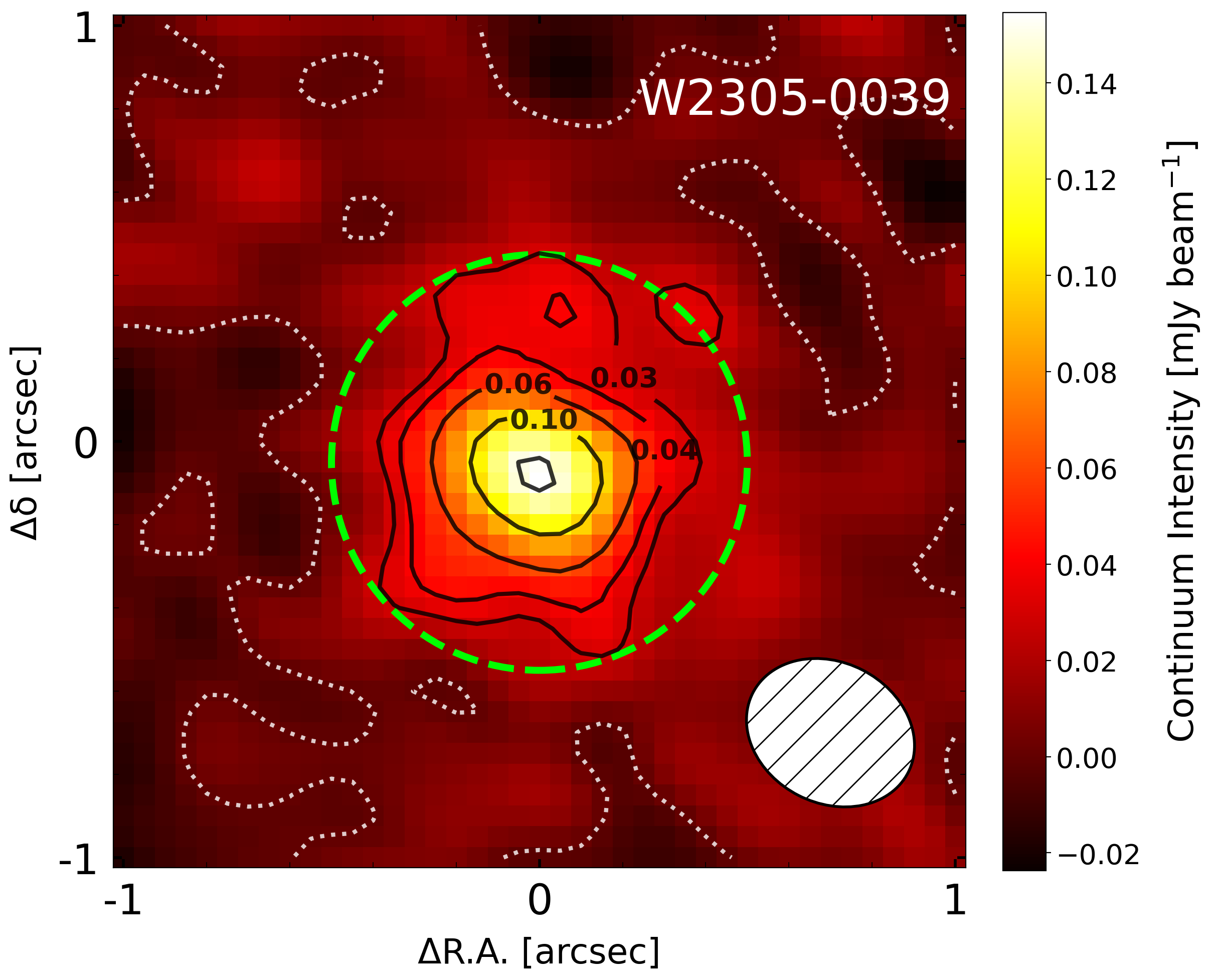}} \\
 \end{tabular}
 \caption{Continuum intensity maps of nine Hot DOGs. W0615--5716 is excluded on account of its uncertain detection. Each map is composed of three available continuum spws. The contours represent the root mean square (RMS) of the combined continuum cubes. Positive contours are shown as solid white lines and denote 3, 4, 6, and 10$\sigma$. Contours of zero intensity are represented by dots. The circular aperture used to measure the total continuum flux is shown by the dashed green lines. The clean beam is shown as a white ellipse in the lower-right corner of each image.}
 \label{fig:continuum}
 \end{figure*}

Continuum emission was detected in nine out of ten sources observed in the sample: see Table~\ref{table:continuum}. 
\begin{table*}
    \centering
    \caption{Dust continuum properties of the sample. (1) Central frequency of the continuum spws, each with 2\,GHz bandwidth; (2) Average (moment-0) flux of the continuum using three continuum spws, extracted using a circular aperture shown in Fig.~\ref{fig:continuum}; (3) Equivalent width of the CO line; (4) CO line-to-continuum ratio, considering the average CO and continuum flux; (5) The flux in the residuals after a beam-shaped signal matched to the peak of the image has been subtracted.}
    \label{table:continuum}
    \setlength\extrarowheight{2pt}
    \begin{tabular}{c c c c c c}
        \hline 
         Source & $\nu$  & \textit{S}\textsubscript{continuum} & \textit{W}\textsubscript{equivalent} & \textit{S}\textsubscript{peak}/\textit{S}\textsubscript{continuum} & \textit{f}\textsubscript{ext} \\
        & [GHz] & [$\mu$Jy] & [$\mu$m]  & &  \\
         & (1) & (2) & (3) & (4) & (5)\\
        \hline
          W0116--0505 & [96.5], [98.4], [108.5] & 99.1 $\pm$ 12.4 & 16.6 $\pm$ 2.6 & 7.7 $\pm$ 2.6 & 26\% \\
          W0134--2922 & [100.0], [101.8], [112.0] & 86.5 $\pm$ 13.2 & 11.5 $\pm$ 2.3 & 4.7 $\pm$ 2.4 & 49\% \\
          W0615--5716 & [92.7], [94.7], [106.7] & $\leq$~50 & $\geq$~20 & $\geq$~9.4 & \dots \\
          W0831+0140 & [96.2], [106.5], [108.3] & 525 $\pm$ 39.9 & 11.8 $\pm$ 1.1 & 4.1 $\pm$ 0.8 & 43\% \\
          W1248--2154 & [93.1], [94.8], [105.0] & 126 $\pm$ 12.4 & 5.3 $\pm$ 1.0 & 2.0 $\pm$ 1.5 & 16\% \\
          W1322--0328 & [100.1], [102.0], [112.1] & 118 $\pm$ 12.1 & 30.3 $\pm$ 3.4 & 16.2 $\pm$ 3.1 & 11\% \\
          W2042--3245 & [94.8], [104.8], [106.7] & 62.5 $\pm$ 12.5 & 9.4 $\pm$ 2.3  & 7.0 $\pm$ 3.6 & 21\% \\
          W2246--0526 & [91.0], [92.9], [104.9] & 68.9 $\pm$ 10.6 & 30.5 $\pm$ 5.2 & 16.9 $\pm$ 4.8 & 6\% \\
          W2246--7143 & [91.4], [93.3], [105.3] & 172 $\pm$ 17.3 & 19.8 $\pm$ 2.5 & 9.4 $\pm$ 2.0 & 12\% \\
          W2305--0039 & [100.3], [102.2], [114.2] & 277 $\pm$ 19.7 & 21.1 $\pm$ 2.1 & 6.3 $\pm$ 1.3 & 24\% \\
         \hline
    \end{tabular}
\end{table*}
The strength of the continuum emission spans an order of magnitude, ranging from $\leq$~50--525\,$\mu$Jy. This shows that the dust is heated to different degrees in each target, and the amount of dust varies from source to source. Continuum emission maps are shown in Fig.~\ref{fig:continuum} and are composed of the combined continuum spws (a total of three spws in each case). \\
\indent The continua are morphologically varied; ranging from compact beam-sized emission (e.g. W2246--0526) to more extended emission (e.g. W0831+0140, W2305--0039). By performing a 2-D elliptical Gaussian subtraction, representative of the clean beam, we examined the fraction of extended flux (f\textsubscript{ext}: Table \ref{table:continuum}). These subtractions consider the maximum unresolved contribution from a point-source, as the peak of the elliptical Gaussian matches the data: see an example of the subtraction in Fig.~\ref{fig:beam_subtraction_example}. On average, 23 per cent of the total continuum flux is recovered from extended components with 3$\sigma$ confidence, though given that we subtracted the maximum unresolved point source, the emission from extended dust could be greater. Nevertheless, the dust in these sources is overall marginally extended. The continuum structure of W0134--2922 is extended over 1$^{\prime\prime}$ and is unlike any other source in the sample. The dust observed at a wavelength around 3\,mm [close to CO(4--3): rest-frame 651\,$\mu$m] is on average a factor of 7 more luminous than observed in the same targets around a wavelength of 1\,cm [close to CO(1--0): rest-frame 2.60\,mm] \citep{penney2020cold}, consistent with a Rayleigh-Jeans spectrum. The dust in (inevitably) less-luminous, high-redshift (\textit{z}~$\approx$~6) optically-selected quasars around the same wavelength of 3\,mm have an average flux of 100\,$\mu$Jy \citep{wang2010molecular}. This is consistent with the majority of Hot DOGs in this sample, though W0831+0140 and W2305--0039 each have a considerably brighter dust continuum (525~\&~277\,$\mu$Jy respectively). This suggests that W0831+0140 and W2305--0039 may harbour greater quantities of cooler dust than most Hot DOGs or optically-selected quasars at higher redshift, or that the dust is considerably hotter than other sources. This could be a consequence of galactic mergers, supplying, spreading and heating dust \citep{ricci2017growing,appleton2018jet}. Notably, a galactic merger in W0831+0140 has been suggested using [CII] data \citep{diaz2021kinematics}. The CO line-to-continuum ratio is a predictor of galactic shocks \citep{meijerink2012evidence}. This is because shocks can compress and excite gas while not affecting dust (unless the shock velocities and densities are high). W2246--0526 has the largest CO line-to-continuum ratio in the sample (Table~\ref{table:continuum}), and is known to be comprised of a merger of at least three galaxies \citep{diaz2018multiple}, likely resulting in galactic-scale shocks.
\subsection{Redshift Estimates} \label{sec:redshift}
\begin{table*}
    \centering
    \caption{Redshifts of the Hot DOGs in the sample. (1) Redshift of CO emission line (this work), with a typical uncertainty of 0.002. CO(5--4) was observed in W2246--0526 and CO(4--3) in all other targets. The (*) symbol denotes that the redshift for this source is the median of a double peak (see Fig.~\ref{fig:spectra}); (2) Redshift derived from rest-frame UV (observed-frame optical) observations \citep{tsai2015most}; (3) Velocity offset between CO and UV redshifts; (4) Redshift derived from near-infrared (NIR) observations \citep{finnerty2020fast}; (5) Velocity offset between CO and NIR redshifts.}
    \label{table:redshift}
    \setlength\extrarowheight{2pt}
    \begin{tabular}{c c c c c c}
        \hline 
         Source & \textit{z}\textsubscript{CO}  & \textit{z}\textsubscript{UV} & $\delta$\textit{v}\textsubscript{(CO--UV)} & \textit{z}\textsubscript{NIR} & $\delta$\textit{v}\textsubscript{(CO--NIR)} \\
        & & & [km~s$^{-1}$] & & [km~s$^{-1}$]\\
        & (1) & (2) & (3) & (4) & (5) \\
        \hline
          W0116--0505 & 3.191 & 3.173 & 1300 $\pm$ 200 & 3.191 & 0 $\pm$ 200 \\
          W0134--2922 & 3.058 & 3.047 & 820 $\pm$ 210 & \dots & \dots \\
          W0615--5716 & 3.347 & 3.399 & $-$3500 $\pm$ 200 & \dots & \dots   \\
          W0831+0140  & 3.916* & 3.888 & 1700 $\pm$ 170 & 3.915 & 60 $\pm$ 170 \\
          W1248--2154 & 3.323 & 3.318 & 350 $\pm$ 200 & \dots & \dots \\
          W1322--0328 & 3.045 & 3.043 & 150 $\pm$ 210 & 3.043 & 150 $\pm$ 210 \\
          W2042--3245 & 3.970* & 3.963 & 420 $\pm$ 170 & \dots & \dots \\
          W2246--0526 & 4.601 & 4.593 &  430 $\pm$ 150 & 4.602 & 50 $\pm$ 150 \\
          W2246--7143 & 3.463 & 3.458 & 340 $\pm$ 190 & \dots & \dots \\
          W2305--0039 & 3.111 & 3.106 & 370 $\pm$ 210 & 3.108 & 220 $\pm$ 210 \\
         \hline
    \end{tabular}
\end{table*}
A CO-based redshift of each source (Table~\ref{table:redshift}) was calculated using the observed central frequency of the CO line and comparing with the rest-frame frequency [$\nu$\textsubscript{CO(4--3)}~=~461.04\,GHz, $\nu$\textsubscript{CO(5--4)}~=~576.27\,GHz]. These values largely agree with the redshifts derived from rest-frame UV data \citep{tsai2015most}, which are typically derived from Ly--$\alpha$ and [CIV] lines. The most noteworthy velocity offset is for W0615--5716, where the CO emission is blueshifted $\delta$\textit{v}~=~$-$3500\,km\,s$^{-1}$. W0116--0505 and W0831+0140 also have CO velocity offsets $\delta$\textit{v}~$\geq$~1000\,km\,s$^{-1}$, though Ly--$\alpha$ and [CIV] often show large velocity offsets on account of their complex dynamics. When comparing our CO redshifts with near-infrared spectroscopic redshifts, derived from [OIII] and H$\alpha$ \citep{finnerty2020fast}, we see good agreement. \\
\indent There is disagreement in the literature as to the redshift of W2042--3245: rest-frame UV observations gave a redshift \textit{z}~=~3.963 \citep{tsai2015most}, whereas \textit{z}~=~2.958 has been reported subsequently from Ly--$\alpha$ and MgII emission lines \citep{jun2020spectral}. The CO data agrees with the former.
\subsection{CO Line Detections} \label{sec:resolution}
\begin{figure*}
 \begin{tabular}{ccc}
 \noindent\subfloat{\includegraphics[width=6cm]{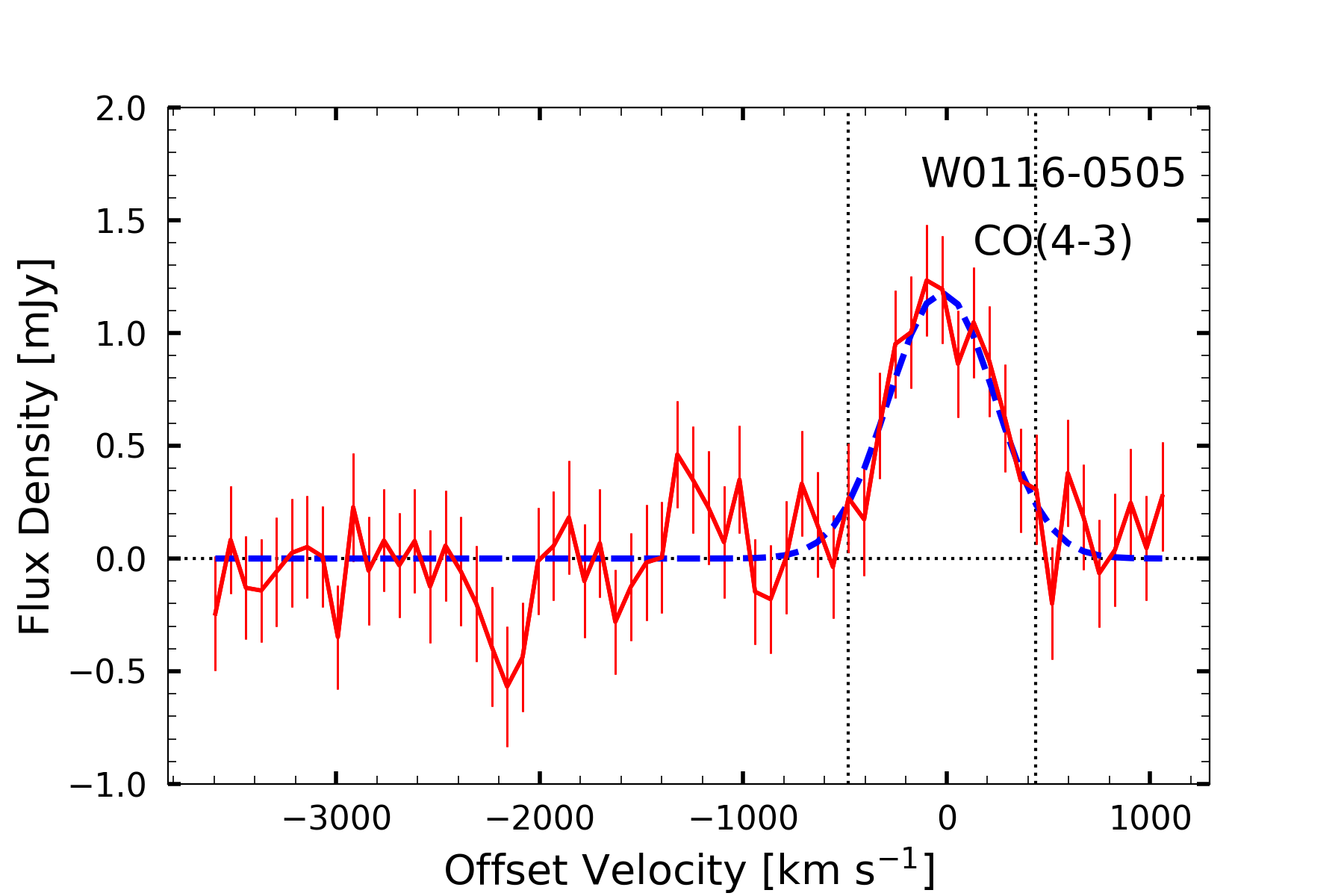}}
 \subfloat{\includegraphics[width=6cm]{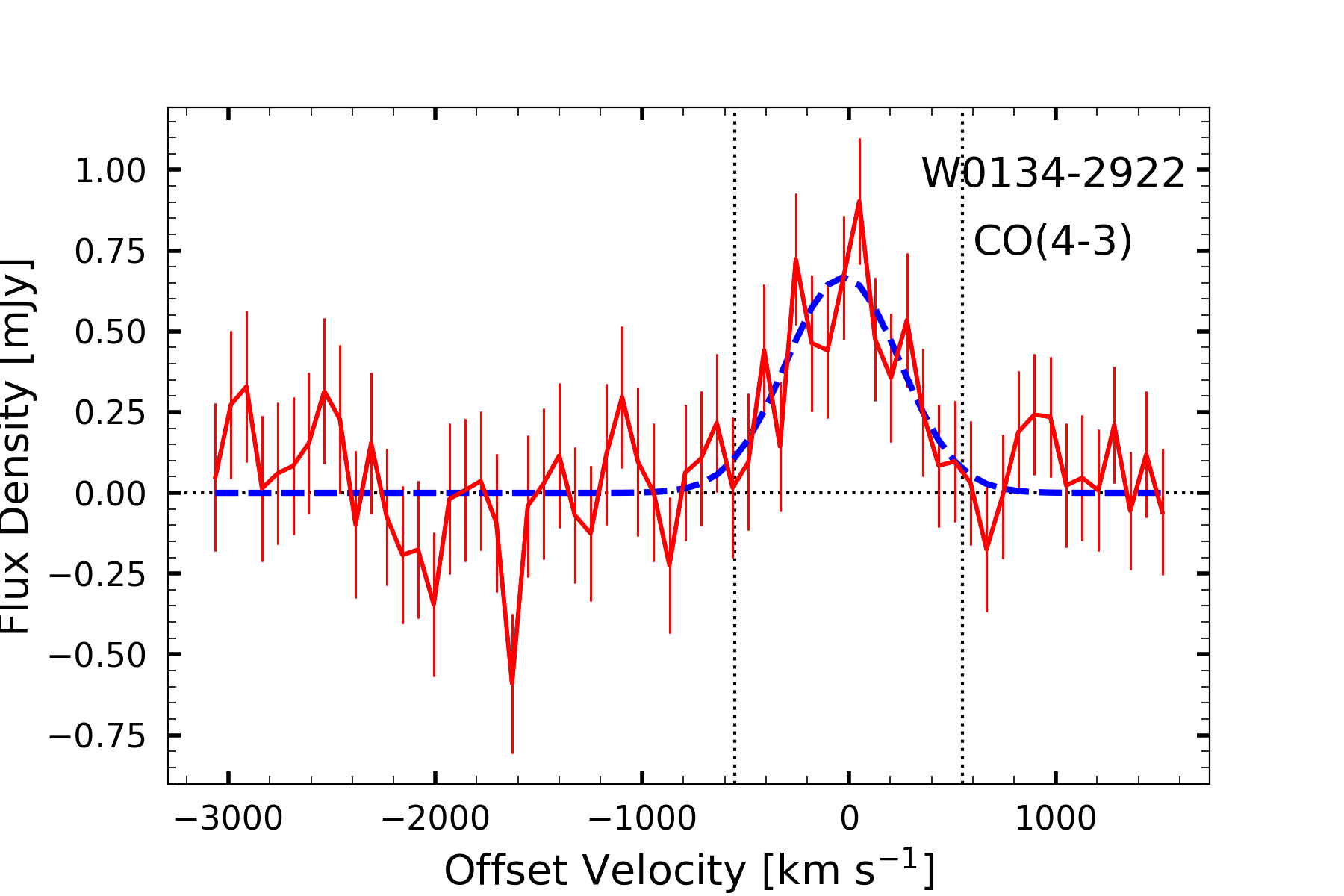}}
 \subfloat{\includegraphics[width=6cm]{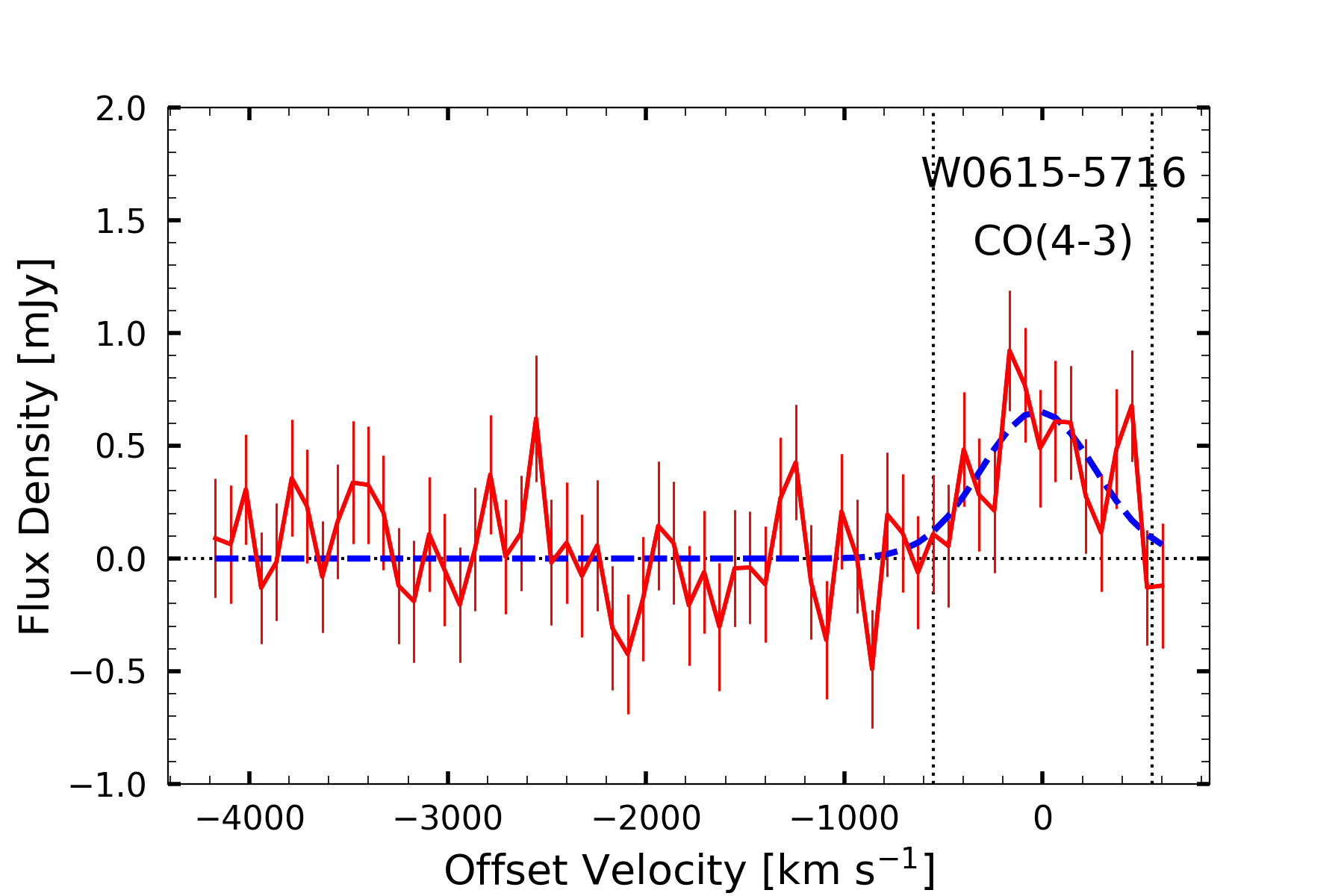}}\\
 \subfloat{\includegraphics[width=6cm]{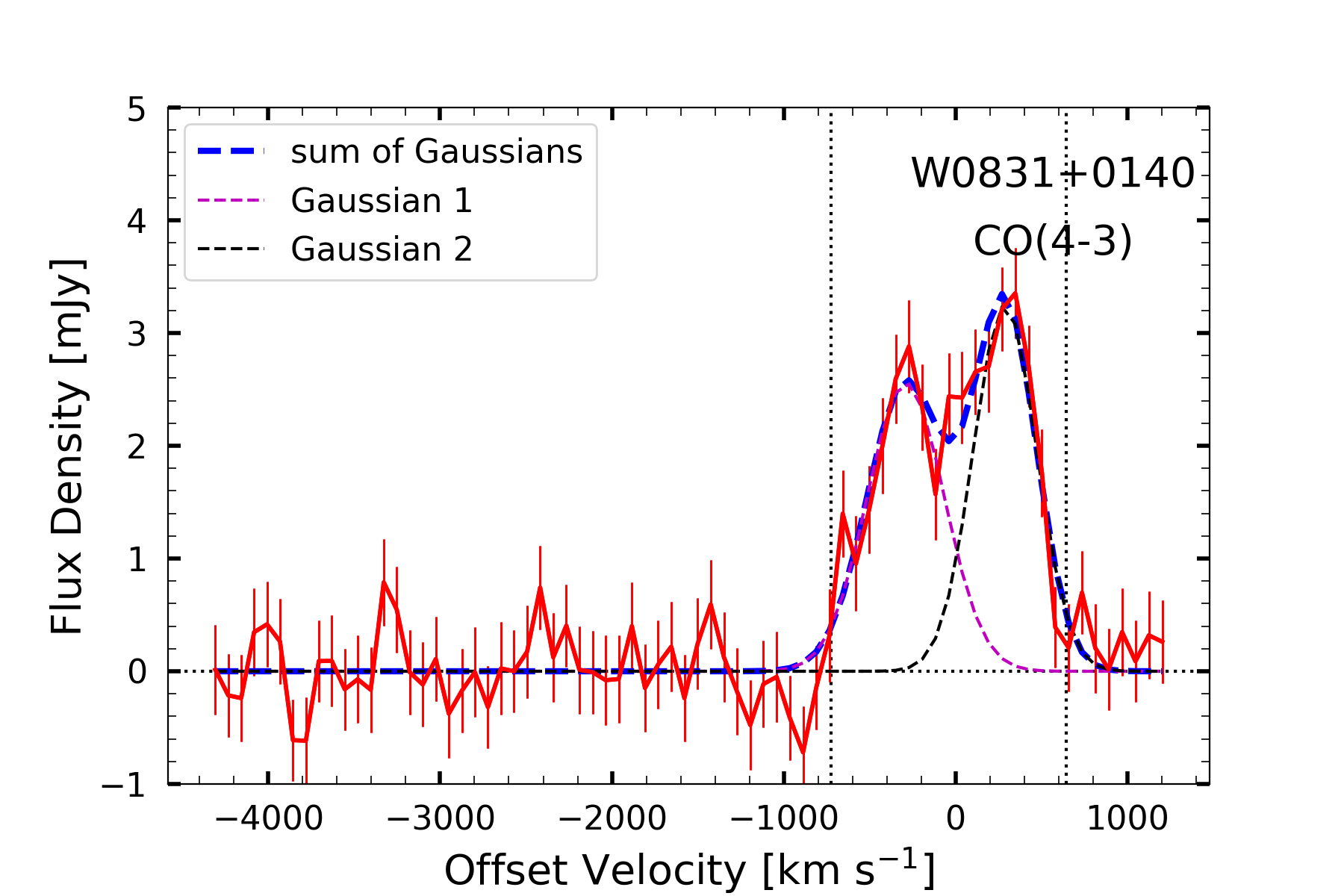}}
 \subfloat{\includegraphics[width=6cm]{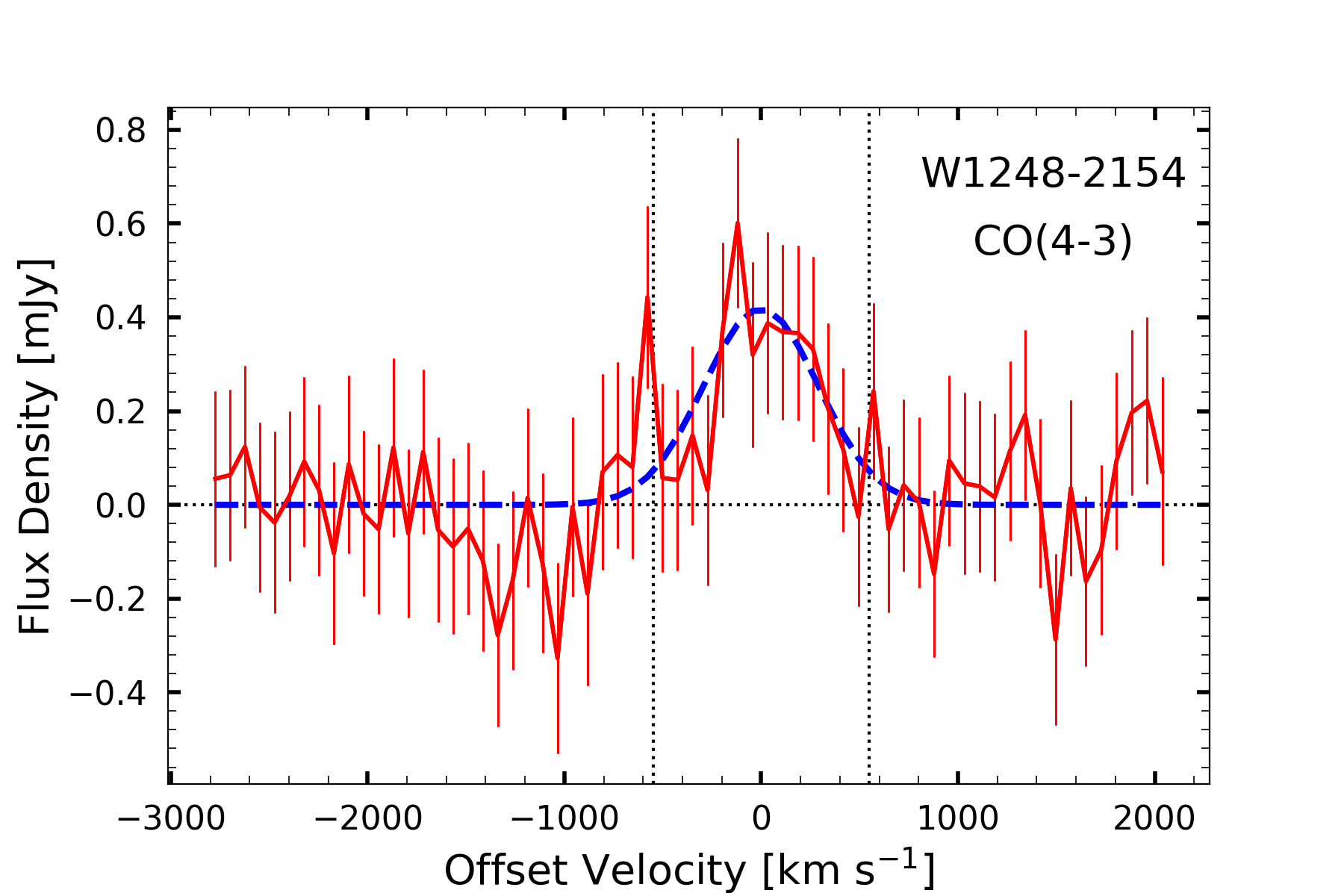}}
 \subfloat{\includegraphics[width=6cm]{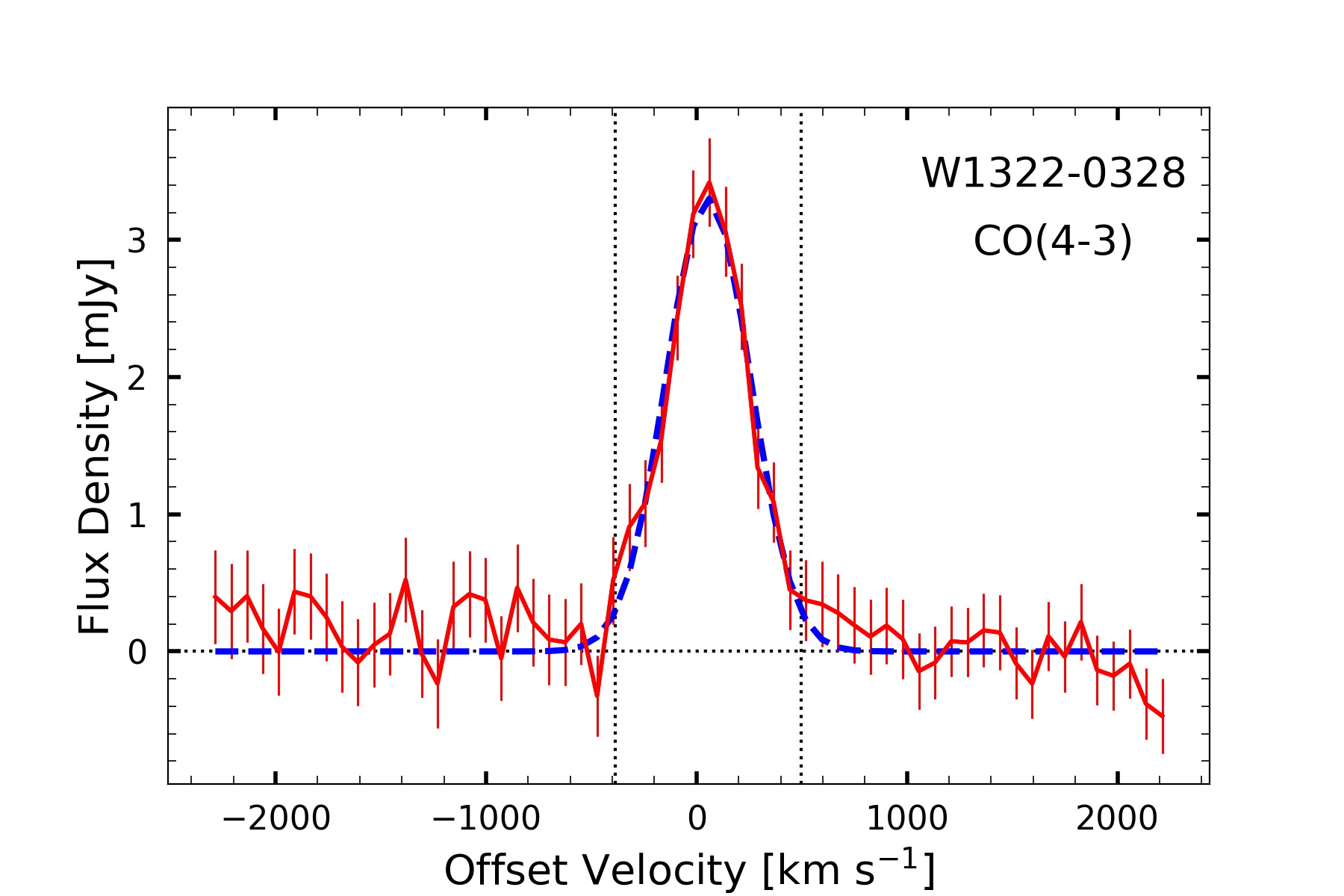}}\\
 \subfloat{\includegraphics[width=6cm]{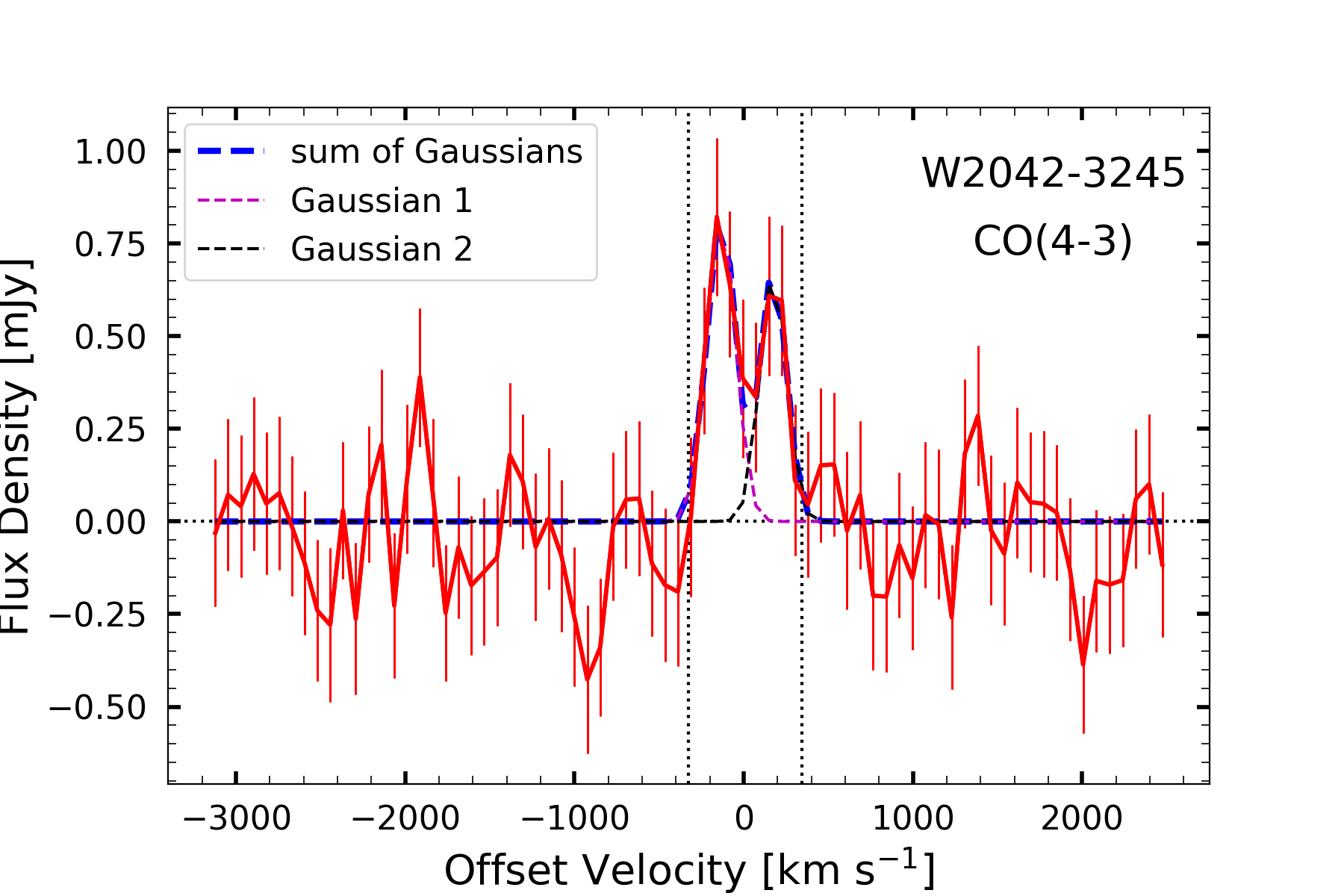}}
 \subfloat{\includegraphics[width=6cm]{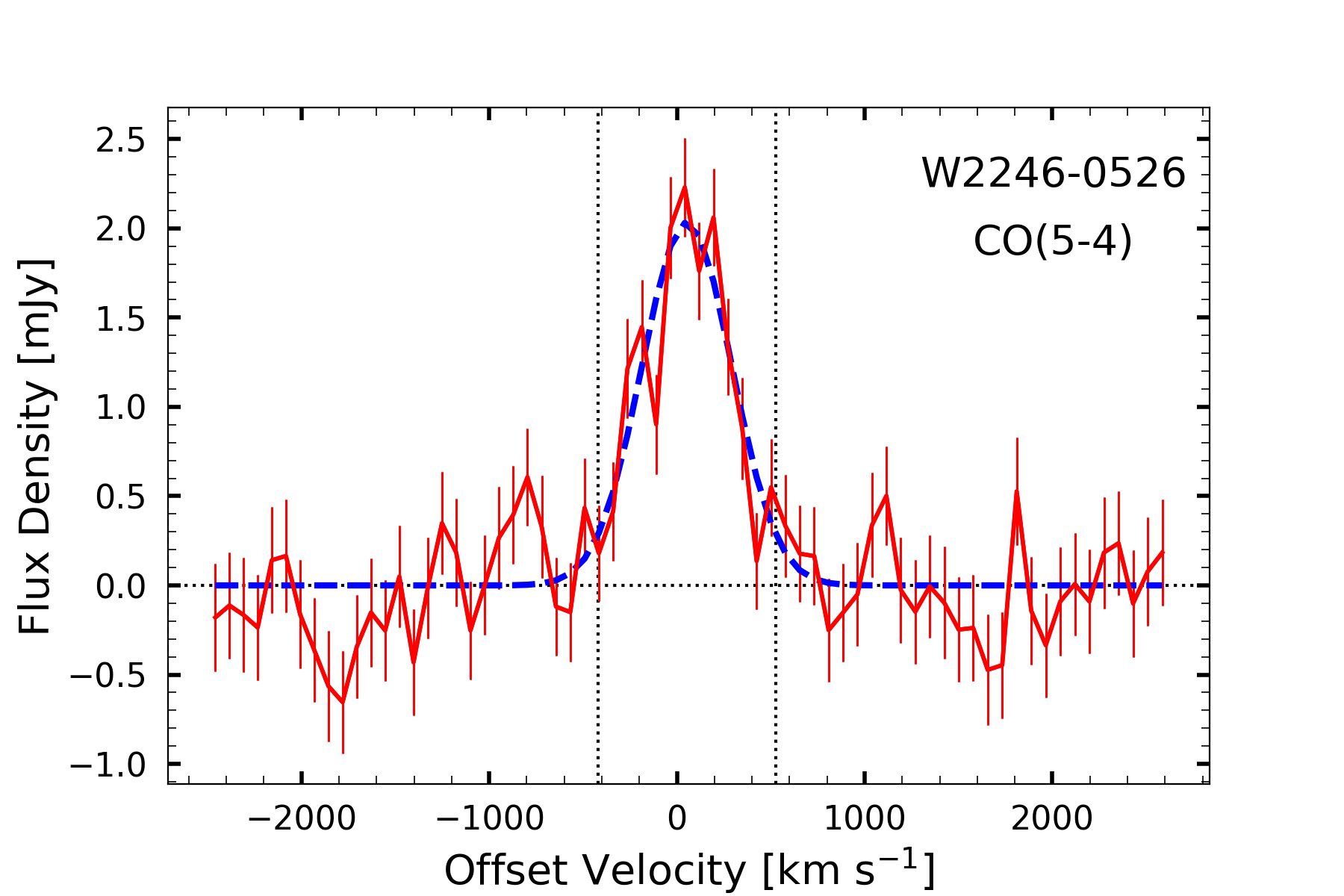}}
 \subfloat{\includegraphics[width=6cm]{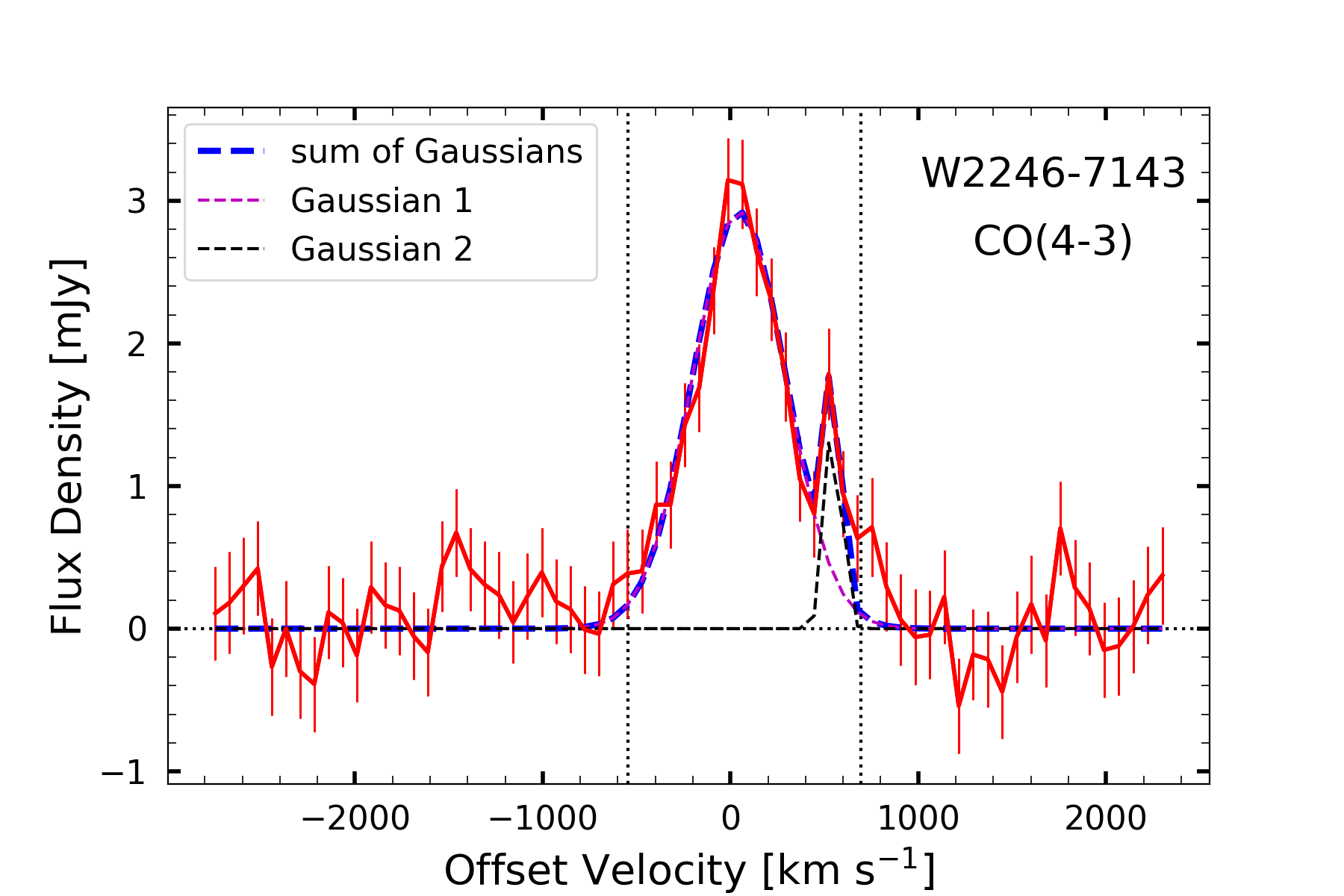}}\\
 \subfloat{\includegraphics[width=6cm]{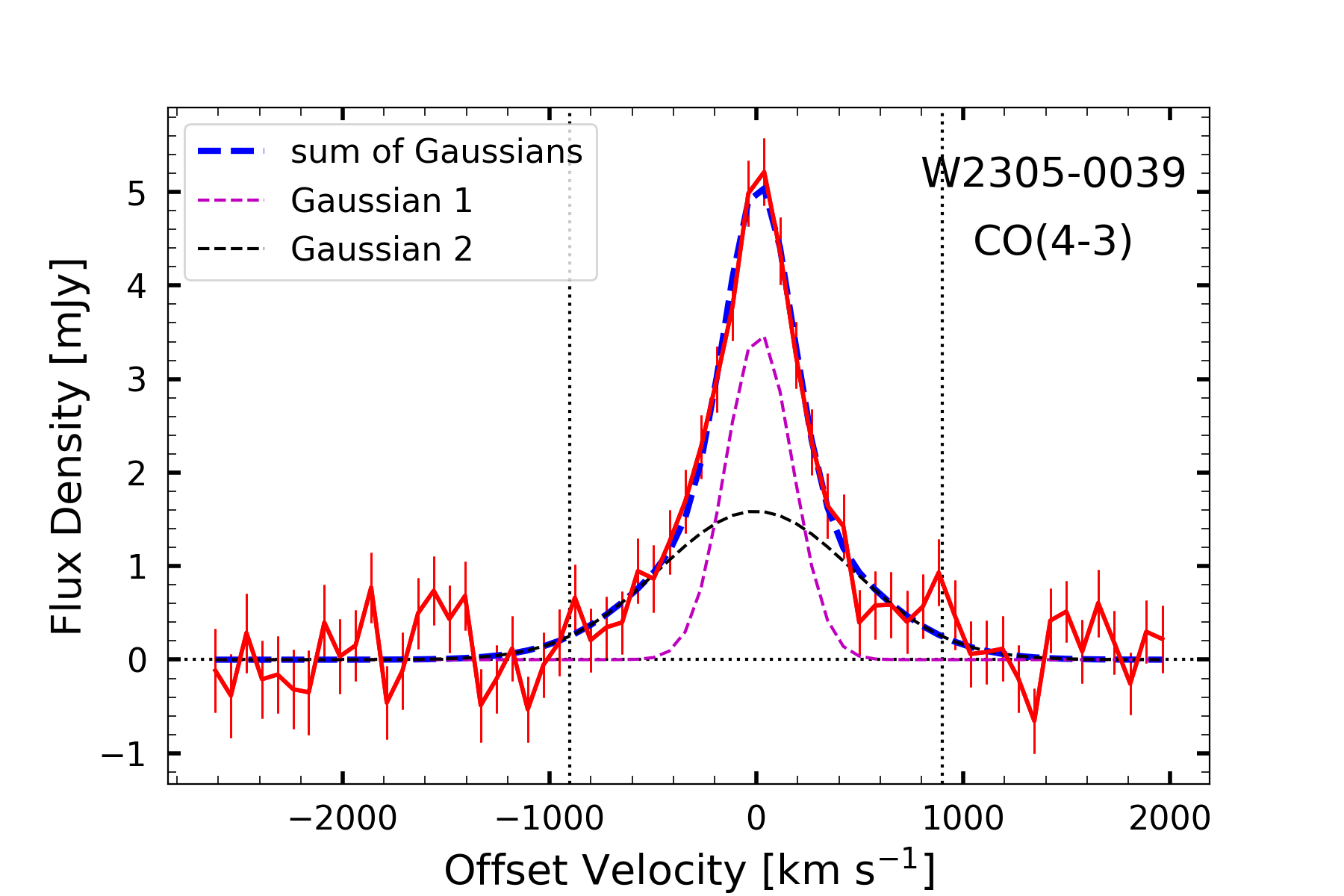}}
 \end{tabular}
 \caption{Continuum-subtracted CO line emission spectra of ten CO-detected Hot DOGs in the sample. The spectra are centred on the CO line. The red lines represent the data points and the red vertical bars represent the standard error per channel. The blue dashed curves are resulting Gaussian fits to the spectra. Four spectra are fit with the double Gaussian functions. The horizontal dashed line shows zero flux and the vertical dashed lines show the extent of the line.}
 \label{fig:spectra}
 \end{figure*}

\begin{figure*}
 \begin{tabular}{ccc}
 \noindent\subfloat{\includegraphics[width=6cm]{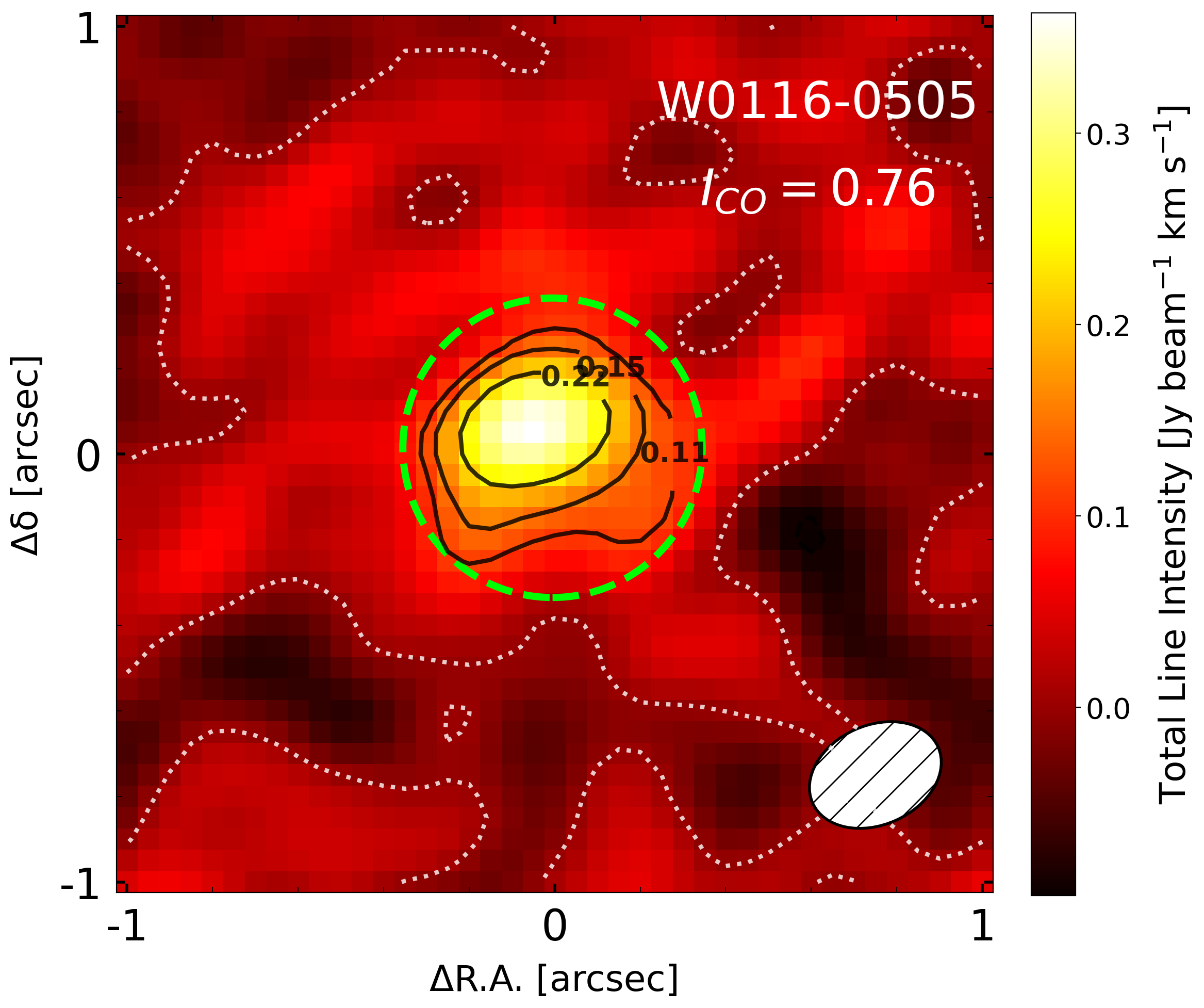}} \hspace{0.2cm}
 \subfloat{\includegraphics[width=6cm]{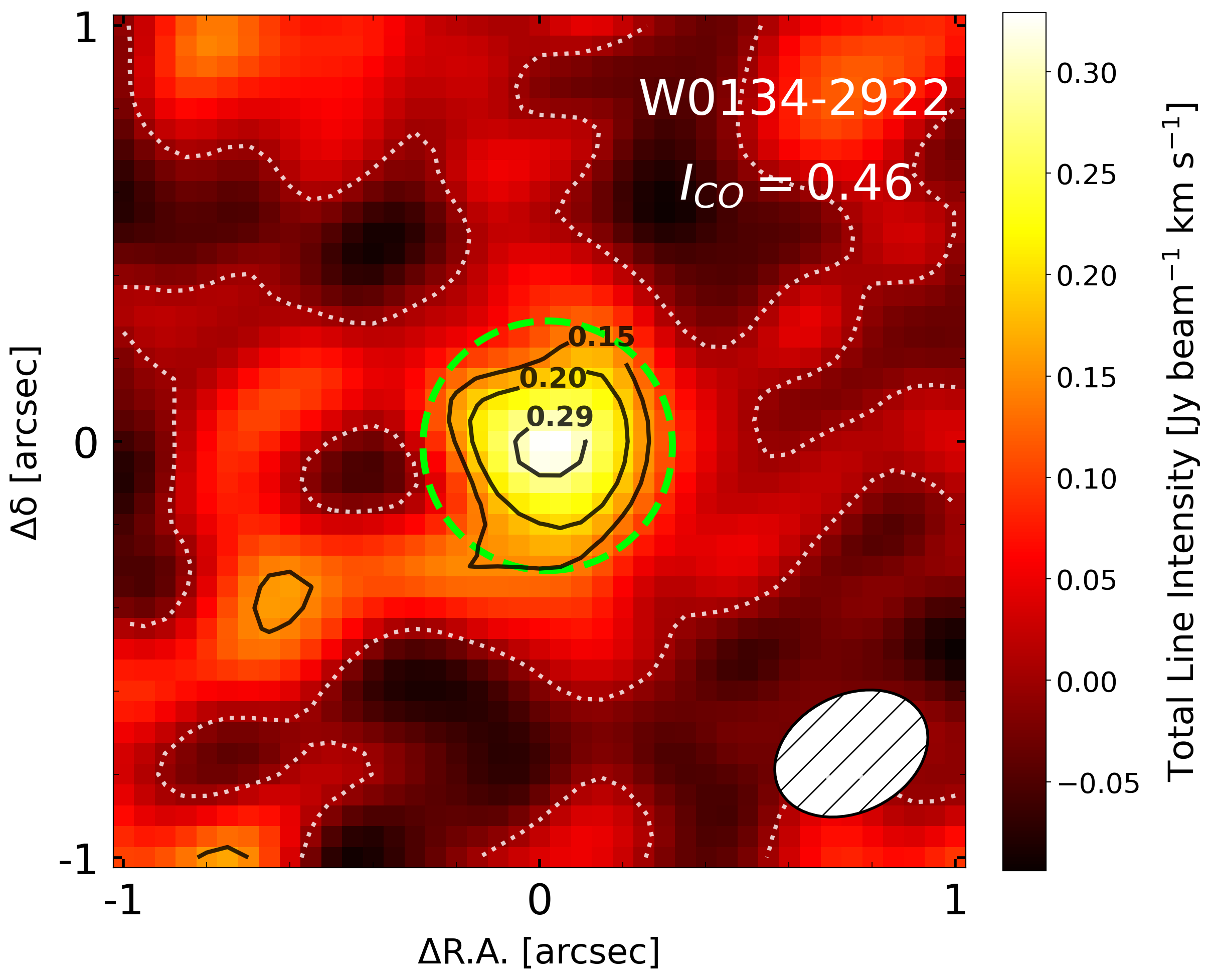}} \hspace{0.2cm}
 \subfloat{\includegraphics[width=6cm]{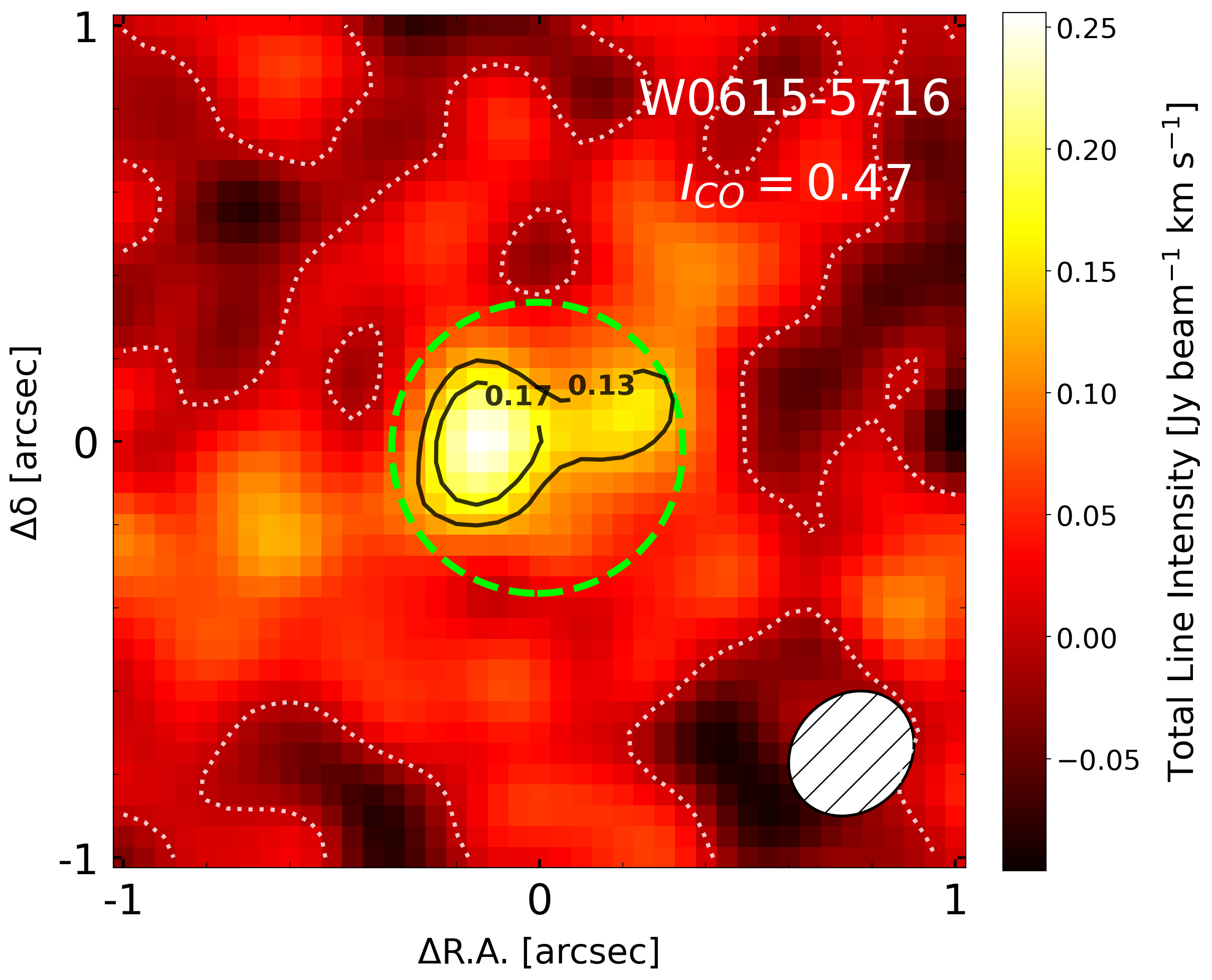}} \\
 \subfloat{\includegraphics[width=6cm]{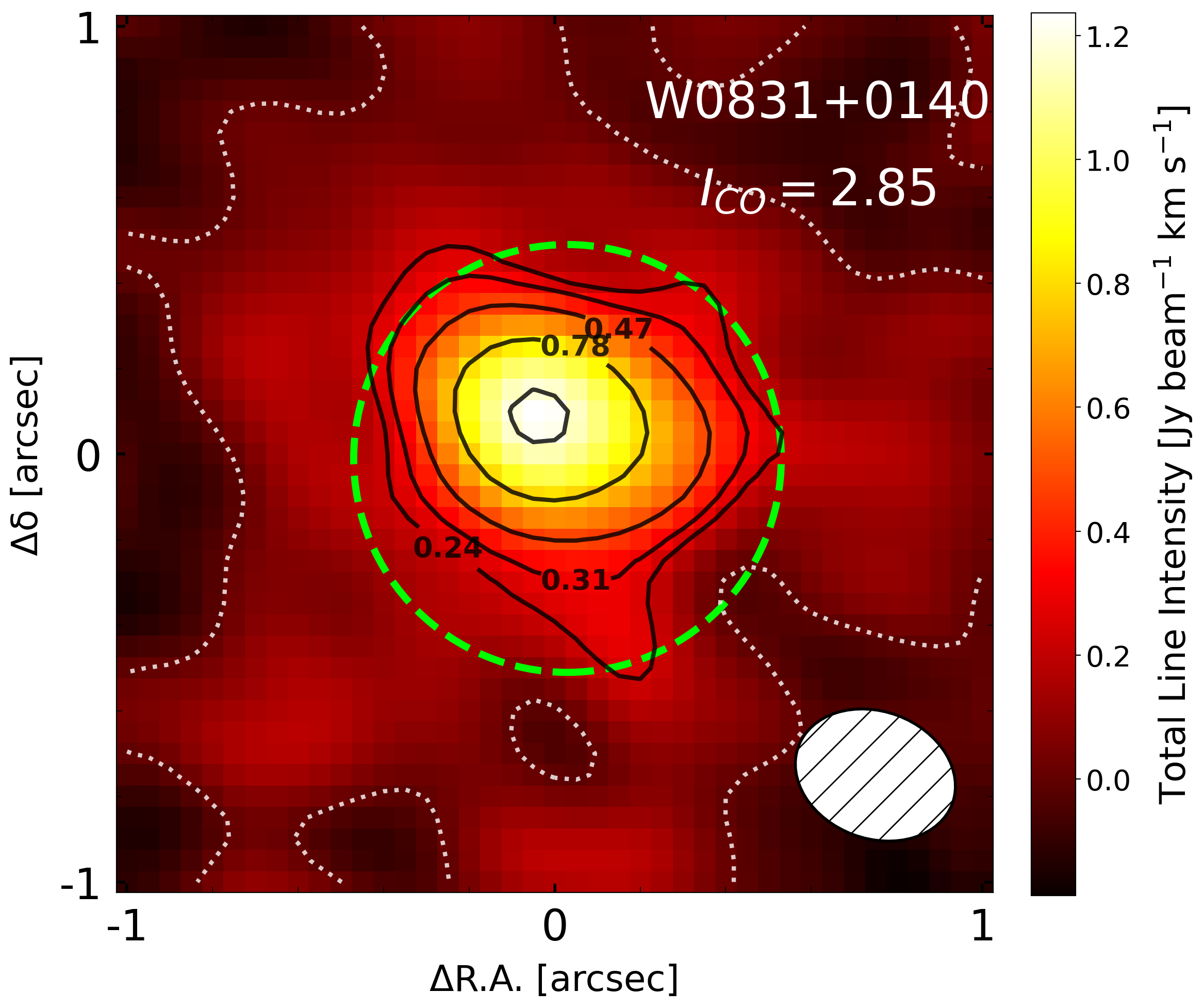}} \hspace{0.2cm}
 \subfloat{\includegraphics[width=6cm]{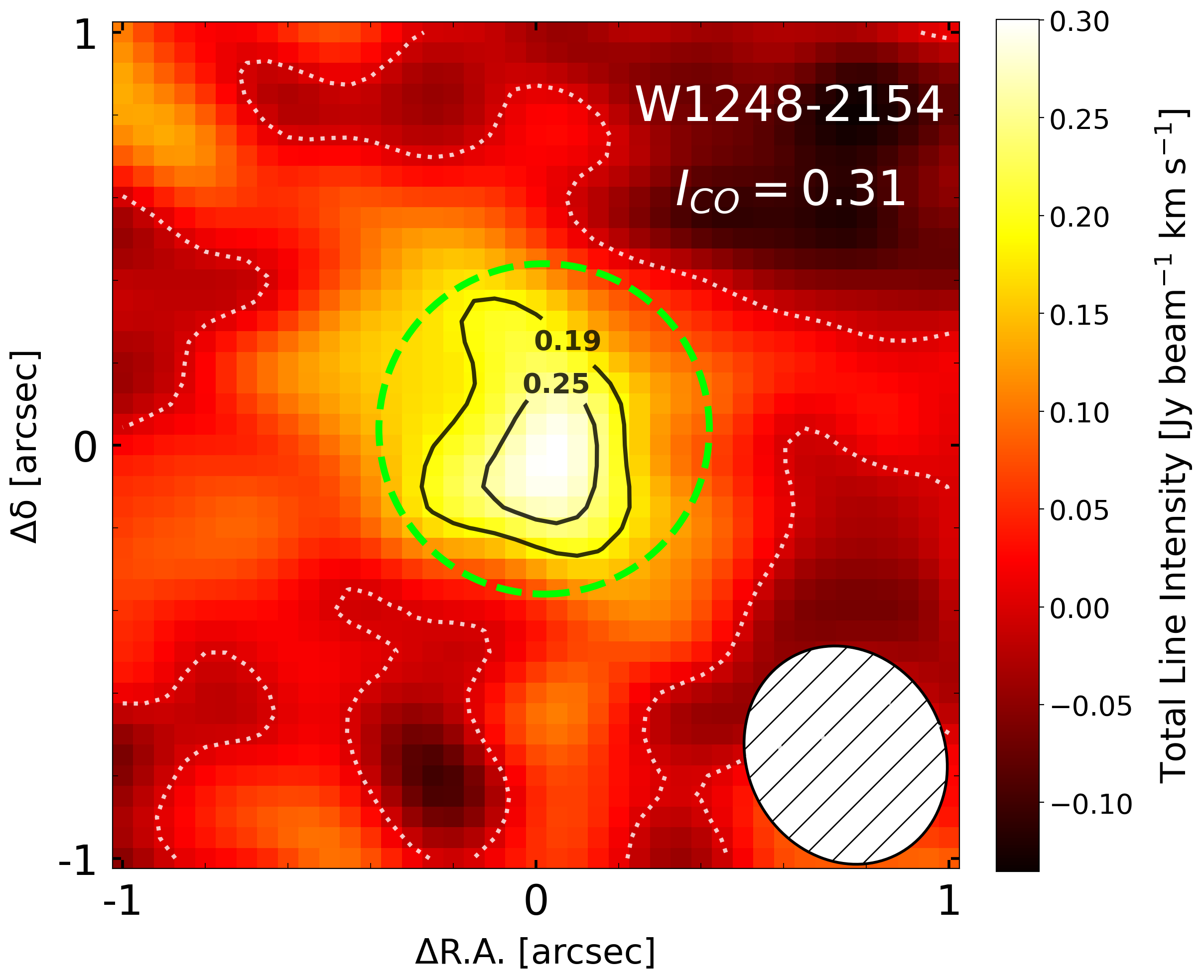}} \hspace{0.2cm}
 \subfloat{\includegraphics[width=6cm]{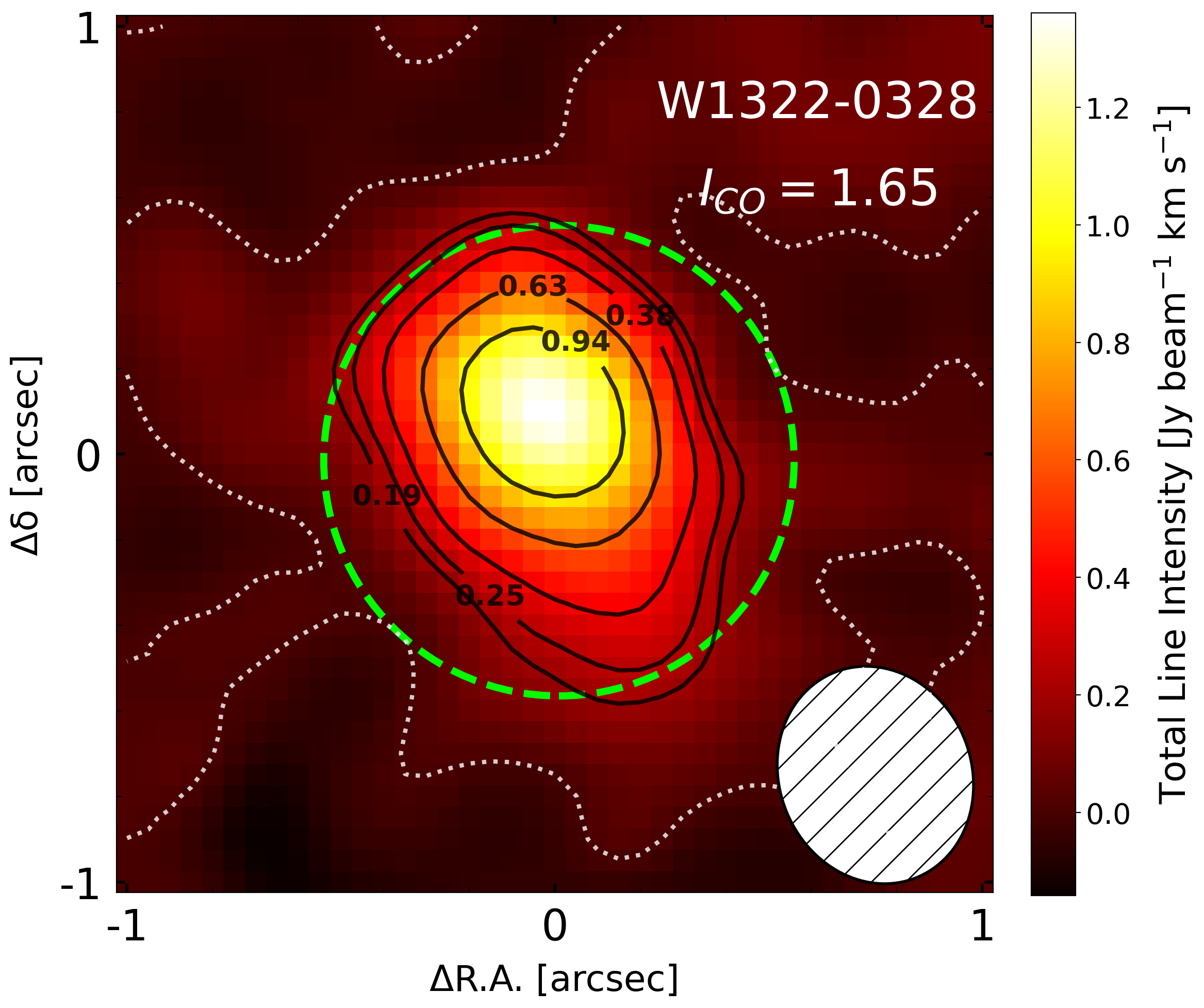}} \\
 \subfloat{\includegraphics[width=6cm]{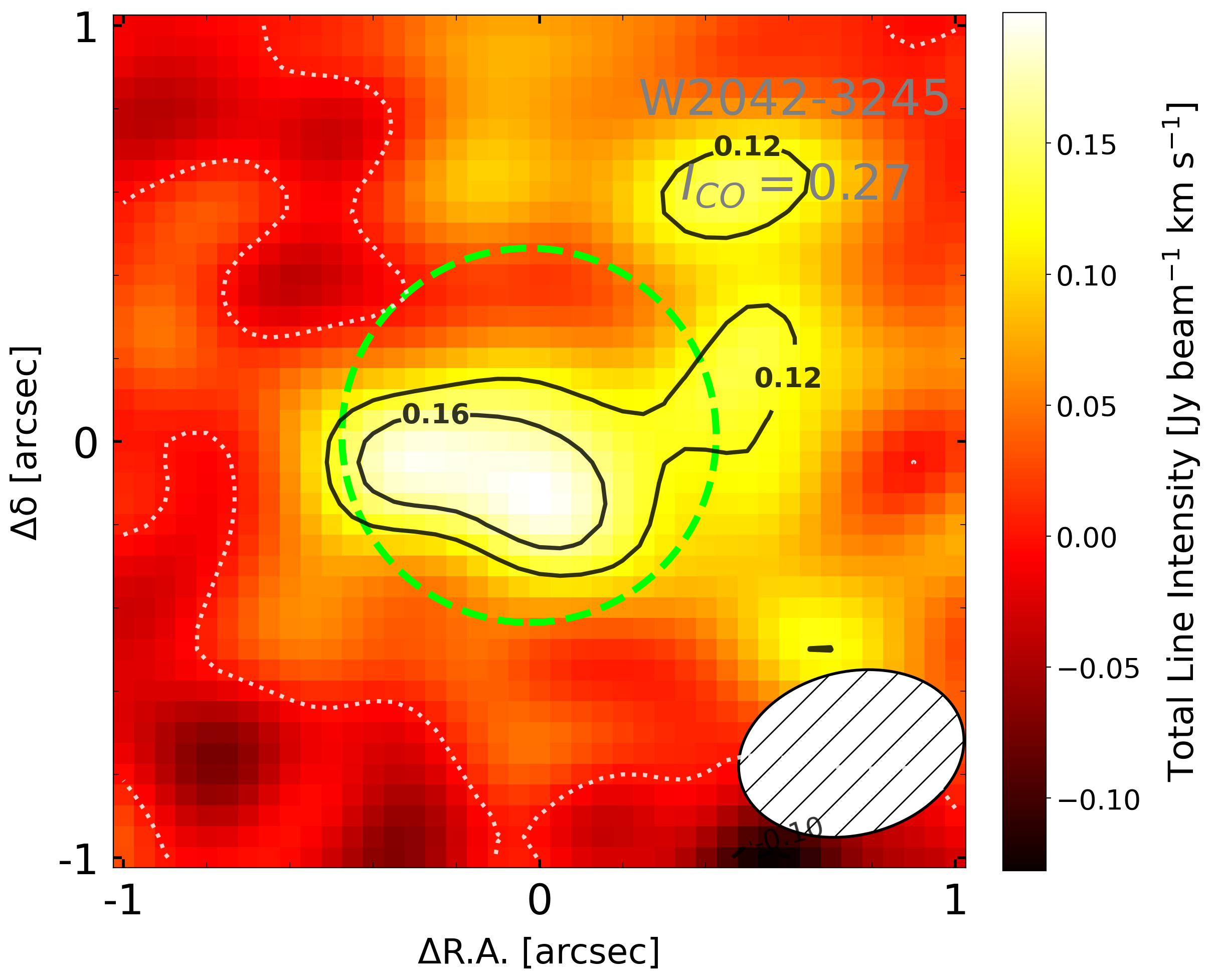}} \hspace{0.2cm}
 \subfloat{\includegraphics[width=6cm]{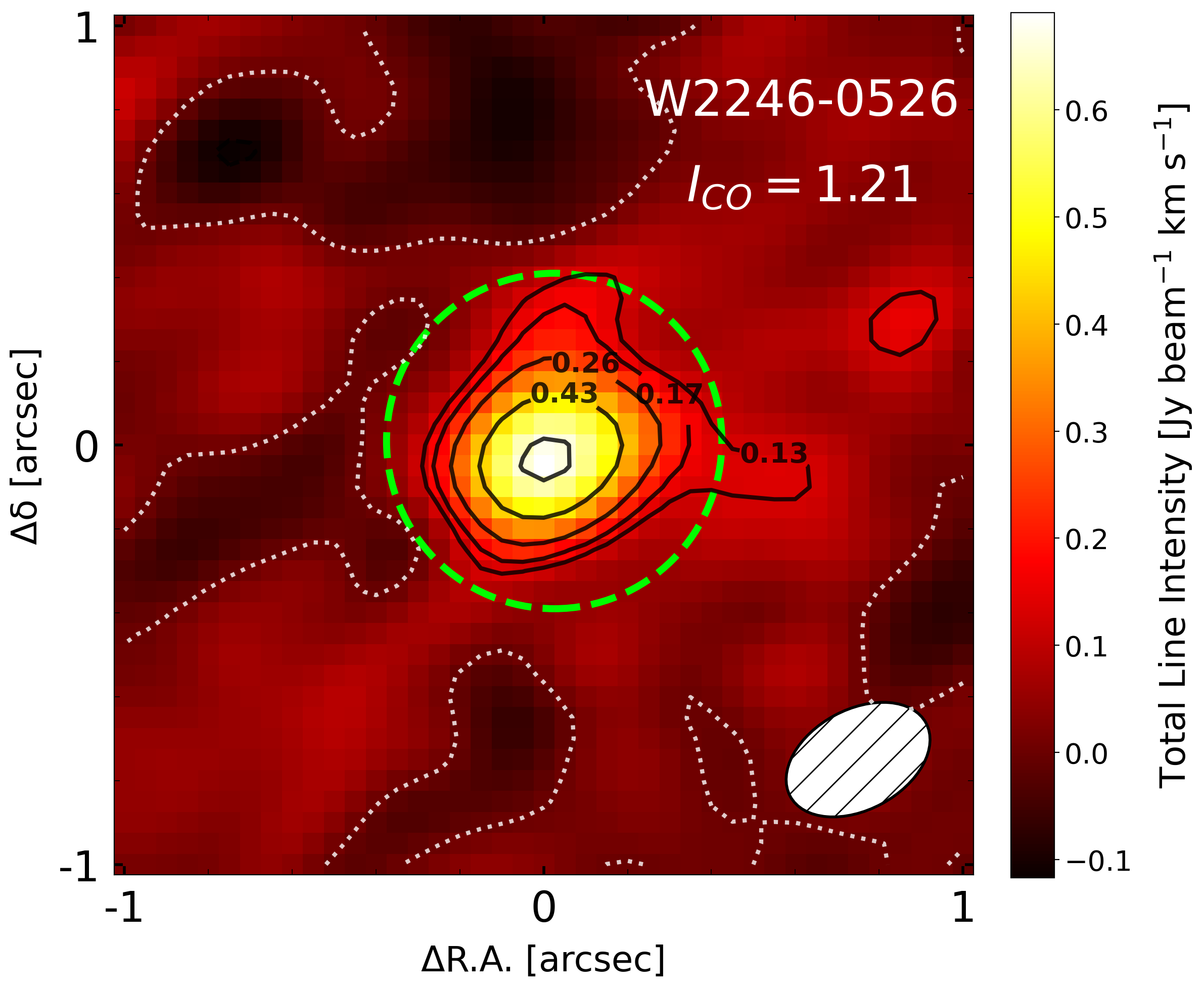}} \hspace{0.2cm}
 \subfloat{\includegraphics[width=6cm]{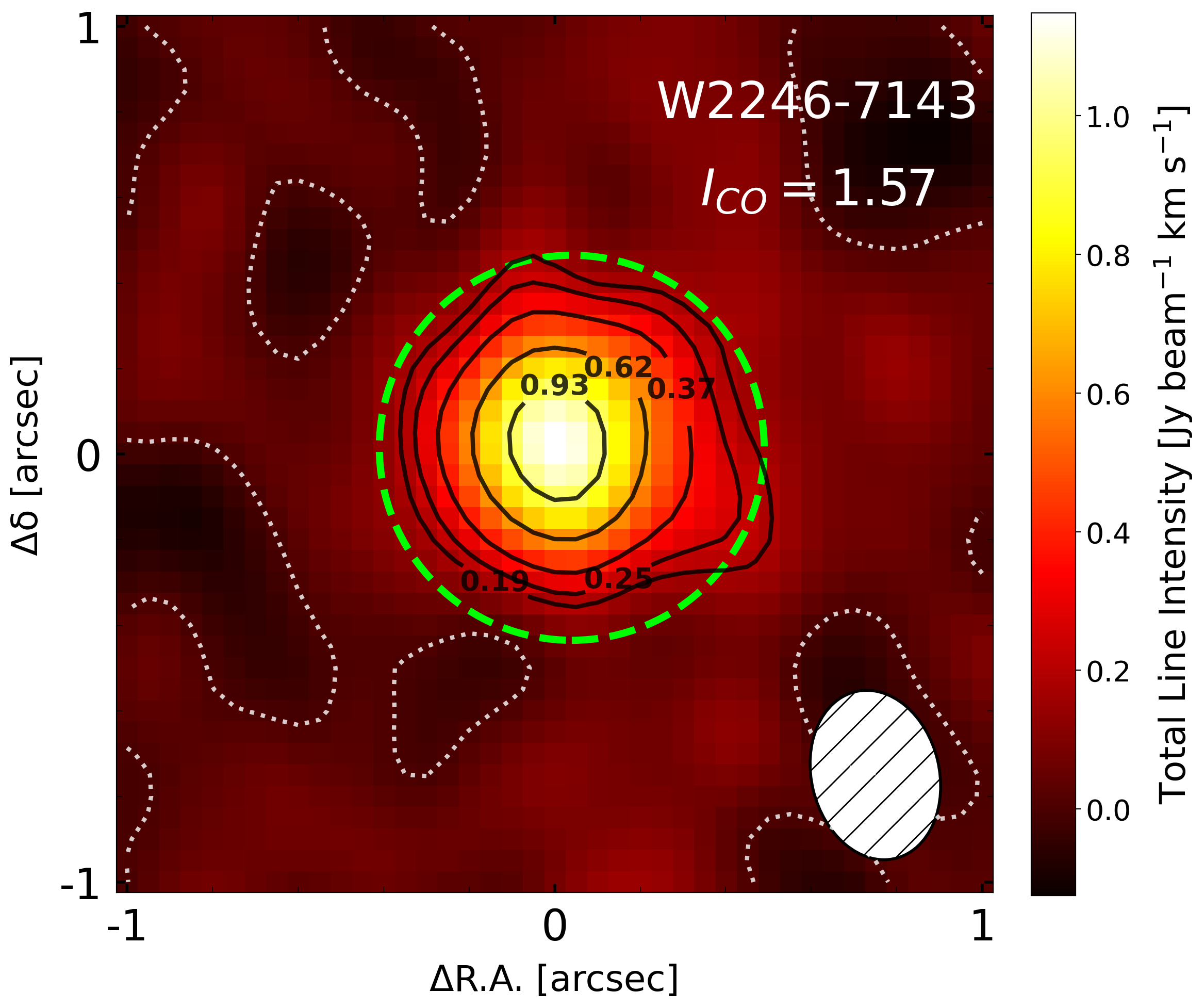}} \\
 \subfloat{\includegraphics[width=6cm]{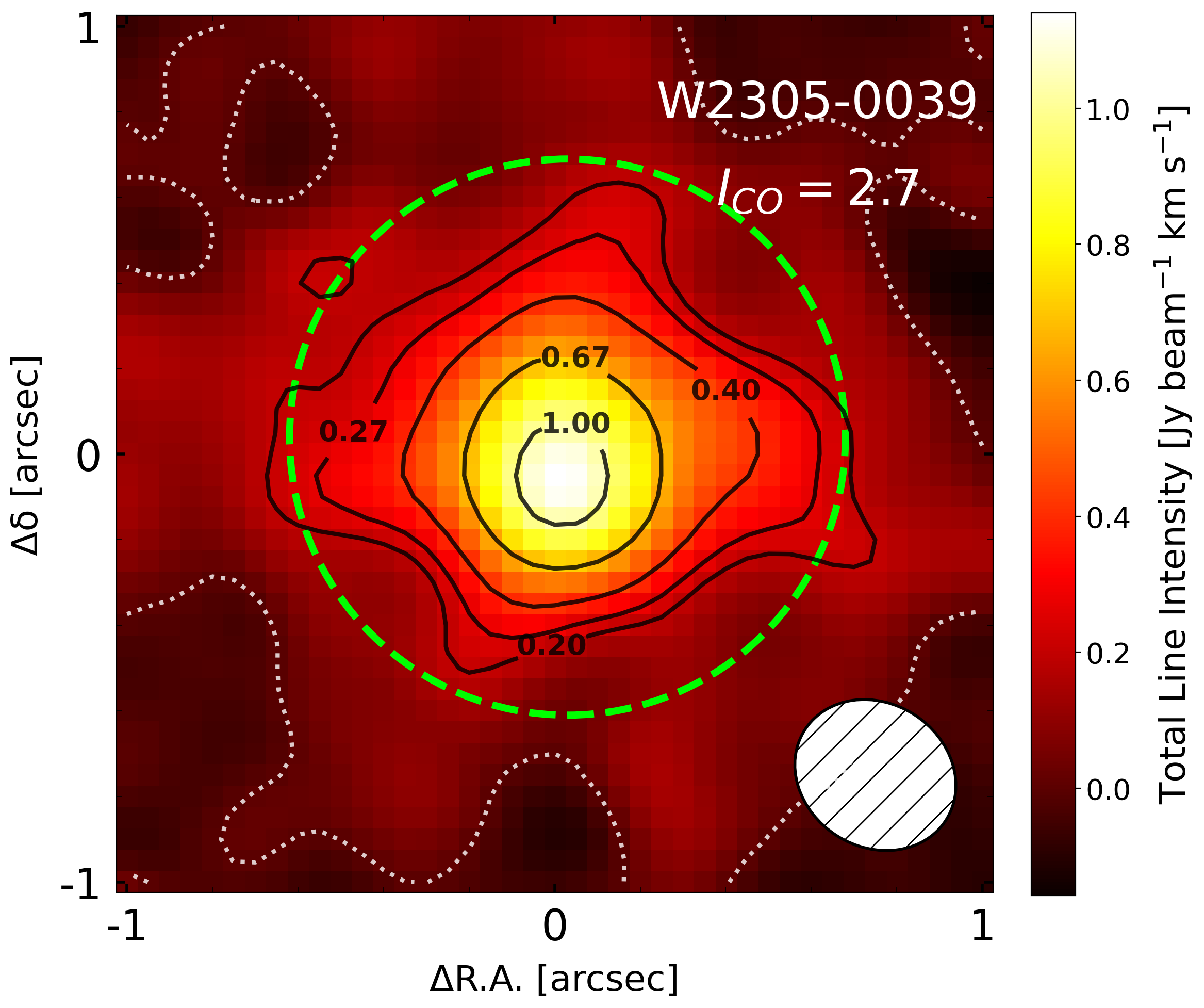}}
 \end{tabular}
 \caption{CO line intensity maps of the ten detected targets. Contours are plotted at 3, 4, 6, 10 and 15$\sigma$. Positive contours are plotted as solid lines, contours of zero intensity are shown with dots, while negative contours ($-$3$\sigma$) are shown with dashed lines. The maps are centred on the peak of the CO emission. The circular aperture used to extract the spectra is shown via the dashed green lines. The velocity-integrated flux (in units of Jy~km~s$^{-1}$) is shown in the upper-right corner of each image. The clean beam is displayed as a white ellipse in the lower-right corner of each image.}
 \label{fig:moment0}
 \end{figure*}

CO was detected in all ten sources with the root mean square (RMS) signal per channel, measured from the line-free regions of the reference spw, of the observations ranging from 0.19--0.39\,mJy\,beam$^{-1}$ (see Table \ref{table:sources}). CO is detected at S/N~$\geq$~3 in six out of ten cases, with W0134--2922, W0615--5716, W1248--2154, and W2042--3245 showing more tentative detections at S/N~$<$~3. Four observations recovered S/N~$\geq$~5, and so 40 per cent of the sources were robustly detected in mid-\textit{J} CO. The spectra and velocity-integrated CO flux (moment-0) images of each target are shown in Fig.~\ref{fig:spectra} and Fig.~\ref{fig:moment0} respectively.\\
The moment-0 maps only show the range of emission within a velocity range defined by the extent of the line, shown in Fig.~\ref{fig:spectra}. The data were spatially smoothed using a normalised 2-D Gaussian kernel ($\mu$~=~0, $\sigma$\textsubscript{x}~=~1, $\sigma$\textsubscript{y}~=~1) which resampled the data in each pixel and convolved structures on scales smaller than the beam. The brightest four sources (S/N~$\geq$~5) show at least some spatially-resolved detail. When performing a 2-D elliptical Gaussian subtraction, representative of the clean beam, an average 38 per cent of the CO flux is associated with extended components at 3$\sigma$ confidence, showing that CO is somewhat extended in these sources. The four fainter sources are unresolved, hence deeper observations may resolve these targets. \\
\indent W2305--0039 contains the most extended CO observation in the sample, with 60 per cent of the flux originating outside a central beam. The source shows a compact, beam-sized, central component, surrounded by a diffuse structure extending out to a radius of 5\,kpc from the centre of the emission (S/N~$\geq$~3). This is substantially larger than the typical 3\,kpc size of \textit{z}~$\approx$~3 galaxies \citep{costantin2023expectations}, consistent with [CII] observations of the most luminous Hot DOGs \citep{diaz2021kinematics}. This extended structure is therefore plausibly associated with the halo of the galaxy.
\subsection{Best-fitting Gaussian Spectral Lines} \label{sec:fits}
To investigate the most appropriate fits to the spectra (Fig.~\ref{fig:spectra}), we used the Bayesian information criterion (BIC) of single, double, and triple Gaussian fits to the spectra. Five sources are best fit with a single Gaussian and four sources are better fit with double Gaussian functions (see Table~\ref{table:FWHM}). The BIC value, along with the relevant Gaussian properties of each fit is detailed in Table~\ref{table:BIC}. We chose to investigate BIC over the Akaike information criteria (AIC), as BIC more harshly penalises overfitting. The fits chosen for these spectra are therefore more conservative. The lowest BIC score (in this case the most negative number) is considered the most favoured model. Models are compared through their difference in BIC value ($\Delta$BIC). Differences of 0--2, 2--6, 6--10, and $\geq$~10 are considered weak evidence, positive evidence, strong evidence, and very strong evidence in favour of the model with the lower BIC value respectively \citep{neath2012bayesian}.\\
\indent Nearly half of the sources in this sample plausibly contain multiple CO components, with very strong evidence ($\Delta$BIC~$\geq$~10) of two Gaussian components: W0831+0140, W2042--3245, W2246--7143, and W2305--0039. In addition, triple Gaussians were investigated in W0831+0140 and W2246--7143. The double Gaussian fit is weakly preferred ($\Delta$BIC~=~1.9) over a triple Gaussian in W0831+0140. In W2246--7143, \texttt{LMFIT} was not able to estimate the accuracy of the triple Gaussian parameters, which is often a consequence of overfitting. While deeper observations may reveal additional CO components, we use the double Gaussian fits in both cases. Using the most statistically significant Gaussian fits, we list the properties of the CO lines in Table~\ref{table:FWHM}.\\
\begin{table*}
    \centering
    \caption{Properties of CO lines observed in the sample. (1) Peak flux density; (2) Full-width at half-maximum; (3) Velocity-integrated flux of the CO line; (4) Velocity separation between Gaussian components in cases where a double Gaussian fit was applied; (5) Difference in Bayesian Information Criterion (BIC) values of double and single Gaussian fits; (6) The integrated flux in the residuals after a beam-shaped signal matched to the peak of the image has been subtracted.}
    \label{table:FWHM}
    \setlength\extrarowheight{2pt}
    \begin{tabular}{c c c c c c c c c c}
        \hline
        Source & \textit{S}\textsubscript{peak}  & FWHM & \textit{I}\textsubscript{CO}  & $\Delta$\textit{v} & $\Delta$BIC & \textit{f}\textsubscript{ext} \\
        & [mJy] &  [km\,s$^{-1}$] & [Jy\,km\,s$^{-1}$] & [km\,s$^{-1}$] & & \\
        & (1) & (2) & (3) & (4) & (5) & (6)\\
        \hline
        W0116--0505 & 1.18 $\pm$ 0.10 & 611 $\pm$ 59 & 0.76 $\pm$ 0.07 &\dots & 0.2 & 40\% \\
        W0134--2922 & 0.67 $\pm$ 0.08 & 647 $\pm$ 91 & 0.46 $\pm$ 0.06  &\dots & \dots & 33\% \\
        W0615--5716 & 0.65 $\pm$ 0.11 & 679 $\pm$ 134 & 0.47 $\pm$ 0.09 & \dots & \dots & 34\% \\
        W0831+0140 & [2.56 $\pm$ 0.20], [3.25 $\pm$ 0.23] & [528 $\pm$ 80], [434 $\pm$ 50] & 2.85 $\pm$ 0.17 & 580 & 11.8 & 53\% \\
        W1248--2154 & 0.42 $\pm$ 0.06 & 688 $\pm$ 117 & 0.31 $\pm$ 0.05 & \dots & \dots & 15\% \\
        W1322--0328 & 3.30 $\pm$ 0.15 & 473 $\pm$ 23 & 1.65 $\pm$ 0.08 & \dots & \dots & 20\% \\
        W2042--3245 & [0.83 $\pm$ 0.16], [0.68 $\pm$ 0.16] & [199 $\pm$ 50], [186 $\pm$ 57] & 0.27 $\pm$ 0.04 & 310 & 18.1 & 45\% \\
        W2246--0526 & 2.03 $\pm$ 0.14 & 562 $\pm$ 46 & 1.21 $\pm$ 0.09 &\dots & \dots & 40\% \\
        W2246--7143 & [2.92$\pm$ 0.14], [1.54 $\pm$ 0.60] & [584 $\pm$ 38], [100 $\pm$ 52] & 1.57 $\pm$ 0.12 & 500 & 12.5 &  41\% \\
        W2305--0039 & [3.50 $\pm$ 0.50], [1.59 $\pm$ 0.51] & [379 $\pm$ 51], [1100 $\pm$ 188] & 2.70 $\pm$ 0.19 & 20 & 15.8 & 60\% \\
        \hline
    \end{tabular}
\end{table*}
\indent \textbf{W0831+0140:} The two Gaussian components with widths of 434~\&~528\,km~\,s$^{-1}$ are separated in velocity space by $\Delta$\textit{v}~=~580\,km\,s$^{-1}$. This separation suggests that either outflowing CO gas is present, or there are two (or three: Fig.~\ref{fig:W2246-7143_fits}) galaxies merging. An [OIII] outflow is already confirmed in this source \citep{finnerty2020fast}, and a similar merging conclusion was reached via analysis of [CII] images, with a third galaxy similarly plausibly lying just beneath the detection threshold \citep{diaz2021kinematics}. However, the double-peaked profile could imply the presence of a rotating disc. \\
\indent \textbf{W2246--7143:} The functions fit to the spectrum consist of a broad (584\,km\,s$^{-1}$) component and a narrow (100\,km\,s$^{-1}$) component. This could plausibly be two different regions of CO. In the case where three Gaussian functions were tentatively fit to the spectrum of this source (Fig.~\ref{fig:W2246-7143_fits}), the components have varying FWHM (364, 980, 48)\,km\,s$^{-1}$. The broad 980\,km\,s$^{-1}$ component could be indicative of a molecular outflow.\\
\indent \textbf{W2305--0039:} The fit consists of a narrow component (379\,km\,s$^{-1}$) and a broad component (1100\,km\,s$^{-1}$) with a negligible ($\Delta$\textit{v}~=~20\,km\,s$^{-1}$) velocity separation. The broad component is plausibly an outflow of molecular gas, in addition to the [OIII] outflow in this source \citep{finnerty2020fast}.\\
\indent Nine of the ten sources observed in CO(4--3) and CO(5--4) show at least one broad component with FWHM~$\geq$~400\,km\,s$^{-1}$. The exception to this is W2042--3245, with two FWHM~$\approx$~200\,km\,s$^{-1}$ components. W2305--0039 is the only source with a very broad FWHM~$>$~1000\,km\,s$^{-1}$ component, consistent with a large, bipolar outflow of molecular gas. The general presence of broad CO components suggests that the molecular ISM in most cases is highly turbulent and disturbed; optically-selected \textit{z}~$\approx$~6 quasars show a similar range of mid-\textit{J} CO linewidths 160--860\,km\,s$^{-1}$ \citep{wang2010molecular}, though 60 per cent of these optically-selected quasars show narrow (FWHM~$\leq$~400\,km\,s$^{-1}$) CO lines, showing that line broadening by AGN can vary significantly from source to source, and CO emission in the ISM of optically-selected z~$\approx$~6 quasars is not consistently broad/turbulent as it seems to be in the most luminous Hot DOGs.
\subsection{Kinematic Maps} \label{sec:widths}
\begin{figure*}
 \begin{tabular}{ccc}
 \noindent\subfloat{\includegraphics[width=6cm]{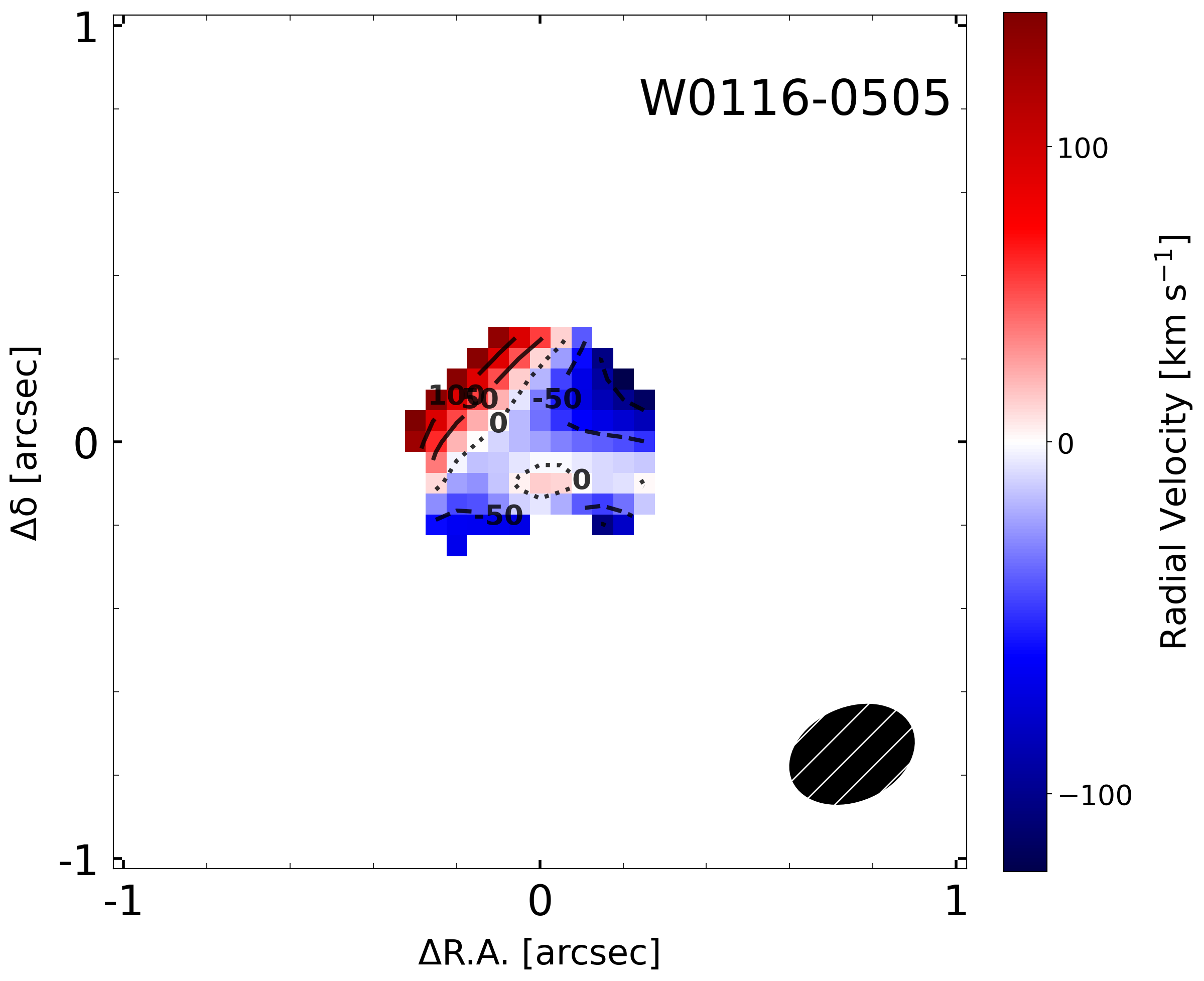}} \hspace{0.2cm}
 \subfloat{\includegraphics[width=6cm]{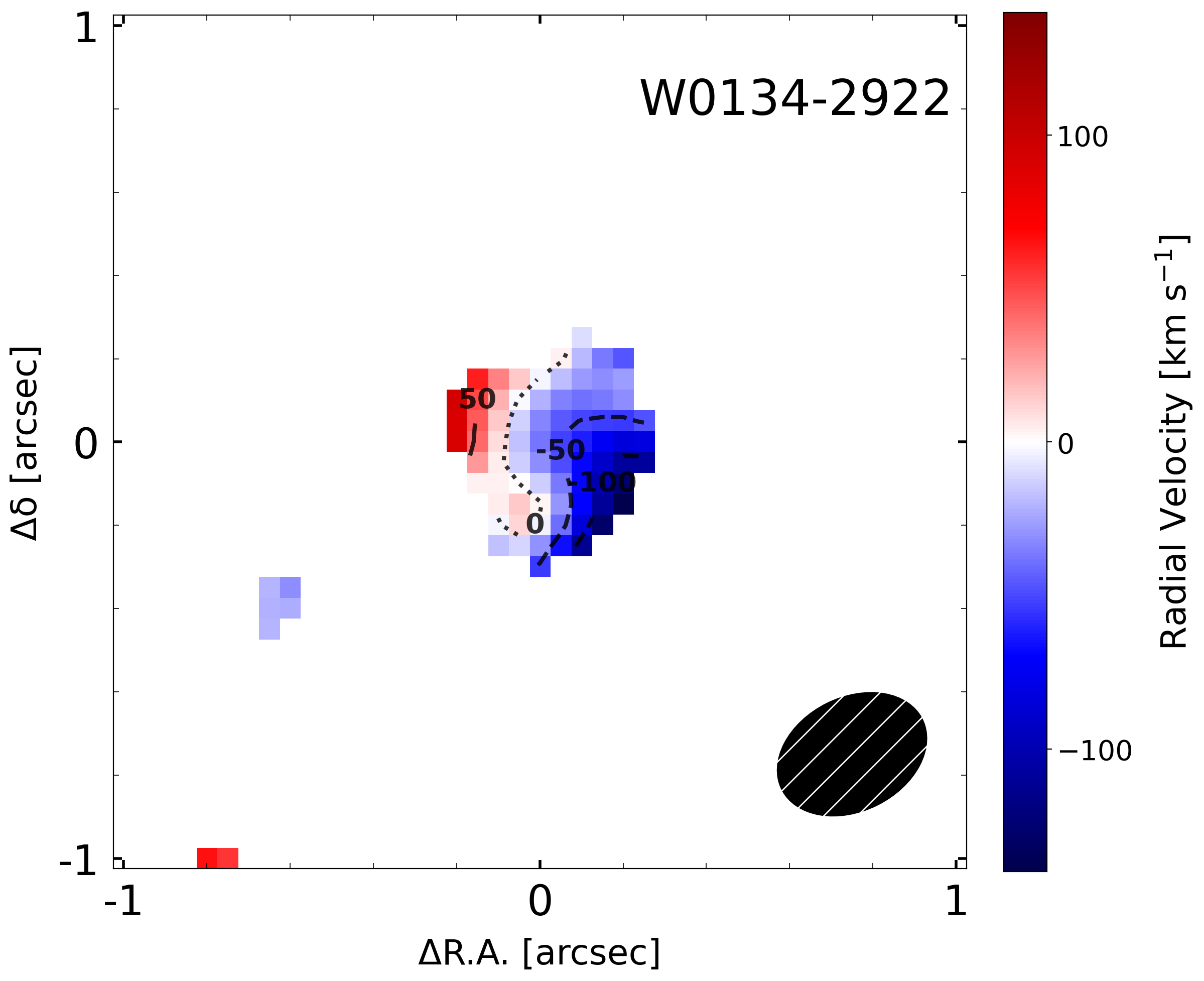}} \hspace{0.2cm}
 \subfloat{\includegraphics[width=6cm]{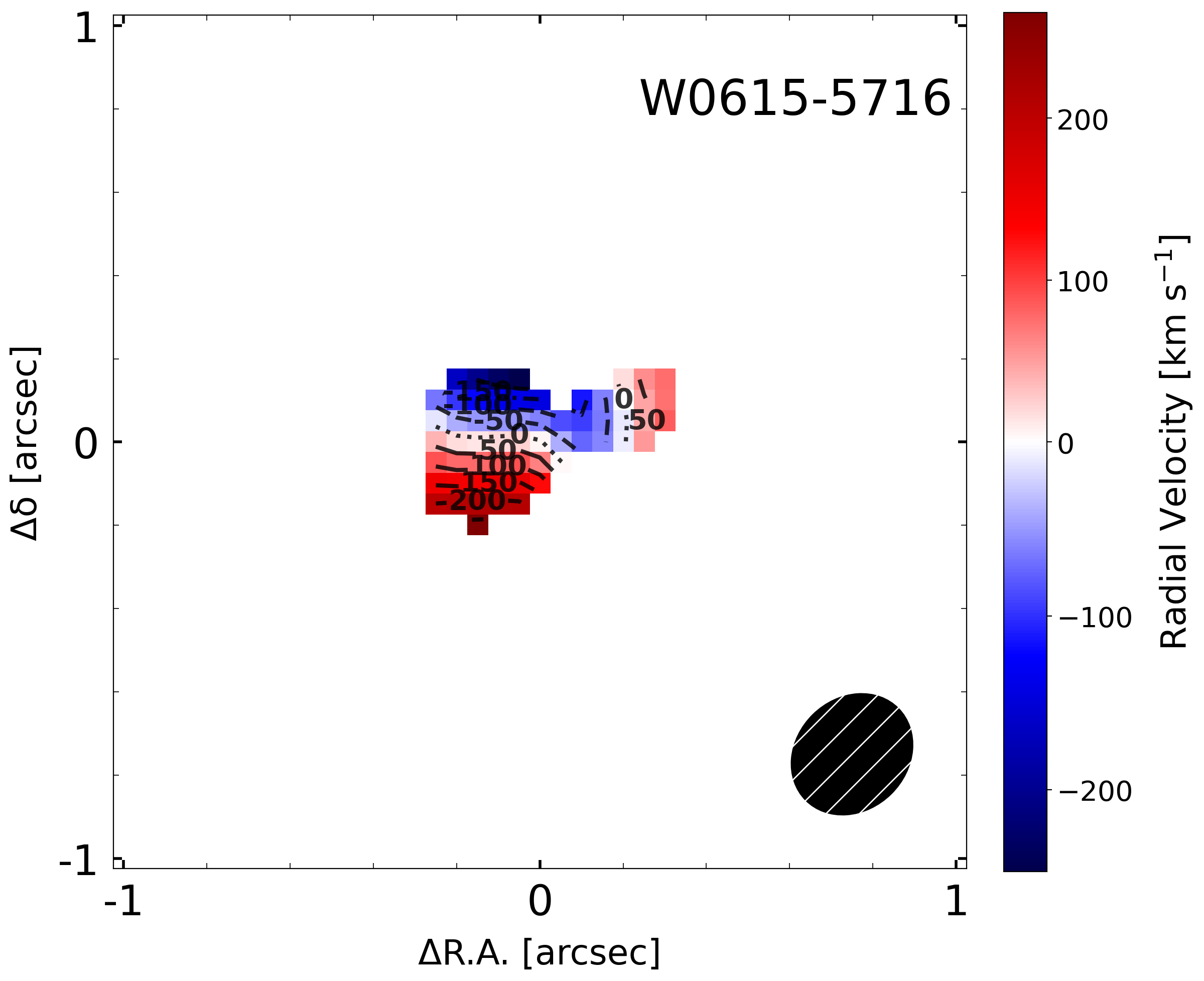}} \\
 \subfloat{\includegraphics[width=6cm]{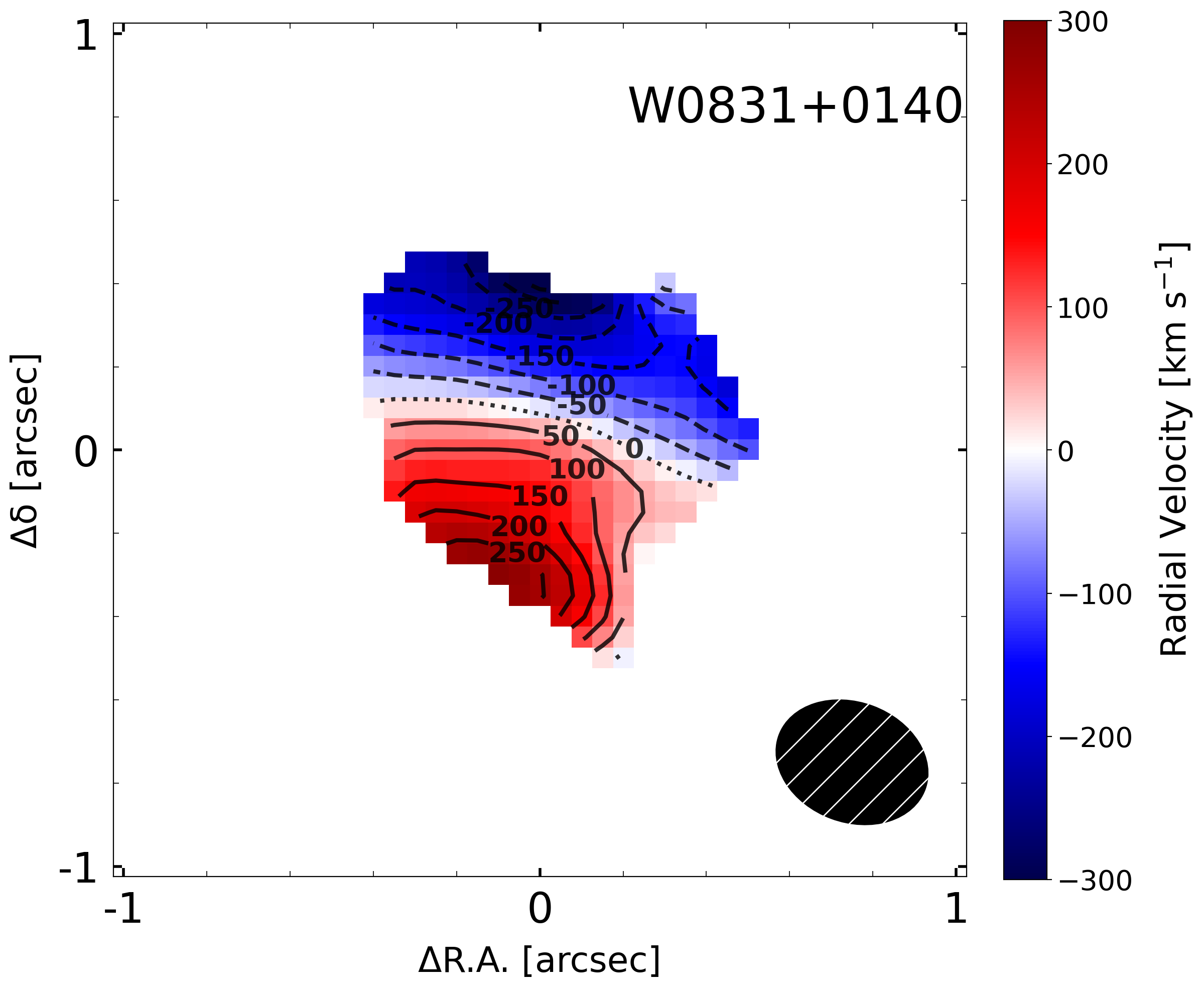}} \hspace{0.2cm}
 \subfloat{\includegraphics[width=6cm]{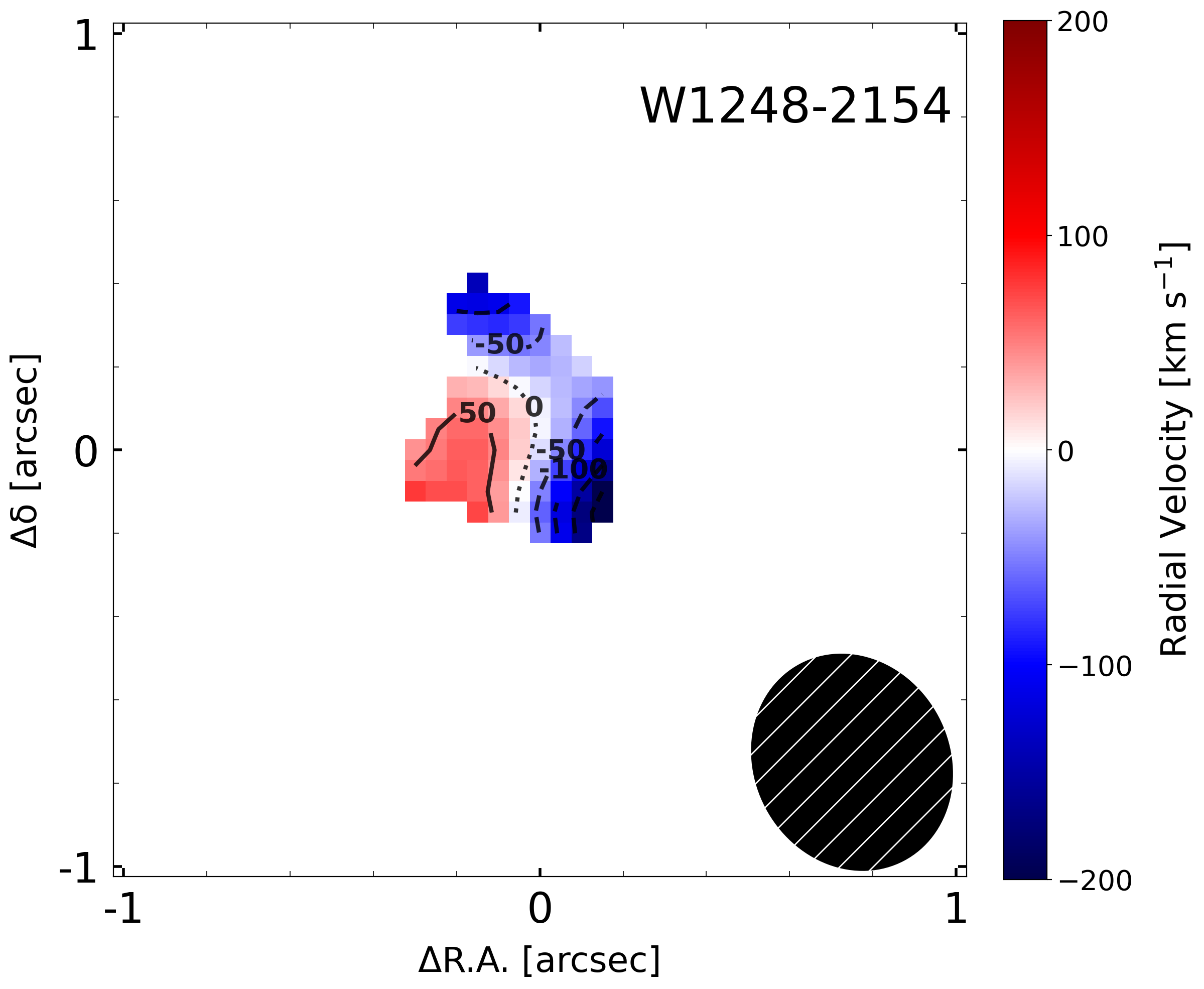}} \hspace{0.2cm}
 \subfloat{\includegraphics[width=6cm]{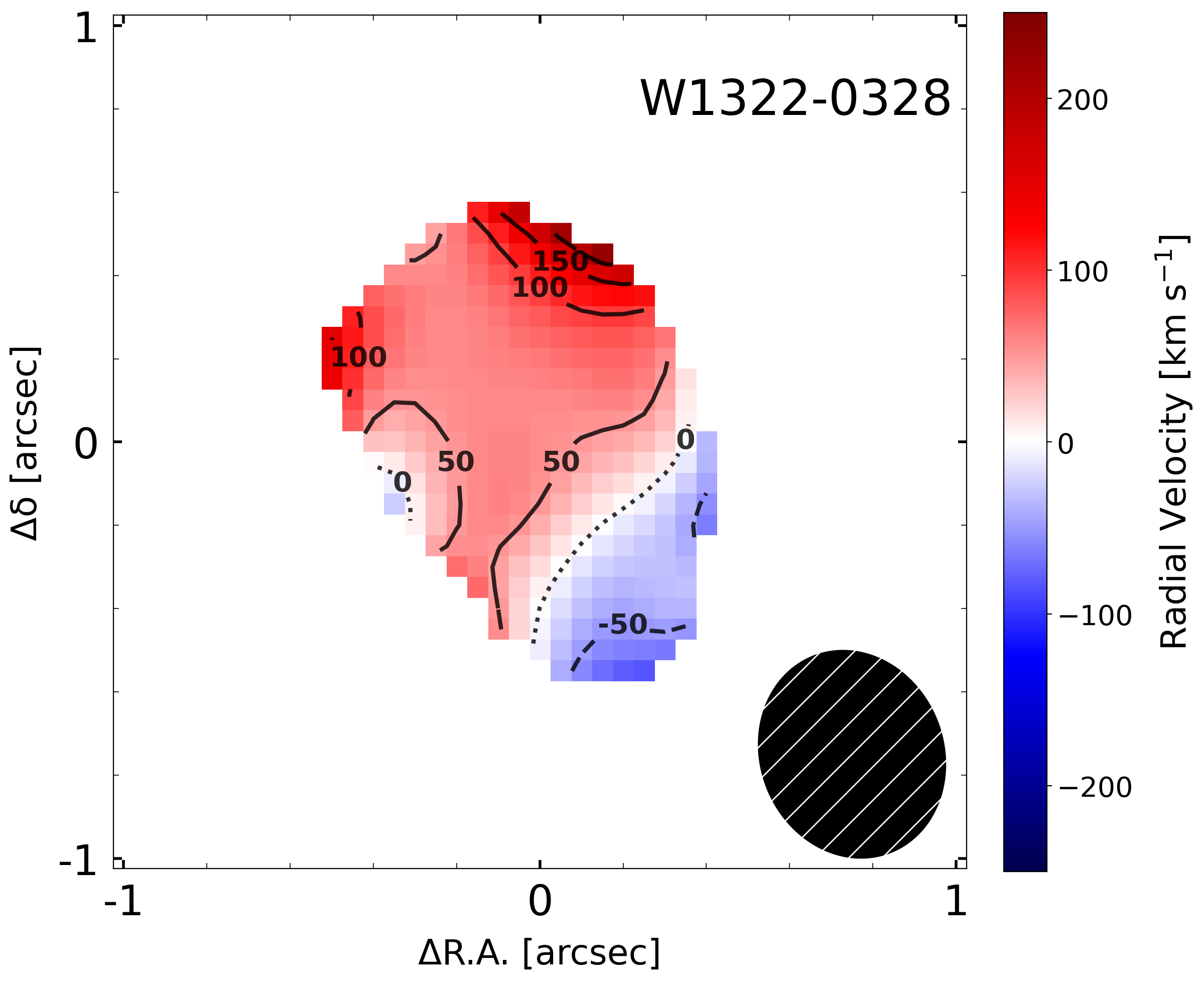}} \\
 \subfloat{\includegraphics[width=6cm]{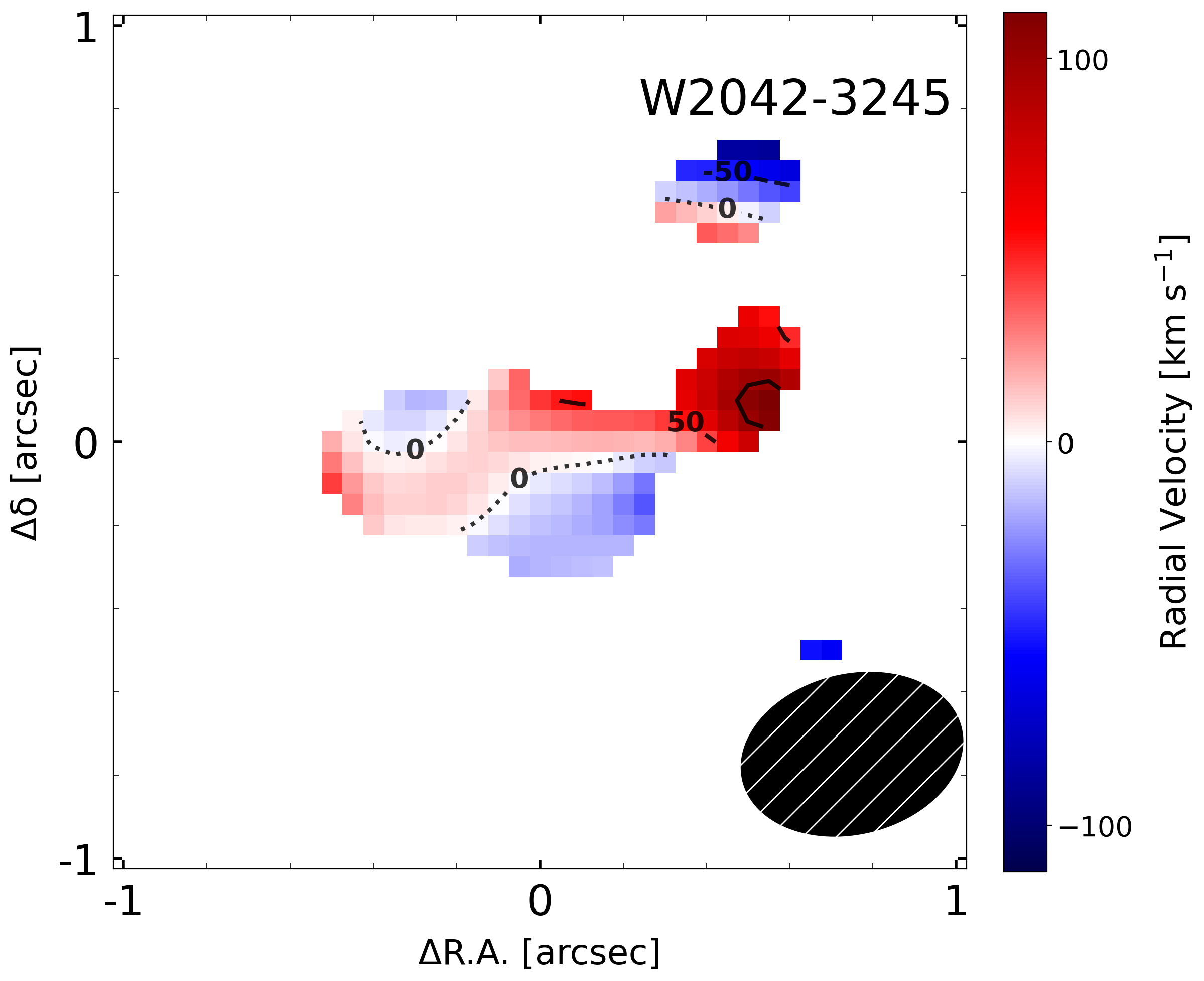}} \hspace{0.2cm}
 \subfloat{\includegraphics[width=6cm]{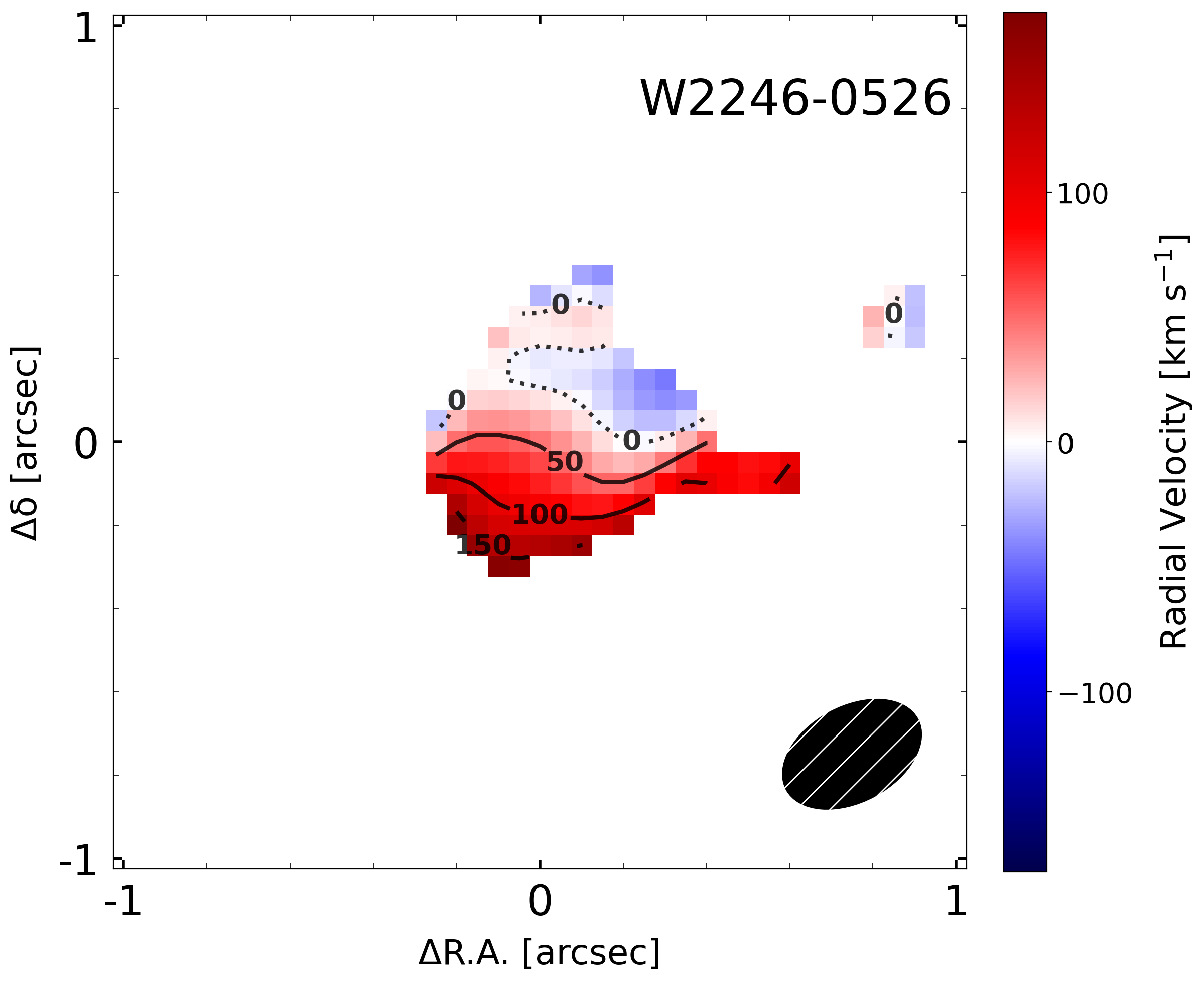}} \hspace{0.2cm}
 \subfloat{\includegraphics[width=6cm]{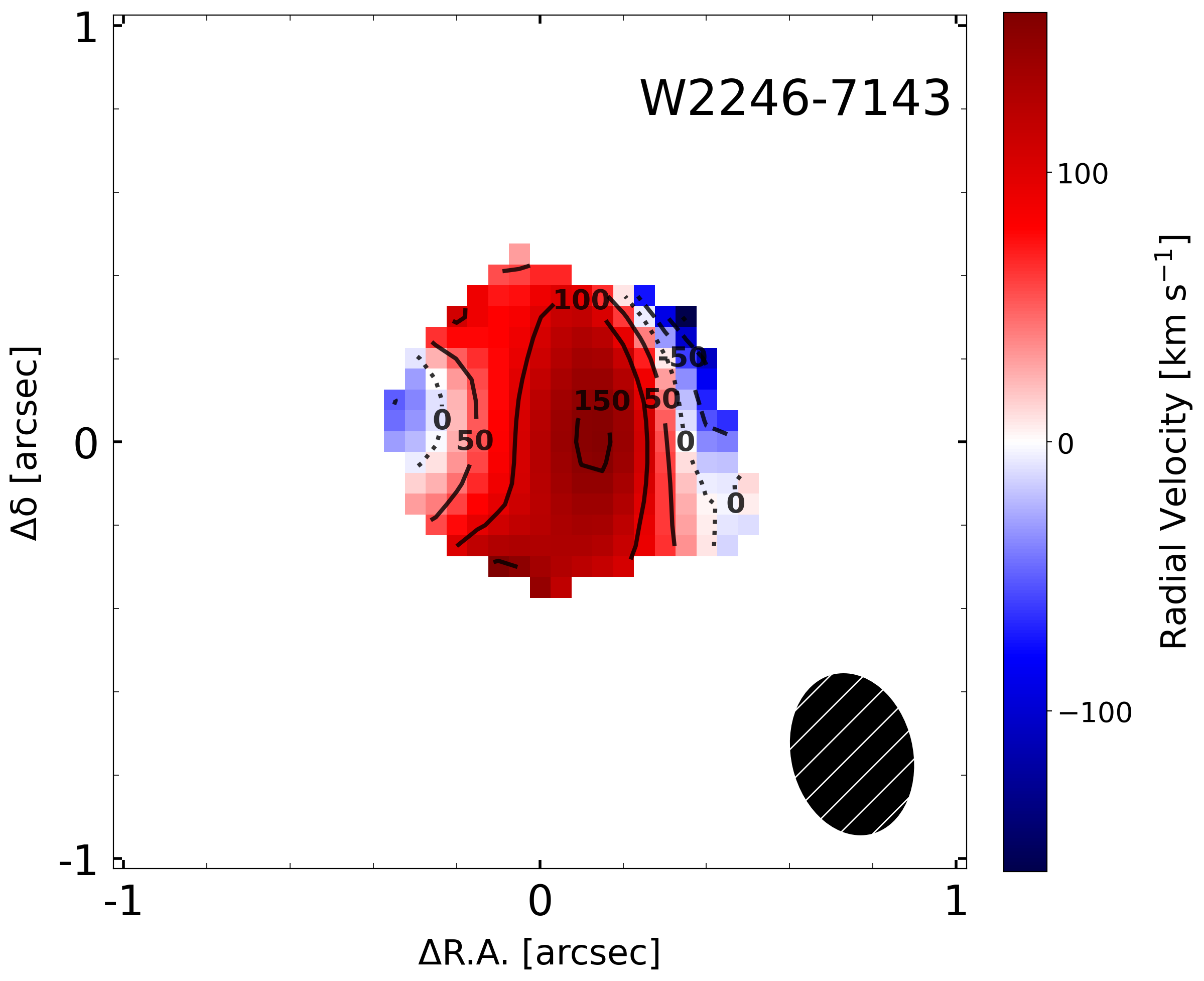}} \\
 \subfloat{\includegraphics[width=6cm]{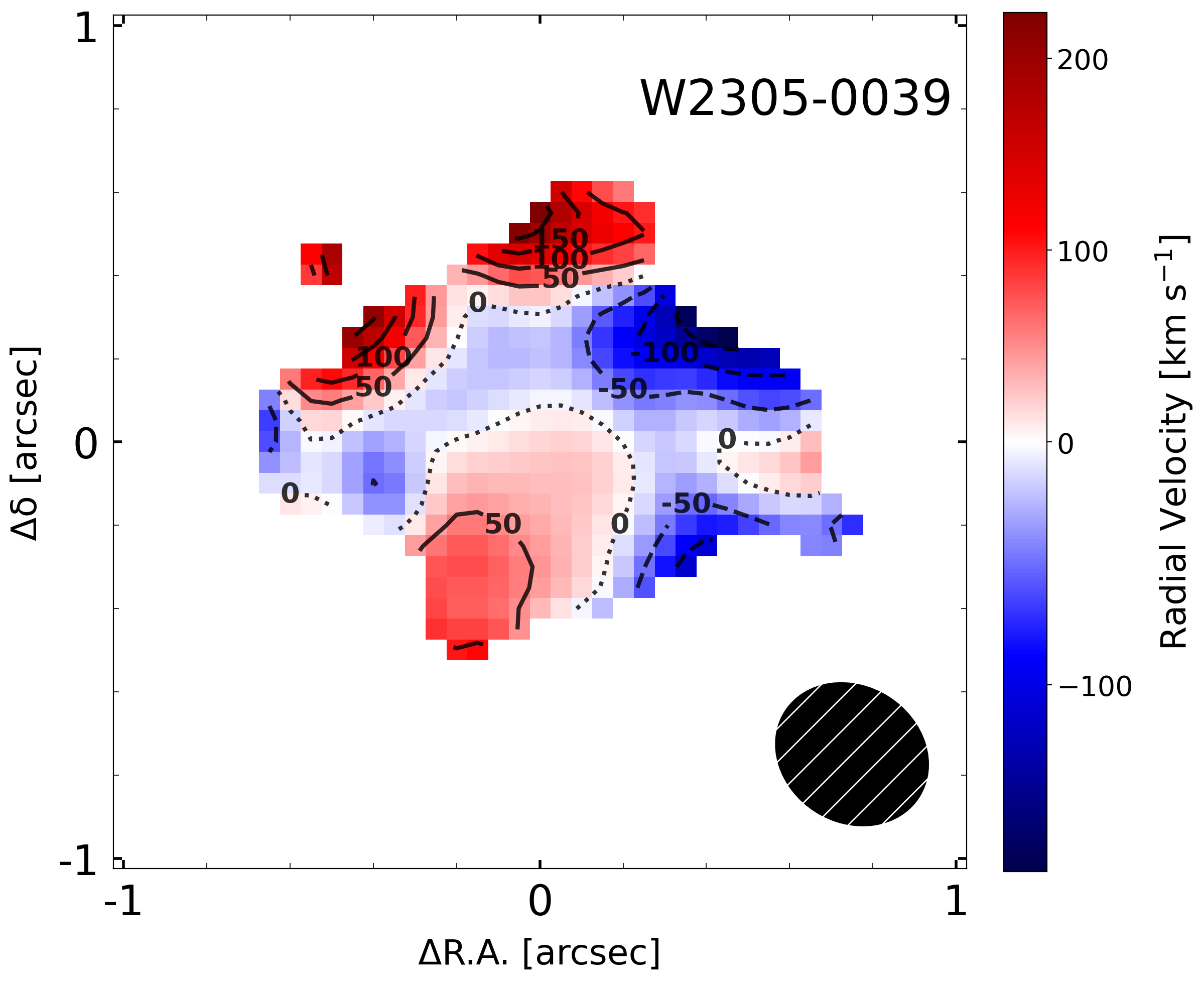}}
 \end{tabular}
 \caption{Moment-1 (velocity) maps of ten sources. Only the data that has S/N~$\geq$~3 is shown in these maps. The velocity is centred on the peak of CO emission. A blue component has a negative velocity and indicates an approaching component, while the red components have a positive velocity and indicate a receding component. A dotted line shows the point of zero velocity and contours are placed at 50\,km\,s$^{-1}$ and then 100\,km\,s$^{-1}$ intervals for both negative and positive velocities. The beam is displayed as a black ellipse in the lower-right corner of each map.}
 \label{fig:moment1}
 \end{figure*}

\begin{figure*}
 \begin{tabular}{ccc}
 \noindent\subfloat{\includegraphics[width=6cm]{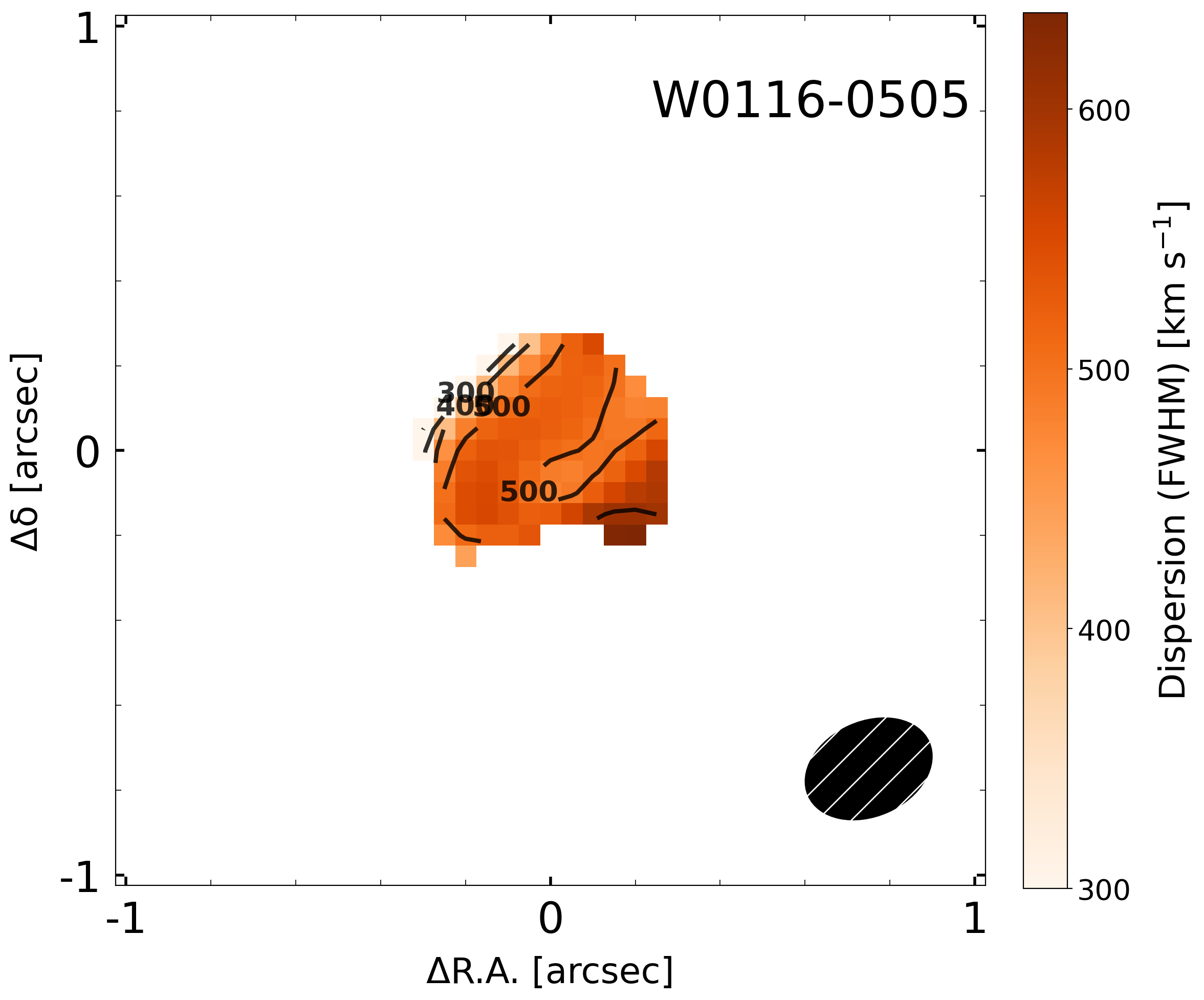}} \hspace{0.2cm}
 \subfloat{\includegraphics[width=6cm]{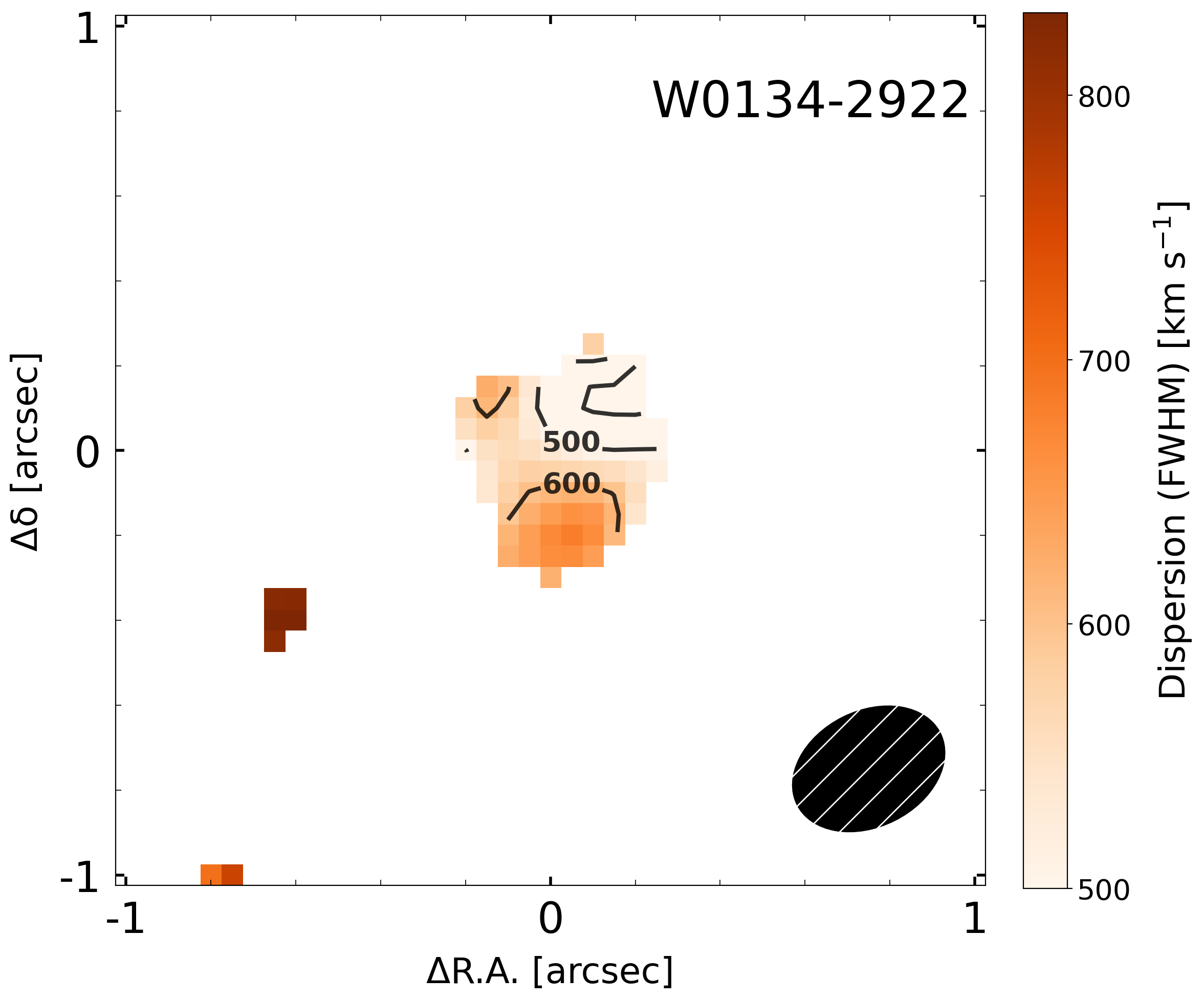}} \hspace{0.2cm}
 \subfloat{\includegraphics[width=6cm]{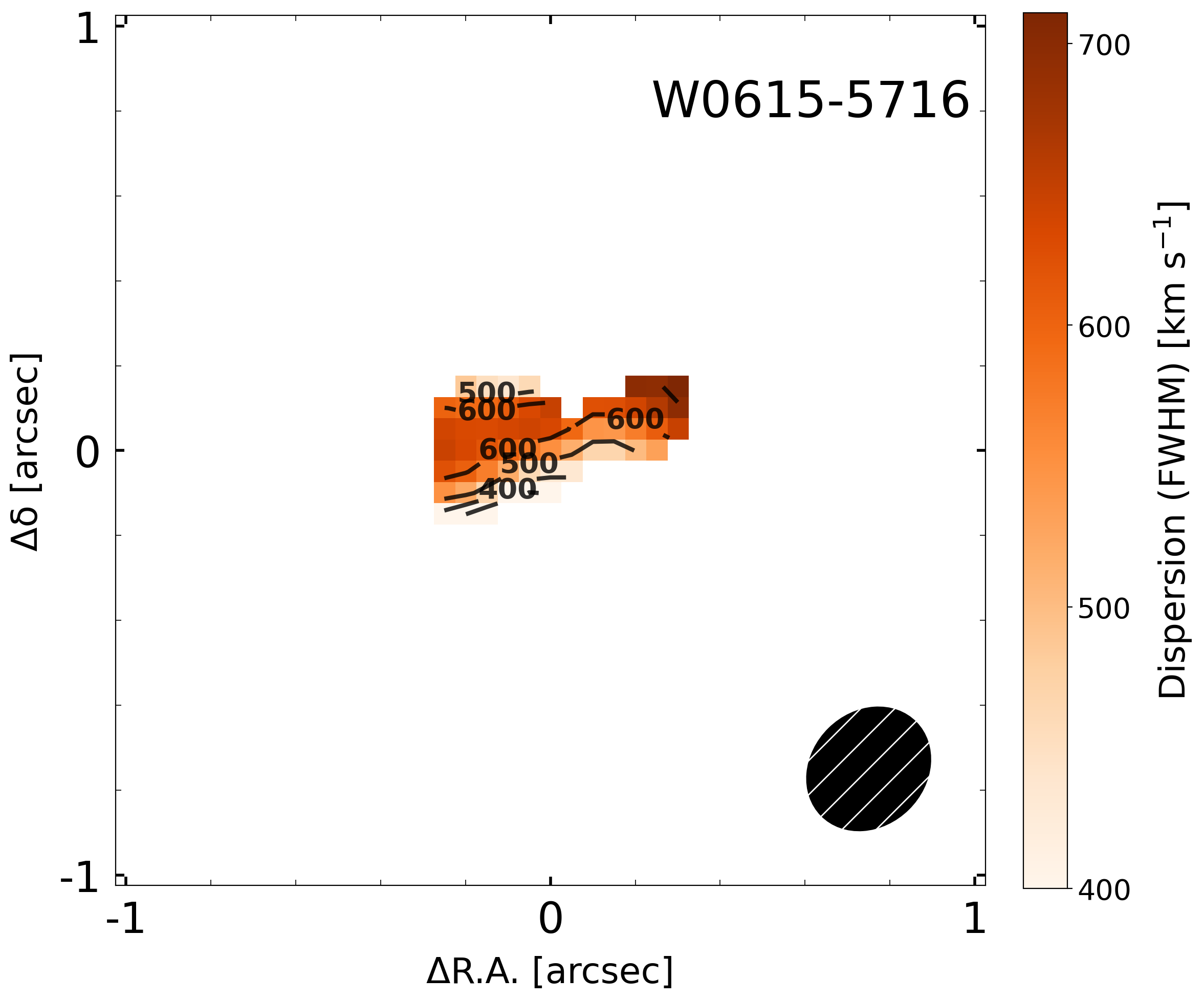}} \\
 \subfloat{\includegraphics[width=6cm]{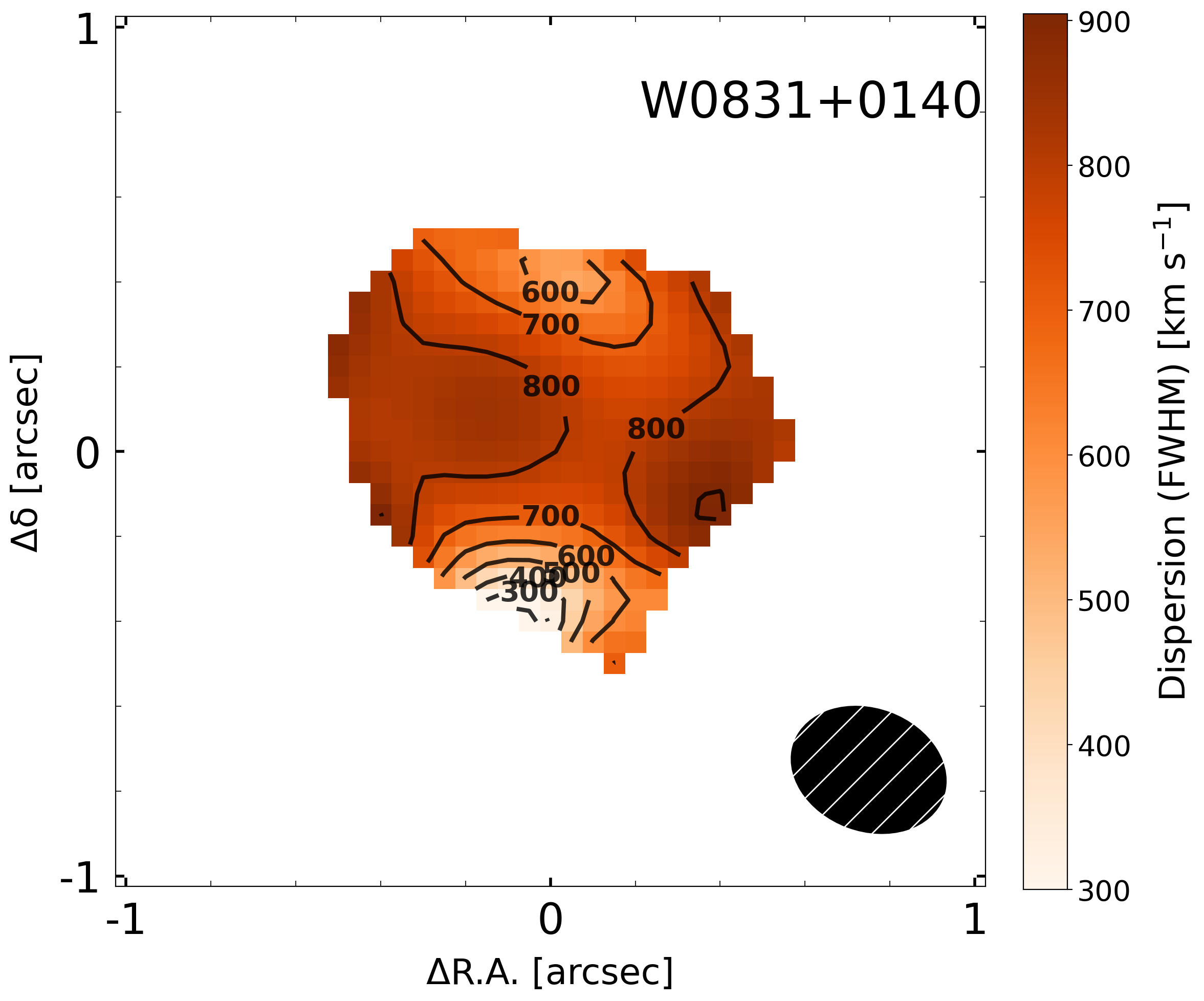}} \hspace{0.2cm}
 \subfloat{\includegraphics[width=6cm]{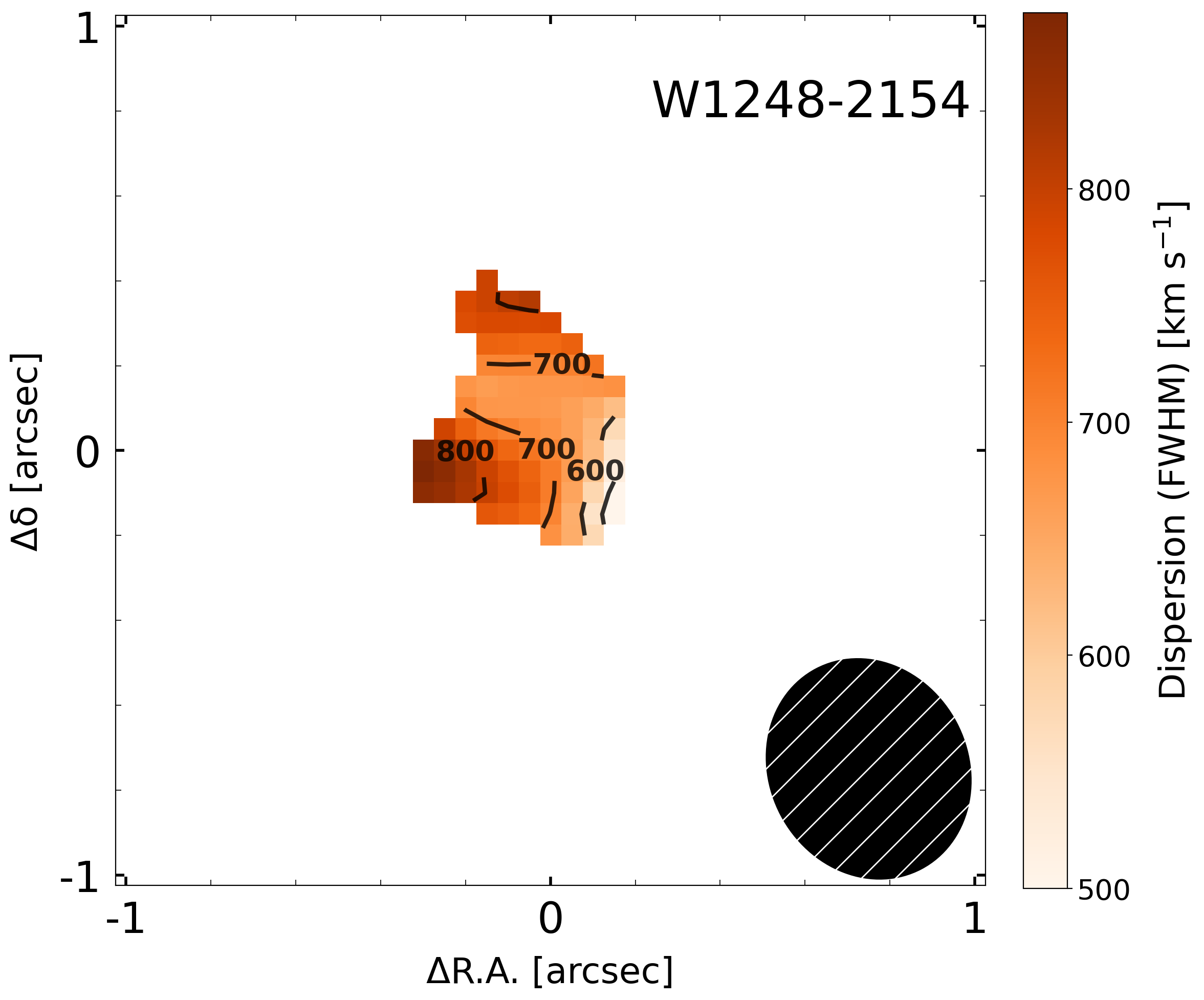}} \hspace{0.2cm}
 \subfloat{\includegraphics[width=6cm]{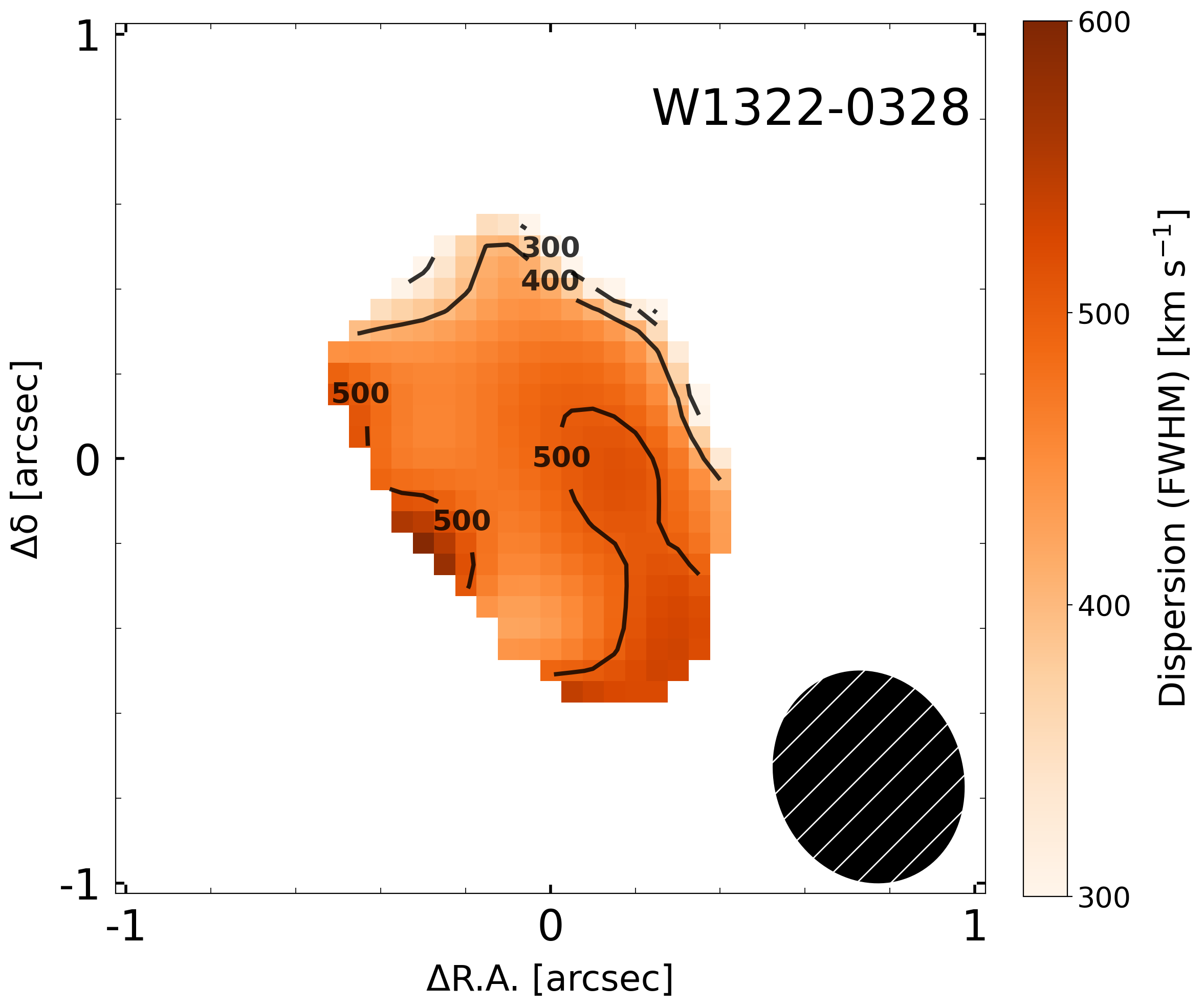}} \\
 \subfloat{\includegraphics[width=6cm]{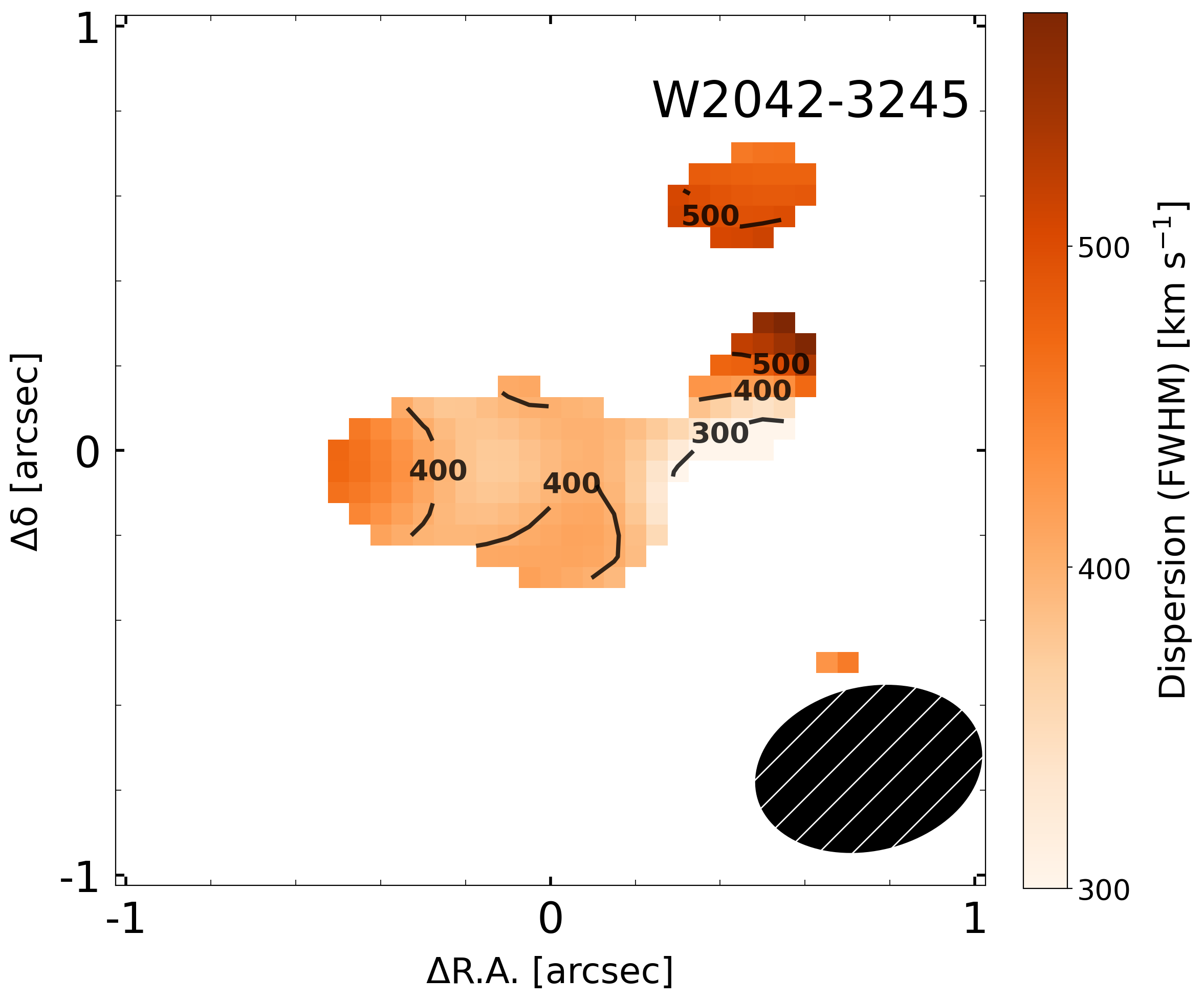}}\hspace{0.2cm}
 \subfloat{\includegraphics[width=6cm]{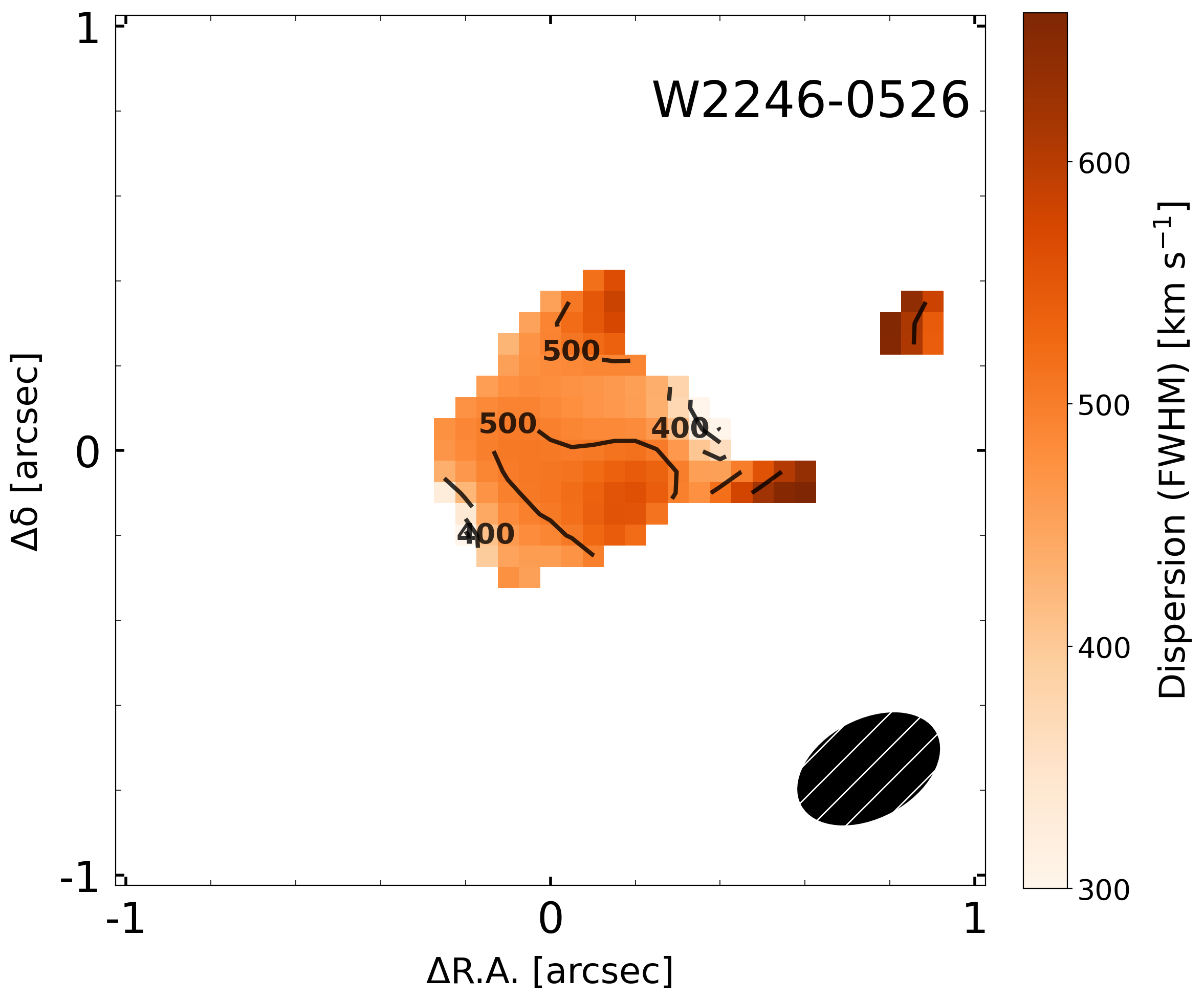}} \hspace{0.2cm}
 \subfloat{\includegraphics[width=6cm]{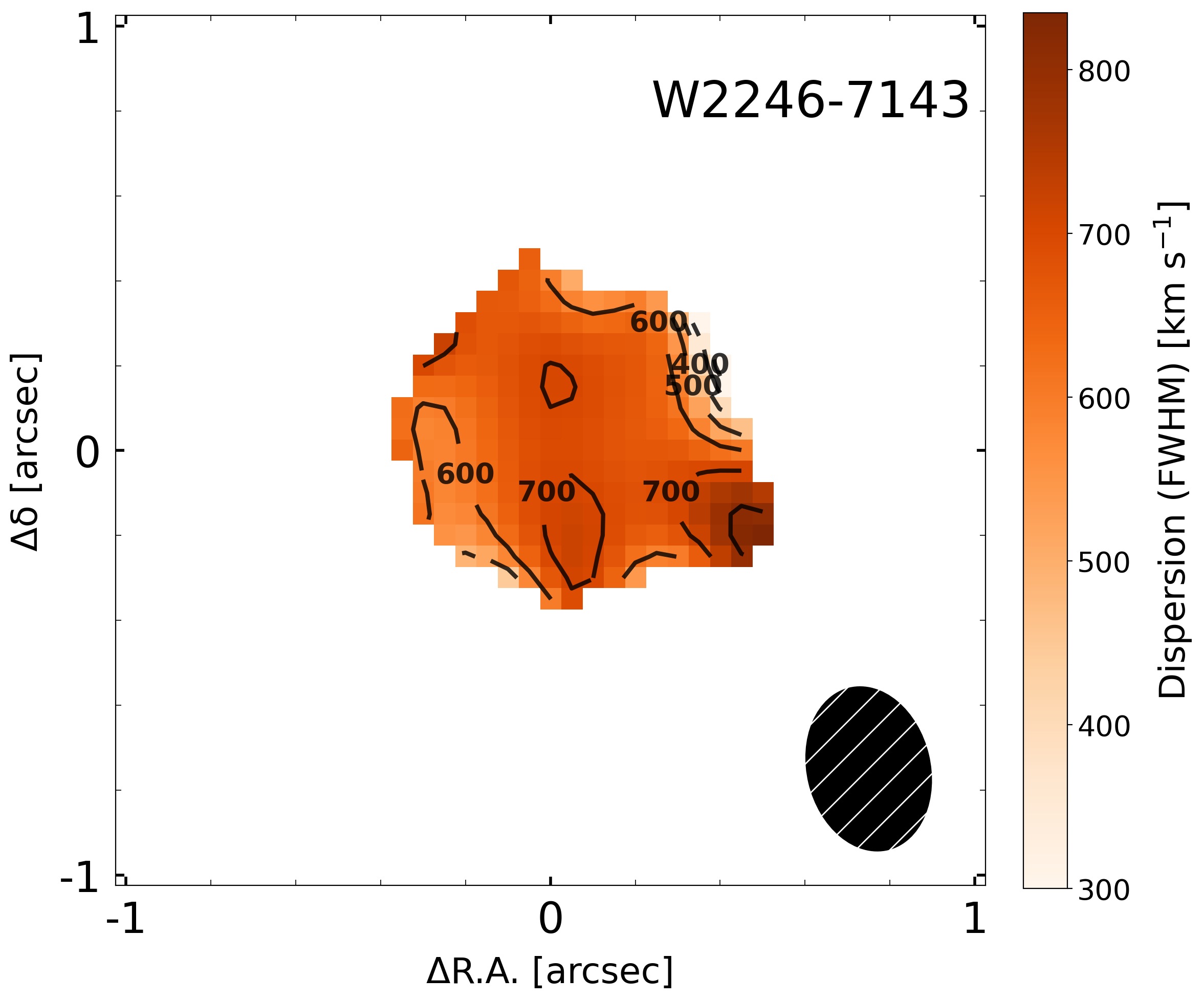}} \\
 \subfloat{\includegraphics[width=6cm]{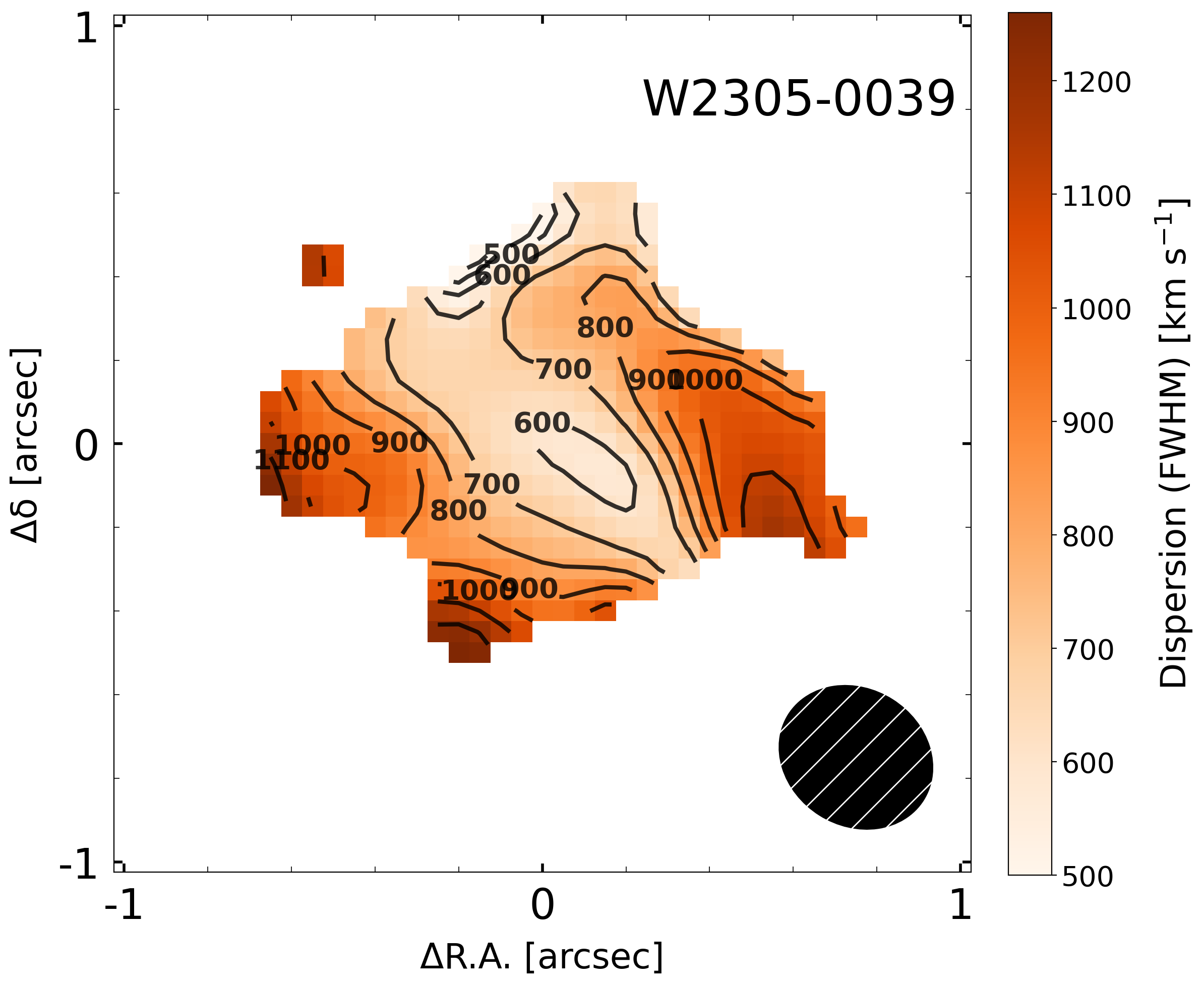}}
 \end{tabular}
 \caption{Velocity dispersion maps derived from moment-2 profiles. Only the data which has S/N~$\geq$~3 is displayed in these maps. The contours are displayed in 100\,km\,s$^{-1}$ intervals. The clean beam is shown as a black ellipse in the lower-right corner of each map. }
 \label{fig:FWHM}
 \end{figure*}

We investigated the spatially-resolved radial velocity (line-of-sight) images of each source to understand the bulk motion of molecular gas. We only analyse data (pixels) within the velocity range defined by the line emission and with S/N~$\geq$~3, masking all other data. The resulting radial velocity images are shown in Fig.~\ref{fig:moment1}.
Based on the propagation of uncertainties in moment maps \citep{teague2019statistical}, the central region of the radial velocity maps has an uncertainty of 50\,km\,s$^{-1}$, increasing to 100\,km\,s$^{-1}$ at the edges of the S/N~$\geq$~3 region. Steep velocity gradients at the edges of the S/N~$\geq$~3 region have a larger uncertainty and are unlikely to represent genuine velocity structure. Due to the relatively low S/N and significant beam sizes in the sample, beam smearing likely blurs any complex velocity structure on smaller scales, smoothing the velocity gradient on scales comparable to the beam. The brighter CO emitters have some resolved velocity components, while the fainter CO emitters have limited velocity information. \\
\indent W0831+0140 has two distinct velocity components; a north (approaching) and south (receding) component. The two Gaussian components separated by $\Delta$\textit{v}~=~580\,km\,s$^{-1}$ in Fig.~\ref{fig:spectra} appear explicitly in the velocity profile. This velocity profile is consistent with an outflow, two merging galaxies, or a rotating disc, consistent with our spectral interpretation. \\
\indent W2305--0039 has a shallow velocity gradient across its centre, with increasing radial velocity towards to edge of the S/N~$>$~3 region. The increased radial velocity could represent the outflowing gas from the broad 1100\,km\,s$^{-1}$ CO component shown in Fig.~\ref{fig:spectra}.\\
\indent The velocity dispersion of each target is shown via dispersion maps in Fig.~\ref{fig:FWHM}, which are derived from moment-2 maps. 
The second order moment is given by
\begin{equation} \label{eq:moment2}
    M_{2} = \frac{\int I_{\rm v}(\rm v-M_{1})^{2}d \rm v}{M_{0}},
\end{equation}
where \textit{I}\textsubscript{v} is the intensity at a given velocity (or frequency), v is the given velocity, \textit{M}\textsubscript{1} and \textit{M}\textsubscript{0} are the first and zeroth order moments. FWHM is the resulting dispersion shown in the dispersion maps;
\begin{equation}
    \text{FWHM} = \sqrt{8\ln{2}~\,M_{2}}.
\end{equation}
Given that dispersion maps are flux weighted, lower S/N regions of the maps may not capture gas dynamics due to their modest sensitivity; they are most reliable in regions where the S/N is high - i.e., the maps' central regions. The typical uncertainty around the central region, using moment uncertainty propagation \citep{teague2019statistical} is 100\,km\,s$^{-1}$, increasing to 150\,km\,s$^{-1}$ at the edges of the S/N~$\geq$~3 region. Similar to the case with radial velocity maps, steep dispersion gradients, particularly in areas at the edges of the maps with relatively low S/N are unlikely to be genuine. Steep dispersion gradients can cause the \texttt{spectral-cube} package (used to create these maps) to disregard the data around these features, and so a normalised 2-D Gaussian kernel ($\mu$~=~0, $\sigma$\textsubscript{x}~=~2, $\sigma$\textsubscript{y}~=~2) was used to smooth the dispersion map of W0831+0140 to avoid this issue. However, in consequence we inevitably lose some spatial resolution in the dispersion map.\\
\indent The dispersion maps show that the velocity dispersions are broad for all sources with dispersions $\geq$~400\,km\,s$^{-1}$. While beam smearing could lead to an increase in perceived velocity dispersion, the dispersion maps are largely consistent with the spectral linewidths (Table~\ref{table:FWHM}), suggesting that beam smearing has not significantly increased the observed dispersion in the majority of cases. The dispersion maps of W0831+0140 and W2042--3245 show dispersions comparable to the sum of their individual spectral linewidths, suggesting the individual CO components may have blended together in these maps. Large dispersion suggests a turbulent ISM, possibly excited by the central AGN \citep{diaz2016strikingly}. For these sources we see uniform dispersion throughout, though this could be contributed to by the lack of spatially-resolved detail and relatively low resolution of some of the data cubes. Hence, the observed uniformity of the dispersion fields could be a consequence of resolution effects and higher S/N observations would be required to investigate this possibility. The dispersion values are diverse across sources, suggesting that either the AGN in each case impacts the ISM to differing degrees, or that the ISM density or structure varies. This non-uniformity of gas in Hot DOGs has previously been observed using [CII] \citep{diaz2016strikingly}.\\
\indent All sources show broad dispersion, in addition to their broad spectral linewidths. This suggests that a turbulent ISM is a common feature, and is likely driven by AGN-ISM interactions, triggered by galactic mergers \citep{gao2020mergers}.\\
\indent The CO velocity fields and dispersion maps of W0134--2922, W0831+0140 and W2246--0526 all show agreement with the [CII] velocity and dispersion maps \citep{diaz2021kinematics}. The [CII] dispersion of W2246--0526 shows remarkable uniformity across the 2.5\,kpc extent of the galaxy \citep{diaz2016strikingly}, and the galactic merger in W0831+0140 is postulated in [CII]. The ionized and molecular gas in the ISM also appear to have similar kinematics, perhaps broadened by the same physical turbulence. 
\subsection{Luminosity and Mass Estimates}\label{sec:lumin}
\begin{figure*}
 \begin{tabular}{cc}
 \noindent\subfloat{\includegraphics[width=8cm]{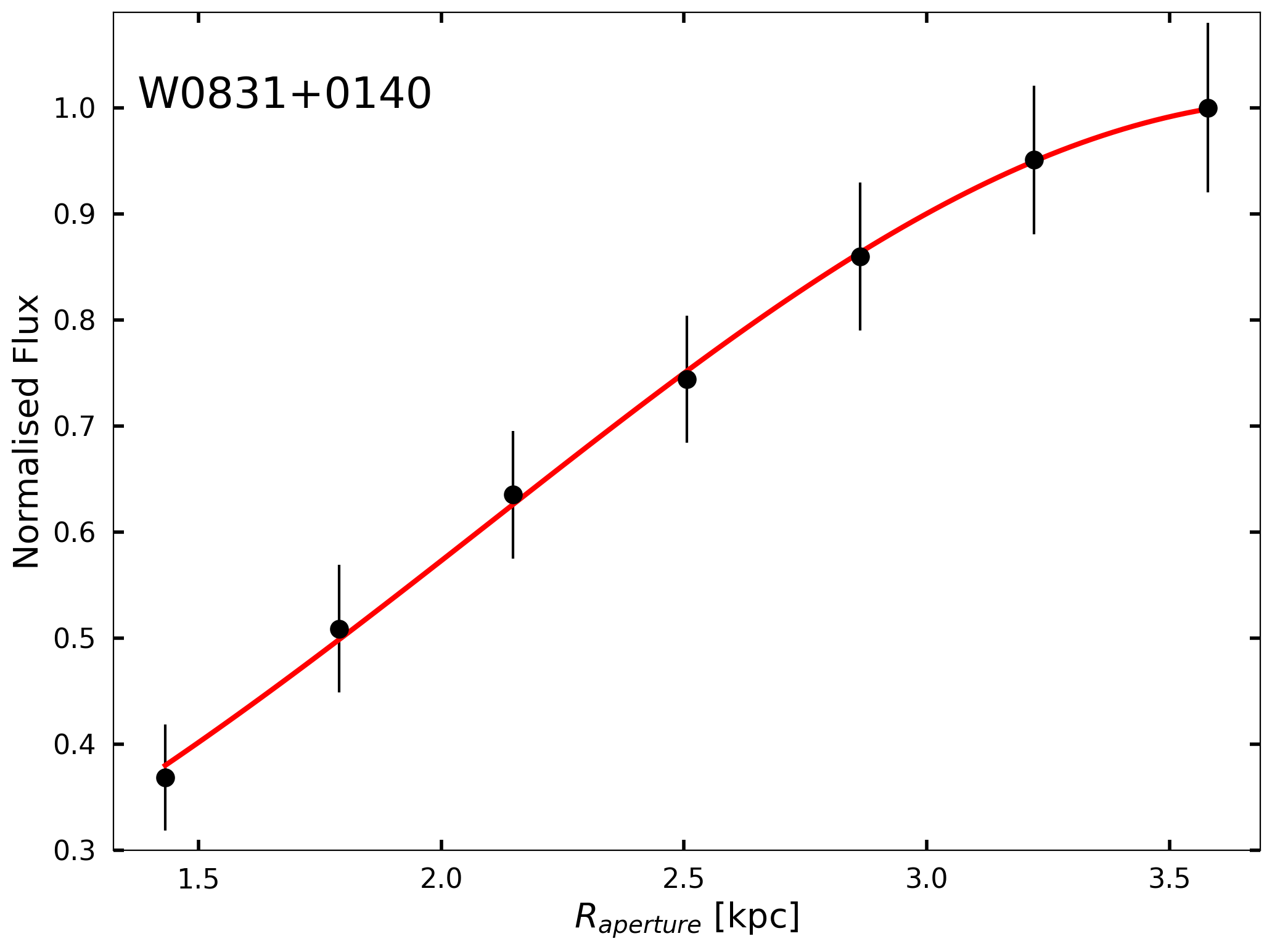}}\hspace{0.2cm}
 \subfloat{\includegraphics[width=8cm]{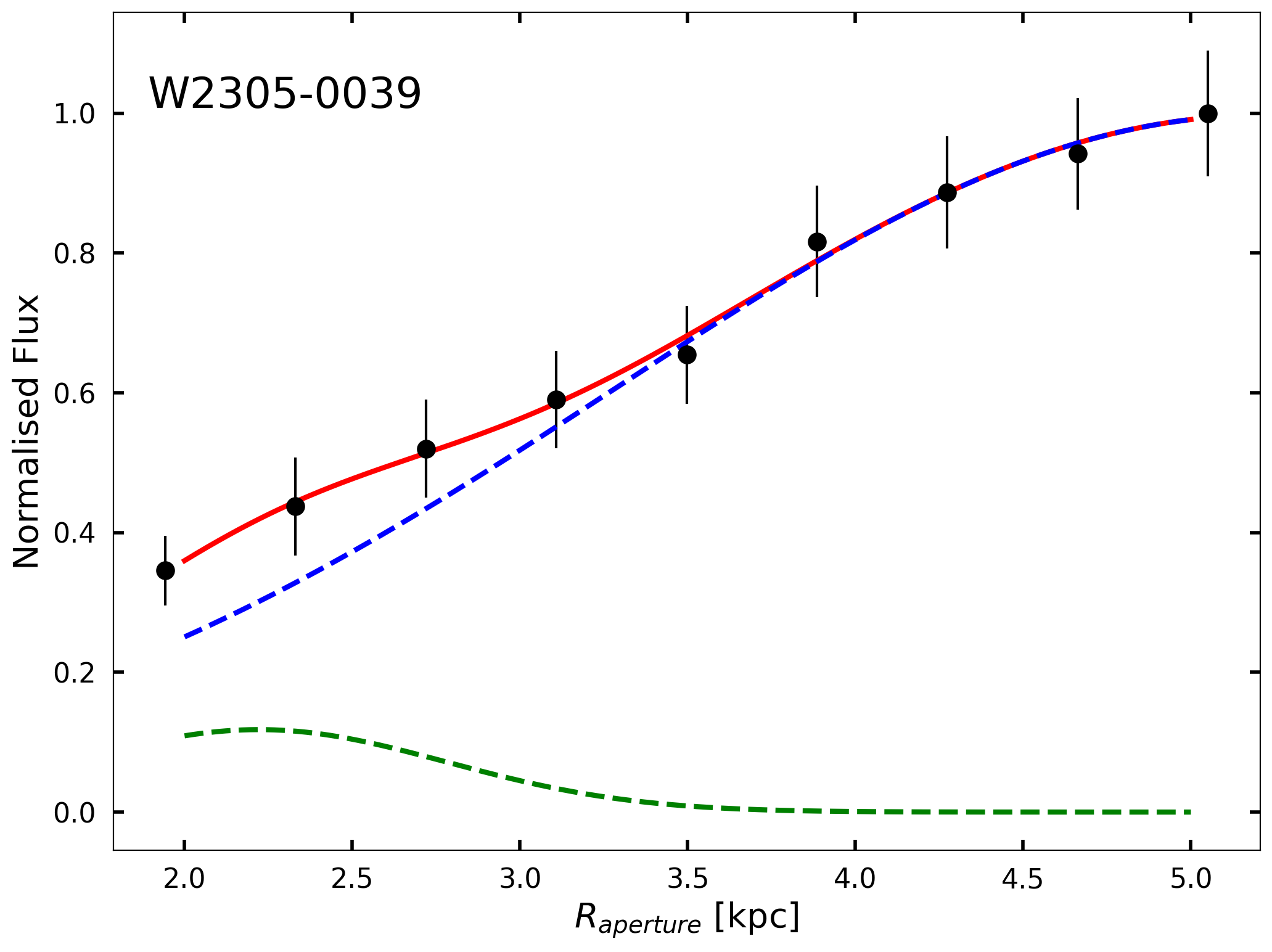}} 
 \end{tabular}
 \caption{Integrated flux profiles of W0831+0140 and W2305--0039, where more than 50 per cent of the total CO flux is recovered after subtracting a central unresolved point source. The profiles were obtained by extracting the total CO flux in iteratively larger circular apertures. The flux has been normalised by the total integrated flux emitted by the source. Gaussian functions are fit to the data and shown via red solid lines. In the case of W2305--0039, the Gaussian fit is the sum of two Gaussian functions (green and blue dashed lines).}
 \label{fig:FluxProfiles}
\end{figure*}

The CO luminosity, molecular gas mass and dynamical mass for each source is shown in Table~\ref{table:lumin_mass}. 
\begin{table*}
    \centering
    \caption{Luminosity and mass derived from mid-\textit{J} CO lines. (1--2) CO(4--3) or CO(5--4) luminosity derived using equation \ref{eq:CO_luminosity}; (3) Molecular hydrogen mass derived using equation \ref{eq:hydrogen_mass}; (4) Estimated radius of the CO emission, defined as half the major axis of the beam, except W0831+0140 and W2305--0039 where we use the half-light radius; (5) Upper limit of dynamical mass derived using equation \ref{eq:dynamical_mass}; (6) Lower limit of fractional molecular mass; (7--8) CO luminosity ratios in units of brightness temperature (7) and solar luminosities (8) with other observed lines \citep[]{diaz2018multiple, penney2020cold,aranda2024benchmark}. Ratios marked with \textit{a} are comparisons between the CO(4--3) and CO(1--0) transition lines, \textit{b} represents a comparison between CO(5--4) and CO(2--1), and \textit{c} is the ratio between CO(7--6) and CO(5--4).}
    \label{table:lumin_mass}
    \setlength\extrarowheight{2pt}
    \begin{tabular}{c c c c c c c c c}
        \hline 
         Source & \textit{L$^{\prime}$}\textsubscript{CO} & \textit{L}\textsubscript{CO}  & \textit{M}(H\textsubscript{2}) & \textit{R} & \textit{M}\textsubscript{dynamical} & \textit{f} & \multicolumn{2}{c}{CO Luminosity Ratios} \\
         & [10$^{10}$\,K\,km\,s$^{-1}$\,pc$^{2}$] & [10$^{7}$\,L\textsubscript{$\rm\odot$}] & [10$^{10}$\,M$_{\rm\odot}$] & [kpc] & [10$^{10}$\,M$_{\rm\odot}$] & &  &  \\
        & (1) & (2) & (3) & (4) & (5) & (6) & (7) & (8)\\
        \hline
          W0116--0505 & 2.18~$\pm$~0.20 & 6.83~$\pm$~0.63 & 1.06~$\pm$~0.25 & $\leq$~1.3 & $\leq$~6.9 & $\geq$~0.15 & 4.7~$\pm$~1.7 $^a$ & 300~$\pm$110 $^a$  \\
          W0134--2922 & 1.22 $\pm$ 0.16 & 3.82 $\pm$ 0.50 & 0.59~$\pm$~0.15 & $\leq$~1.6 & $\leq$~9.5 & $\geq$~0.06 & \dots & \dots \\
          W0615--5716 & 1.44 $\pm$ 0.27 & 4.53 $\pm$ 0.84 & 0.71 $\pm$ 0.20 & $\leq$~1.3 & $\leq$~8.5 & $\geq$~0.08 & \dots & \dots \\
          W0831+0140 & 11.3 $\pm$ 0.69 & 35.4 $\pm$ 2.15 & 5.50 $\pm$ 1.22 & 1.8 $\pm$ 0.2 & 24.1 $\pm$ 4.1 & 0.23 $\pm$ 0.06 & 5.1 $\pm$ 0.5 $^a$ & 330~$\pm$~30 $^a$\\
          W1248--2154 & 0.93 $\pm$ 0.15 & 2.92 $\pm$ 0.46 & 0.45 $\pm$ 0.12 & $\leq$~2.1 & $\leq$~14.2 & $\geq$~0.03 & \dots & \dots\\
          W1322--0328 & 4.36 $\pm$ 0.20 & 13.7 $\pm$ 0.64 & 2.13 $\pm$ 0.46 & $\leq$~2.1 & $\leq$~6.7 & $\geq$~0.32 & 11.5 $\pm$ 2.9 $^a$ & 730~$\pm$~180 $^a$\\
          W2042--3245 & 1.11 $\pm$ 0.16 & 3.39 $\pm$ 0.50 & 0.54 $\pm$ 0.14 & $\leq$~2.0 & $\leq$~6.6 & $\geq$~0.08 &\dots & \dots\\
          W2246--0526 & 3.95 $\pm$ 0.31 & 24.2 $\pm$ 1.87 & 1.39 $\pm$ 0.36 & $\leq$~1.3 & $\leq$~5.9 & $\geq$~0.24 & 0.4 $\pm$ 0.1 $^b$, 1.1 $\pm$ 0.1 $^c$ & 6.6~$\pm$~1.7 $^b$, 2.9~$\pm$~0.2 $^c$\\
          W2246--7143 & 5.11 $\pm$ 0.38 & 16.0 $\pm$ 1.20 & 2.50 $\pm$ 0.56 & $\leq$~1.5 & $\leq$~10.6 & $\geq$~0.24 & \dots & \dots\\
          W2305--0039 & 7.38 $\pm$ 0.53 & 23.2 $\pm$ 1.67 & 3.60 $\pm$ 0.81 & 2.6 $\pm$ 0.3 & 12.7 $\pm$ 1.8 & 0.28 $\pm$ 0.08 & 7.5 $\pm$ 0.9 $^a$ & 480~$\pm$~57 $^a$\\
         \hline
    \end{tabular}
\end{table*}
The CO luminosity is found using
\begin{equation} \label{eq:CO_luminosity} 
    L^{\prime}\textsubscript{CO} = 3.25 \times 10^{7} I\textsubscript{CO}  \frac{D\textsubscript{L}^{2}}{\nu\textsubscript{obs}^{2}(1+z)^{3}} 
\end{equation} \citep{solomon2005molecular, carilli2013cool}.
\textit{L}$^{\prime}$\textsubscript{CO} is the CO luminosity in K\,km\,s$^{-1}$\,pc$^{2}$, \textit{I}\textsubscript{CO} is the velocity-integrated flux of the CO line (Table~\ref{table:FWHM}), $\nu$\textsubscript{obs} is the observed frequency of the transition [$\nu$\textsubscript{rest} = $\nu$\textsubscript{obs}~(1~+~\textit{z})], \textit{z} is the redshift and \textit{D}\textsubscript{L} is the luminosity distance of the target. We find CO luminosities ranging from 0.93--11.3~$\times$~10$^{10}$\,K\,km\,s$^{-1}$\,pc$^{2}$. This is in reasonable agreement with CO(4--3) studies of Hot DOGs \citep{fan2018alma}, where three sources had luminosities (2.4--17.9)~$\times$~10$^{10}$\,K\,km\,s$^{-1}$\,pc$^{2}$. The mean mid-\textit{J} CO luminosity in solar units is a factor $\geq$~300 greater than the CO(1--0) luminosity in the same sources \citep{penney2020cold}. The mid-\textit{J} CO gas in these targets is therefore significantly more luminous than expected from the CO(1--0) emission, assuming thermal excitation.\\
\indent Assuming that abundances of CO and molecular hydrogen are linked \citep{solomon2005molecular}, the mass of molecular hydrogen can then be estimated using
\begin{equation} \label{eq:hydrogen_mass}
    \textit{M(\rm H\textsubscript{2})} = \alpha L^{\prime}\textsubscript{CO(1--0)},
\end{equation}
where $\alpha$ is the conversion factor for ultra-luminous infrared galaxies (ULIRGs), which is taken to be $\alpha$~$\sim$~0.8\,M\textsubscript{$\rm\odot$}~(K\,km\,s$^{-1}$\,pc$^{2}$)$^{-1}$, and \textit{L}$^{\prime}$\textsubscript{CO(1--0)} is the CO(1--0) luminosity. \textit{L$^{\prime}$}\textsubscript{CO(4--3)} and \textit{L$^{\prime}$}\textsubscript{CO(5--4)} are  scaled down to \textit{L}$^{\prime}$\textsubscript{CO(1--0)} using CO excitation ratios in ULIRGs of 0.61~$\pm$~0.13 [CO(4--3) to CO(1--0)] and 0.44~$\pm$~0.11 [CO(5--4) to CO(1--0)] \citep{boogaard2020alma}. The QSO scaling factors of 0.87 and 0.69 \citep{carilli2013cool} have been adopted in other studies of CO in Hot DOGs \citep[e.g.,][]{fan2018alma,sun2024physical}, though these scaling factors are derived from stacked averages of CO observations in quasars at various redshifts, and are larger than those derived from ULIRGs. Previous estimations of molecular gas masses in Hot DOGs may therefore be overestimated by 40--50 per cent. \\
\indent The molecular gas mass of these systems ranges from (0.45--5.50)~$\times$~10$^{10}$\,M\textsubscript{$\rm\odot$}.The average dust mass in DOGs is 3~$\times$~10$^{8}$\,M\textsubscript{$\rm\odot$} \citep{riguccini2015composite}, hence molecular gas appears to be on average a factor of 60 more abundant than dust in these systems. Due to the high abundance of molecular gas, the potential for high levels of star formation is clear and is plausible due to mergers resupplying gas reservoirs. The SFRs of sources in this sample, obtained by examining only resolved dust and [CII] emission; thus excluding any non-thermal emission from the unresolved AGN, ranges from 202--2863\,M\textsubscript{$\rm\odot$}~\,yr$^{-1}$ \citep{diaz2021kinematics}. However, the potential for the AGNs to heat extended dust in these systems could imply that these SFRs are overestimated. These SFRs imply that the depletion times of the molecular gas in these systems is $\approx$~10\,Myr, though this is possibly a lower limit given the uncertainty in calculating SFRs in these sources. \\
\indent The dynamical mass for a dispersion-dominated (\textit{v}$/\sigma<$1) system is
\begin{equation} \label{eq:dynamical_mass}
    M\textsubscript{dyn} = \frac{\alpha \sigma\textsubscript{0}^{2}R}{G},
\end{equation}
where $\sigma$\textsubscript{0} is the width of the line, \textit{R} is the radius of the source, and \textit{G} is the gravitational constant. Due to the broad CO lines, lack of rotational features, turbulence, and previous studies suggesting that the most luminous Hot DOGs are dispersion-supported \citep{diaz2021kinematics}, we give the dynamical mass of each source using this assumption. As in other studies of dispersion-dominated systems, we adopt a value of $\alpha$~=~3.4 \citep{stott2016kmos, diaz2021kinematics}. Due to the unresolved nature of the majority of sources in this sample, we take the semi-major axis of the beam as the upper limit for the physical radius of the source. Two sources have more than 50 per cent of their flux remaining after an unresolved central source is subtracted, taking the central  flux all the way down to zero. These sources (W0831+0140 and W2305--0039) thus have substantial extended emission. We estimate the physical radius by examining the flux profiles: see Fig~\ref{fig:FluxProfiles}. We define the galactic radius as the radius of the aperture which recovers half of the total CO flux.\\
\indent The dynamical mass estimates are ($\leq$5.9--24.1)~$\times$~10$^{10}$\,M\textsubscript{$\rm\odot$} and are in agreement with masses derived in other Hot DOG studies \citep{diaz2018multiple,fan2019alma}. The fraction of molecular to dynamical mass ranges from $\geq$3--32 per cent. W0831+0140, W1322-0328, W2246--0526, W2246-7143, and W2305--0039 show large molecular gas fractions ($\geq$~20 per cent). The mean fraction of molecular mass is 17 per cent, which agrees with [CII] and dust continuum analysis of \textit{z}~$\approx$~6 quasars, where the fraction of molecular to dynamical mass is $\geq$~10--30 per cent \citep{neeleman2021kinematics}. In the host galaxies of local AGN, the molecular mass fraction is typically 10 per cent \citep{popping2014evolution,jarvis2020high}. 
\section{Discussion} \label{sec:discussion}
\subsection{Broad Velocity Dispersion in the Molecular Gas}\label{sec:uniformdispersion}
The dispersion maps in Fig.~\ref{fig:FWHM} show each source with a broad $\geq$~400\,km~\,s$^{-1}$ molecular dispersion. The FWHM of the Gaussian profiles fit to the CO spectra (Fig.~\ref{fig:spectra}) range from 100--1100\,km\,s$^{-1}$ and nine out of ten sources show at least one broad ($\geq$~400\,km\,s$^{-1}$) component. The consistently broad CO lines suggests that a turbulent, molecular ISM is a common property of Hot DOGs in this sample, as it is in the atomic gas phase \citep{diaz2021kinematics}. \\
\indent Turbulence can enhance emission and can be induced through low-velocity shocks in gas colliding via a merger or relativistic acceleration of an outflow \citep{appleton2018jet}. In this sample, W0831+0140 is identified as a potential merger between at least two galaxies. It is the most luminous mid-\textit{J} CO emitter in the sample (\textit{L}\textsubscript{CO}~=~35.4~$\times$~10$^{7}$\,L\textsubscript{$\rm\odot$}). With less than half the CO luminosity of W0831+0140, a tentative merger was postulated in W2246--7143 (\textit{L}\textsubscript{CO}~=~16.0~$\times$~10$^{7}$\,L\textsubscript{$\rm\odot$}). Galactic mergers are likely the triggers of extreme AGN feedback, the Hot DOG phase, and turbulence in the molecular ISM.
\subsection{Outflow Energetics in W2305--0039}
\begin{figure*}
 \begin{tabular}{c}
 \noindent\subfloat{\includegraphics[width=15cm]{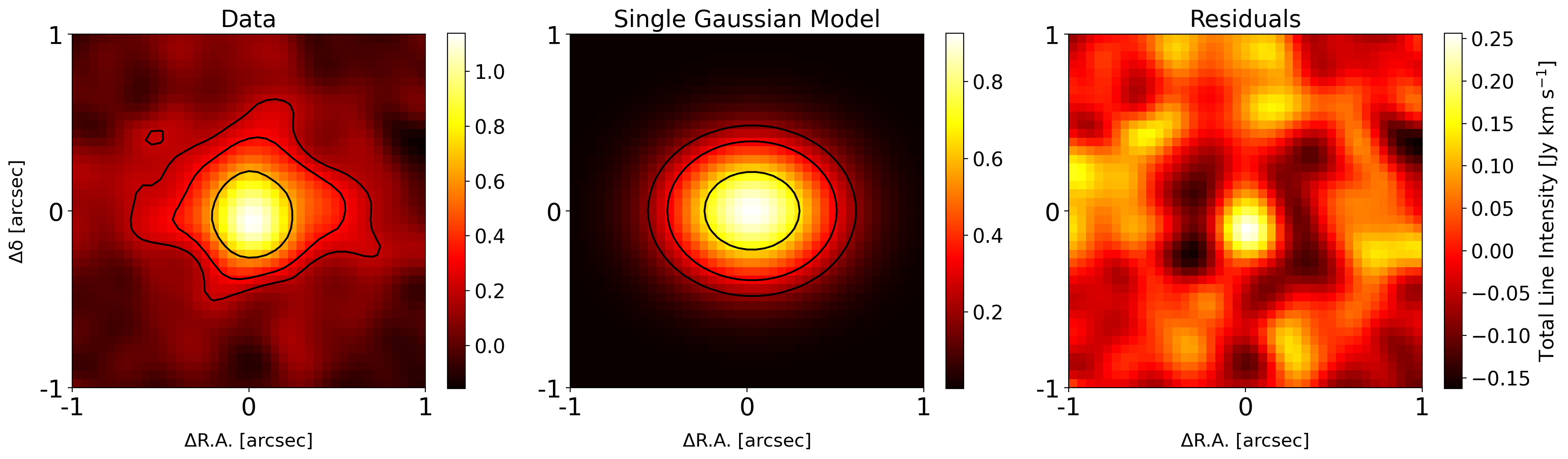}} \\
 \subfloat{\includegraphics[width=15cm]{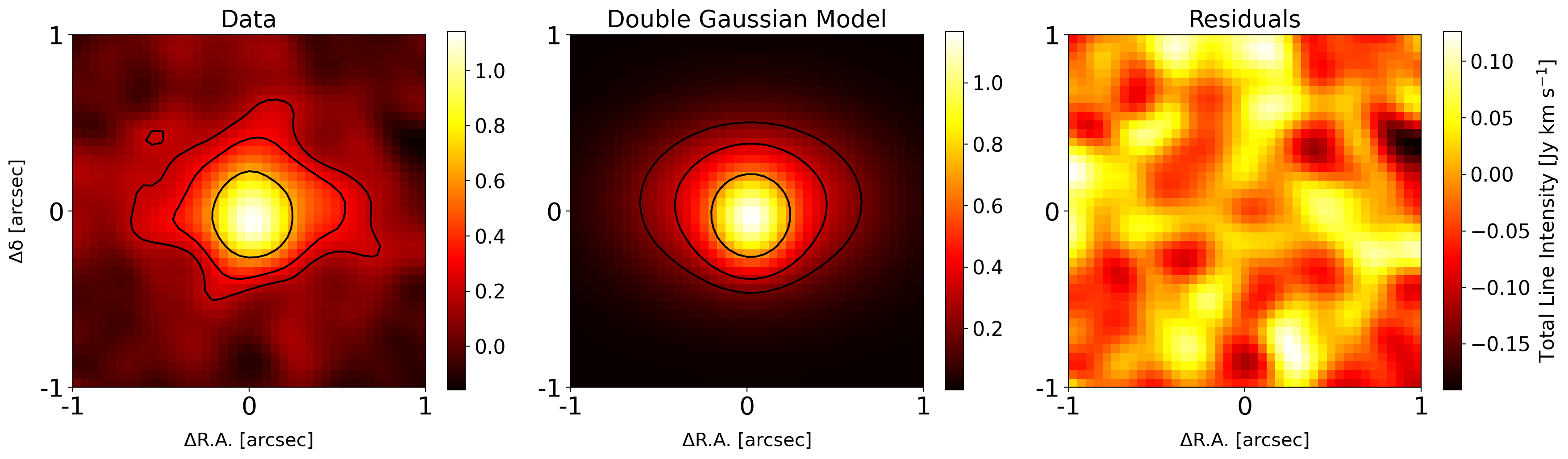}} 
 \end{tabular}
 \caption{Single and Double 2-D Gaussian fits to the moment-0 map of W2305--0039. The residuals are obtained by subtracting the model from the data in each case. The double Gaussian contains a narrow, bright, central point source and a broader, dimmer source.} 
 \label{fig:2Dfits}
\end{figure*}

\begin{figure*}
\begin{tabular}{cc}
\noindent\subfloat{\includegraphics[width=8cm]{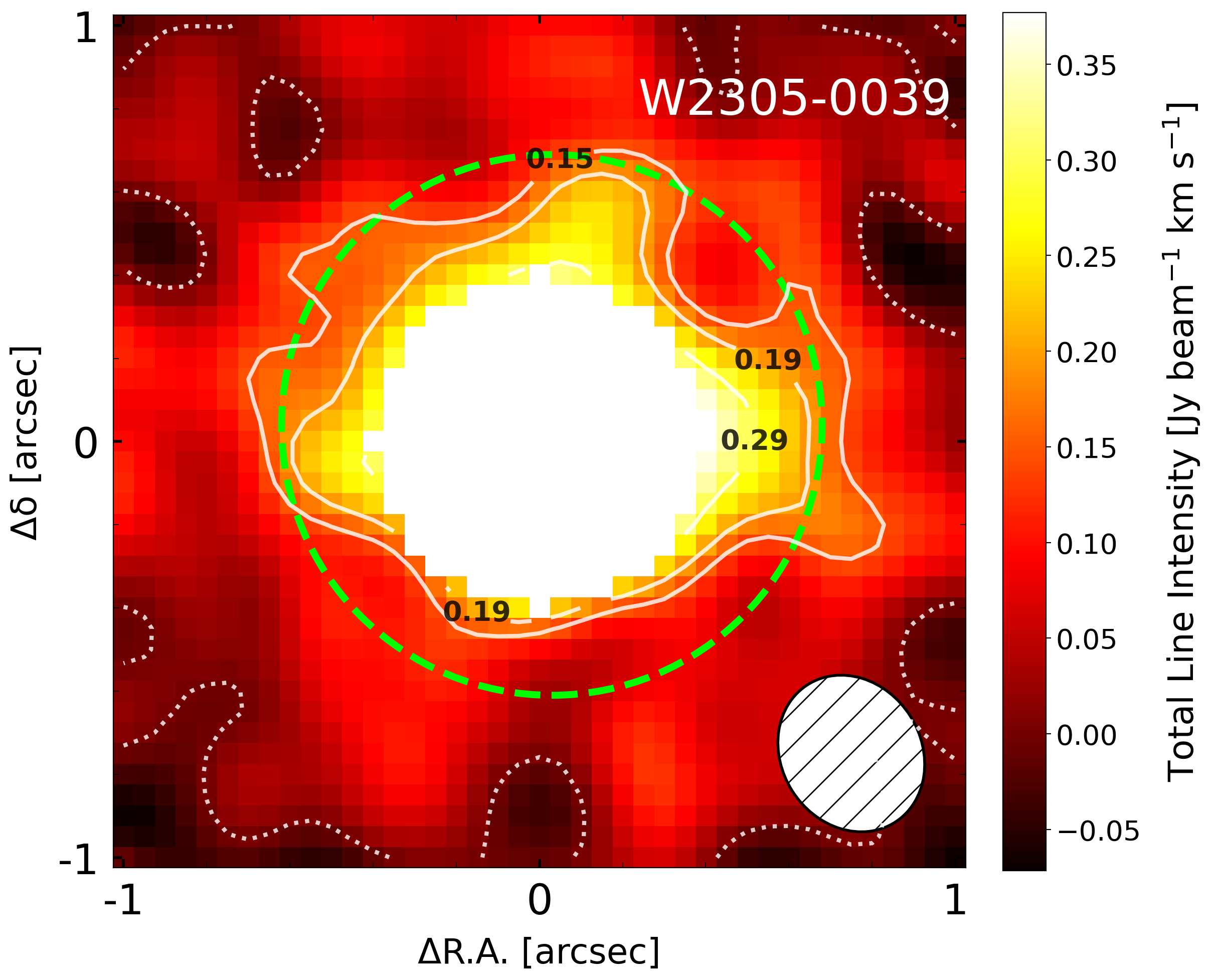}} \hspace{0.2cm}
\subfloat{\includegraphics[width=8cm]{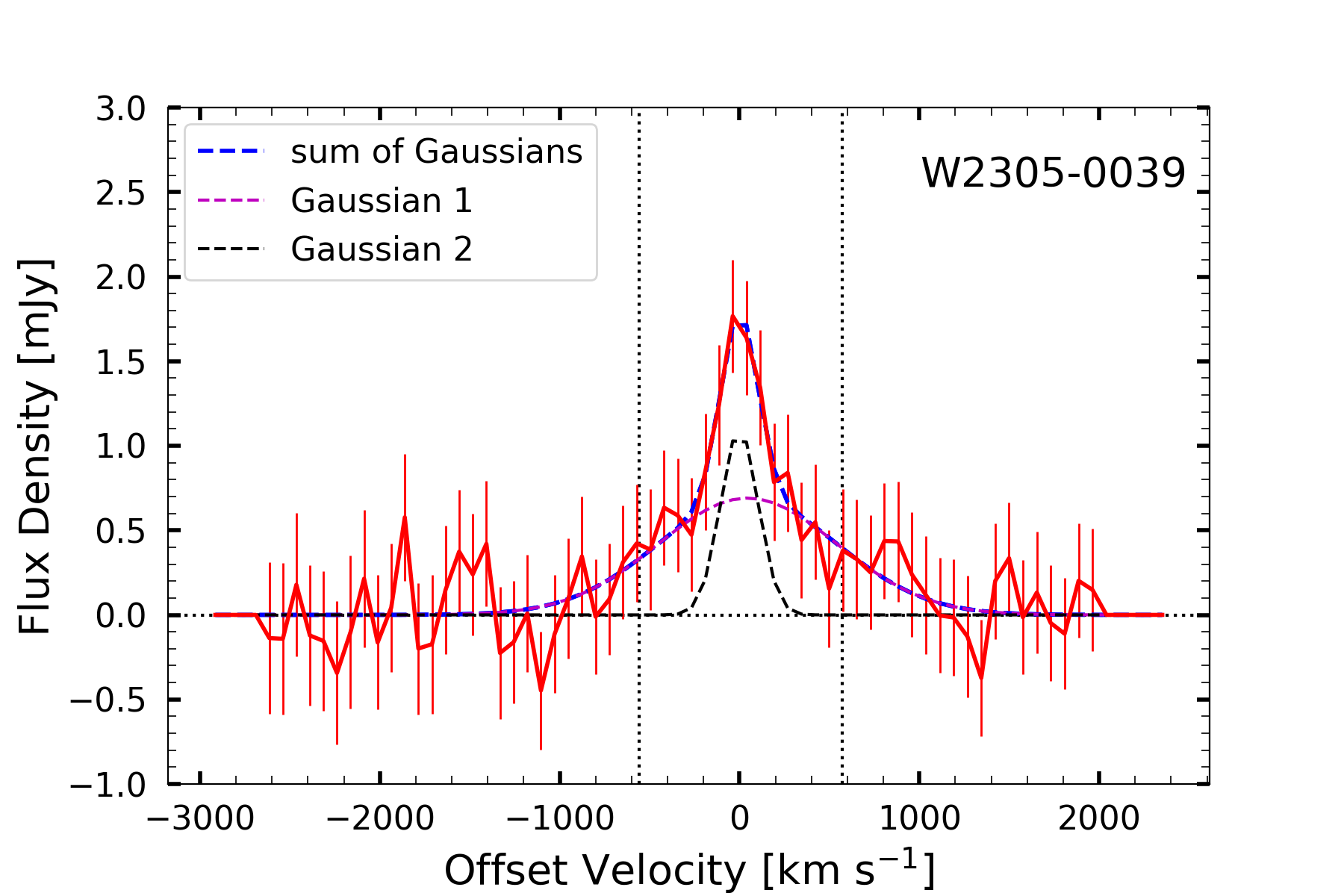}}
\end{tabular}
\caption{\textbf{Left:} Moment-0 map of W2305--0039 with a masked circularised area (white) of radius 3\,kpc. The aperture is shown via the green dashed lines, and is the same as in Fig.~\ref{fig:moment0}. \textbf{Right:} Extracted spectrum within the circular aperture. The spectrum therefore samples only the extended component of W2305--0039. The resulting fit is a double Gaussian with a broad (1170~$\pm$~180)\,km\,s$^{-1}$ component and a narrow (246~$\pm$~52)\,km\,s$^{-1}$ component. The double Gaussian fit is more statistically significant than a single Gaussian fit with $\Delta$BIC~=~9.6.}
\label{fig:beam_subtraction}
\end{figure*}

The moment-0 map of W2305--0039 shows a compact central source with a diffuse surrounding residual structure. We reason that the diffuse structure corresponds to the molecular halo of the source, as it extends out to at least 5\,kpc from the centre of emission.
We also suggest that the broad 1100\,km\,s$^{-1}$ CO component is indicative of a molecular outflow. Previous outflow studies of Hot DOGs have defined outflows as being broad and blueshifted \citep{jun2020spectral,finnerty2020fast}. While the broad CO component in W2305--0039 contains a small velocity offset ($\Delta$\textit{v}~=~20\,km\,s$^{-1}$), it is nevertheless associated with the halo at $\geq$~3\,kpc.\\
\indent We examine 2-D Gaussian fits to the moment-0 map in Fig.~\ref{fig:2Dfits}. When subtracting a single 2-D Gaussian fitted to the data, there are significant residuals; an under-subtracted central peak and an over-subtracted surrounding structure. The single 2-D Gaussian fit therefore does not accurately represent the total line intensity obtained from the moment-0 map. Conversely, the subtraction of a double 2-D Gaussian fit, with a narrow central and broader diffuse source, results in no significant residuals. As with the 1-D spectrum, these fits confirm the multi-component nature of the system.\\
\indent We subtracted a central circularised area (with a radius equivalent to 3\,kpc) from the source. The residual diffuse component is illustrated in Fig.~\ref{fig:beam_subtraction}, along with the extracted spectrum. 
When using the same circular aperture as in Fig.~\ref{fig:moment0}, we recover a broad (1170\,km\,s$^{-1}$) component. The broad CO component is therefore plausibly outflowing molecular gas. The halos of high-redshift quasars have been observed using [CII] \citep{fujimoto2019first,fujimoto2020alpine,lambert2023extended}. The observation of W2305--0039 recovered a relatively high S/N for this sample (S/N~=~5.6), hence deeper observations may reveal similar molecular halos in other sources.\\
Given that the halo is supplied by molecular outflows, we investigate the outflow energetics. The effective outflow velocity can be calculated using:
\begin{equation}
v\textsubscript{out} = 2\sqrt{\sigma^{2}\textsubscript{(broad)} + \Delta v^{2} }
\end{equation}
where $\sigma$\textsubscript{broad} is the width of the broad component, and $\Delta$\textit{v} is the velocity offset of the broad component from the central emission position. The expected CO outflow velocity is therefore (940~$\pm$~160)\,km\,s$^{-1}$.\\
\indent The highly broadened and blueshifted [OIII] components in other Hot DOGs, with with FWHM up to 6600\,km\,s$^{-1}$ \citep{jun2020spectral,finnerty2020fast}, correspond to fast ionized outflows. The molecular outflow in W2305--0039 therefore has a comparatively low velocity.\\
\indent Assuming a spherical outflow geometry \citep[e.g.,][]{maiolino2012evidence}, the mass outflow rate, energy injection rate, and momentum flux can be estimated using:
\begin{equation}
	\begin{aligned}
	\dot{M}\textsubscript{out} = \dfrac{3 M\textsubscript{gas} v\textsubscript{out}}{R\textsubscript{out}} \\
	\dot{E}\textsubscript{out} = \dfrac{3 M\textsubscript{gas} v^{3}\textsubscript{out}}{2 R\textsubscript{out}} \\
	\dot{P}\textsubscript{out} = \dfrac{3 M\textsubscript{gas} v^{2}\textsubscript{out}}{R\textsubscript{out}}.
	\end{aligned}
\end{equation}
We assume an outflow radius (\textit{R}\textsubscript{out}) of 3\,kpc \citep[as in][]{jun2020spectral,finnerty2020fast}, and that \textit{M}\textsubscript{gas} is the mass of the gas in the diffuse component (Fig.~\ref{fig:beam_subtraction}) found using equation~\ref{eq:hydrogen_mass}. The gas mass in the halo is therefore (6.2~$\pm$~1.1)~$\times$~10$^{9}$\,M\textsubscript{$\rm\odot$}. \\
\indent We obtain an outflow rate (5900~$\pm$~1500)\,M\textsubscript{$\rm\odot$}\,yr$^{-1}$, an energy injection rate $\dot{E}$\textsubscript{out}~=~(1.6~$\pm$~0.7)~$\times$~10$^{45}$\,erg\,s$^{-1}$, and momentum flux $\dot{P}$\textsubscript{out}~=~(3.5~$\pm$~1.1)~$\times$~10$^{37}$\,dyn; in agreement with [OIII] outflows in Hot DOGs \citep{jun2020spectral,finnerty2020fast}.\\
\indent Using the bolometric luminosity of W2305--0039 \citep[13.9~$\times$~10$^{13}$\,L\textsubscript{$\rm\odot$};][]{tsai2015most}, we estimate the outflow efficiencies. $\dfrac{{\dot{E}}_{\rm{out}}}{L_{\rm\odot}}$ is 0.47~$\pm$~0.21 per cent, while $\dfrac{\dot{P}_{\rm{out}}c}{L_{\rm\odot}}$ is 1.96~$\pm$~0.62. The CO outflow is therefore more likely to be momentum-driven than energy-driven, meaning that it radiates its thermal energy away. Analysis of W2305--0039's dust continuum revealed that this object likely has compact radio jets \citep{penney2020cold}. A bipolar molecular outflow is consistent alongside this analysis.
\subsection{Comparisons with other dusty galaxies}
\begin{figure*}
\begin{tabular}{cc}
\subfloat{\includegraphics[width=8cm]{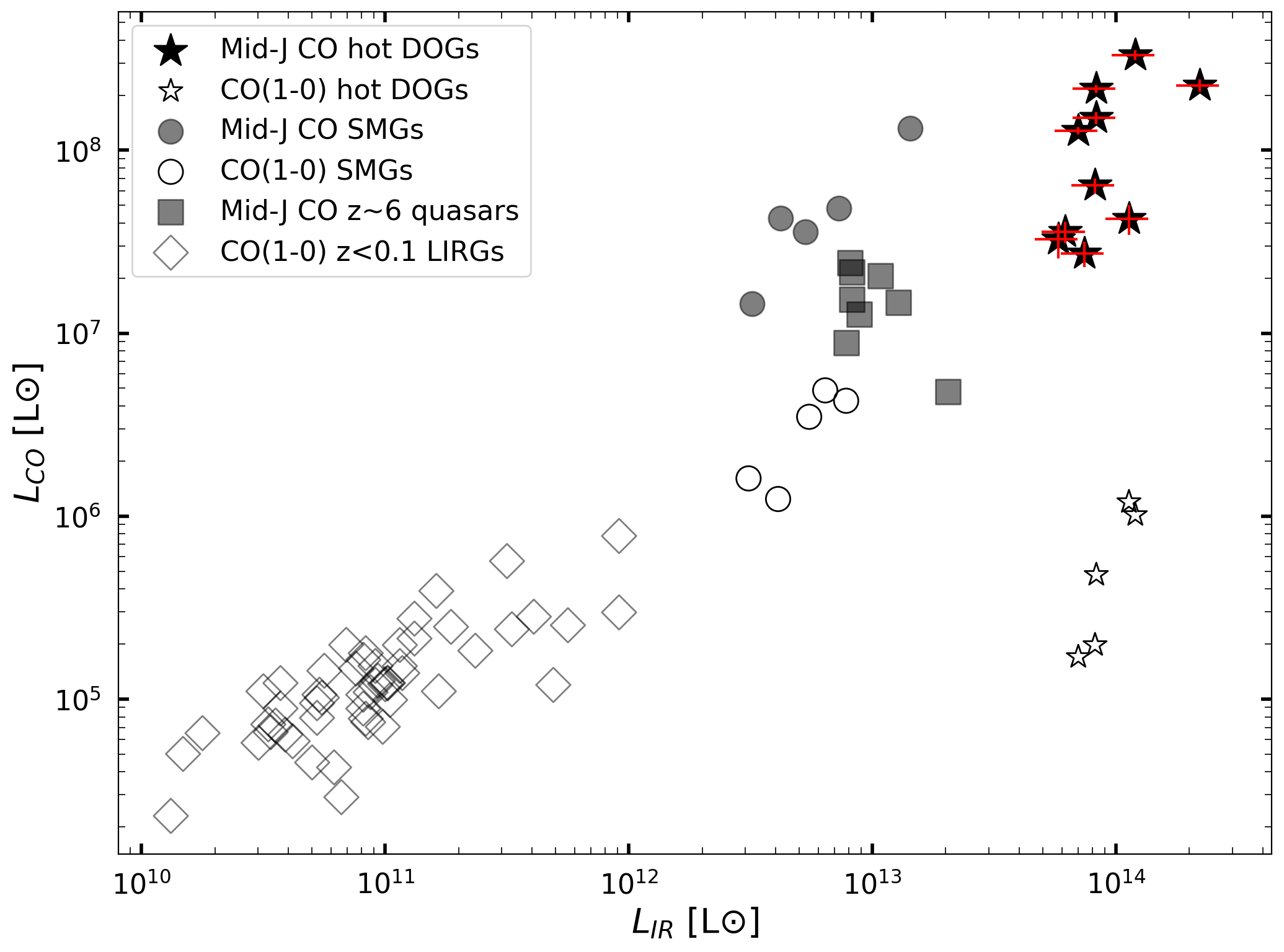}} \hspace{0.2cm}
\subfloat{\includegraphics[width=8cm]{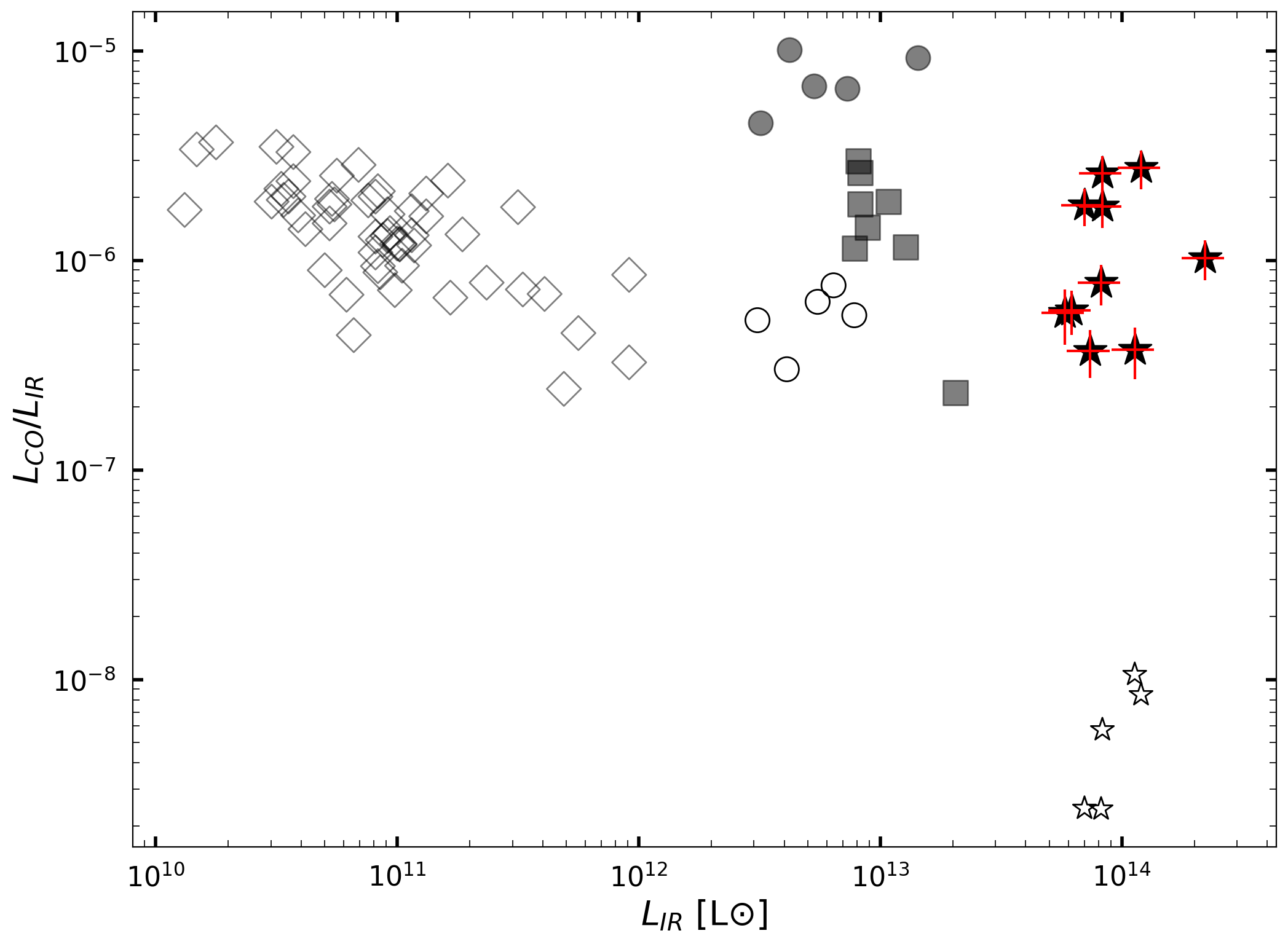}} \\
\subfloat{\includegraphics[width=8cm]{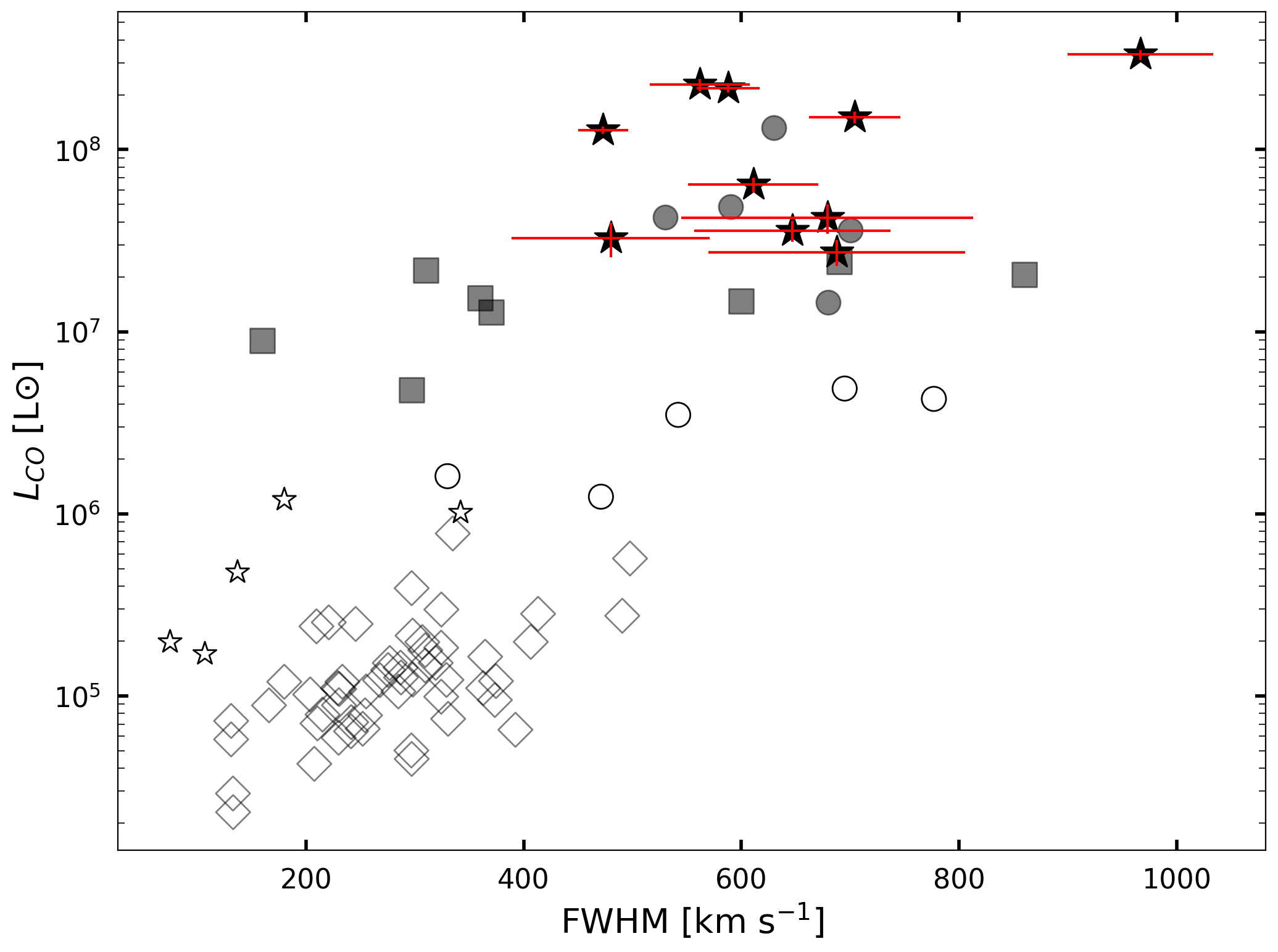}} \hspace{0.2cm}
\subfloat{\includegraphics[width=8cm]{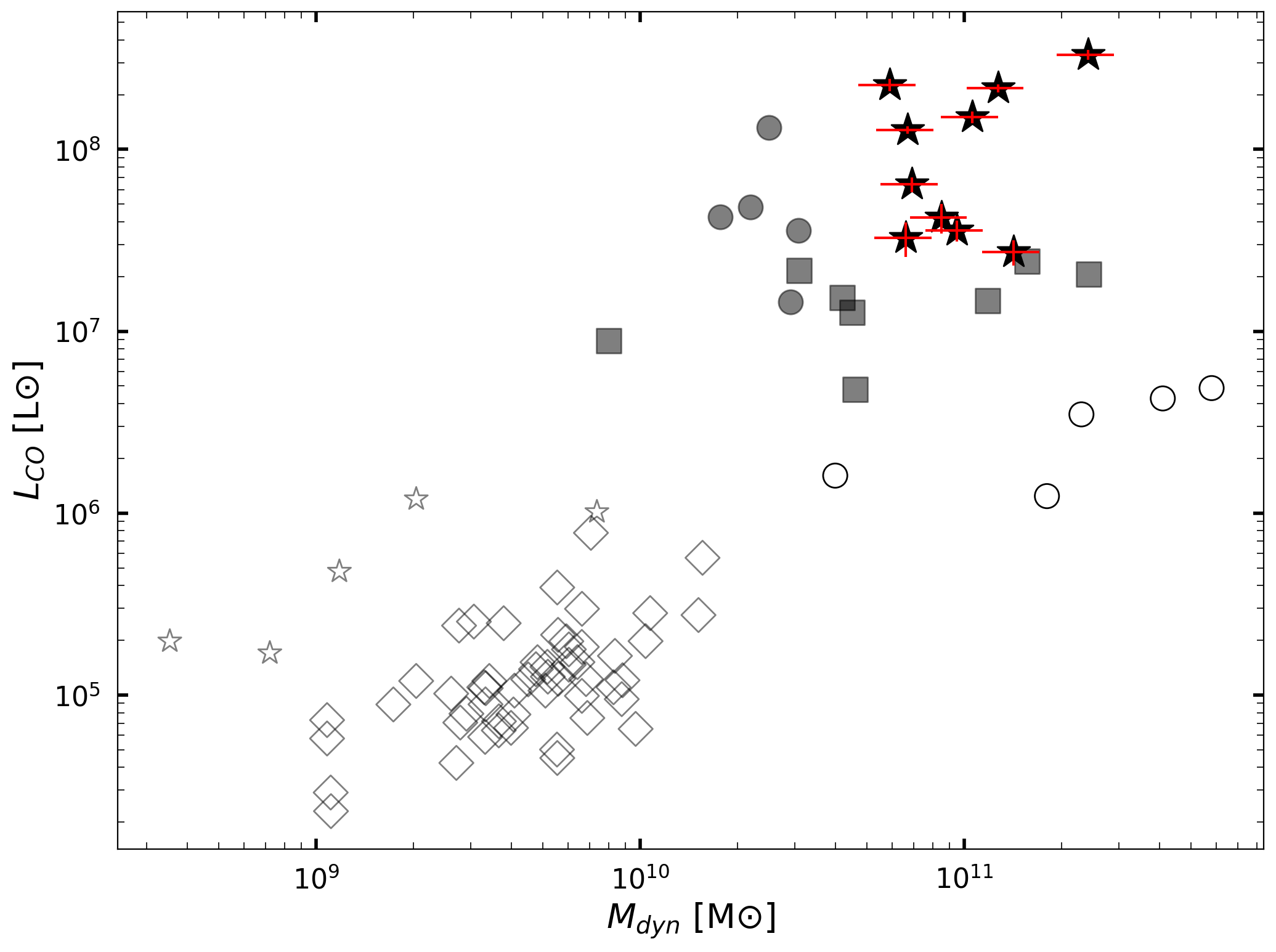}} \\
\end{tabular}
\caption{Properties of mid-\textit{J} CO lines compared with mid-\textit{J} CO emission in SMGs \citep{wardlow2018alma} and high-redshift (\textit{z}~$\approx$~6) optically-selected quasars \citep{wang2010molecular}. Low-\textit{J} CO emission is also shown in Hot DOGs \citep{penney2020cold}, SMGs \citep{ivison2011tracing} and low-redshift (\textit{z}~$\leq$~0.1) luminous infrared galaxies (LIRGs) \citep{herrero2019molecular}. A filled marker represents a mid-\textit{J} CO line, while an empty marker represents CO(1--0) emission. Hot DOGs are shown as stars, SMGs as circles, low-redshift LIRGs as diamonds, and high-redshift optically-selected quasars as squares. The red horizontal and vertical bars represent the uncertainty in the respective variable. \textbf{Upper-left:} Luminosity of CO line vs IR luminosity. \textbf{Upper-right:} Ratio of CO and IR luminosity vs IR luminosity. \textbf{Lower-left:} CO luminosity vs FWHM of the line. The FWHM of single Gaussian fits are used so as to plot the total line luminosity. \textbf{Lower-right:} CO luminosity vs dynamical mass. Where dynamical mass was not readily available in the literature, it has been estimated through a virial argument.}
\label{fig:comparisons}
\end{figure*}

While Hot DOGs are undeniably extreme systems, we compare the mid-\textit{J} CO lines observed in this work with low-redshift (\textit{z}~$\leq$~0.1) quasars, high-redshift (\textit{z}~$\approx$~6) optically-selected quasars, and SMGs in Fig.~\ref{fig:comparisons}.\\
\indent The linewidths of mid-\textit{J} CO in these targets are consistent with SMGs and high-redshift optically-selected quasars (Fig.~\ref{fig:comparisons} lower-left), though other dusty galaxies do not appear ubiquitously turbulent in mid-\textit{J} CO. High merger fractions are found in many dusty galaxies \citep[SMGs, DOGs;][]{bussmann2011hubble,barrows2023census}, and mergers are thus likely the triggers of AGN activity and turbulence in molecular gas. \\
\indent The dynamical mass of Hot DOGs is comparable to both SMGs and high-redshift optically-selected quasars (Fig.~\ref{fig:comparisons}: lower-right) and the luminosity of CO in Hot DOGs is largely consistent with SMGs and optically-selected quasars; Hot DOGs are on average a factor of 3 more luminous in mid-\textit{J} CO than SMGs, and a factor of 10 more luminous than high-redshift optically-selected quasars (Fig.~\ref{fig:comparisons} upper-left). Hot DOGs, SMGs, low-redshift quasars and high-redshift optically-selected quasars all have a consistent CO--IR ratio of $\frac{L\textsubscript{CO}}{L\textsubscript{IR}}$~$\approx$~10$^{-6}$--10$^{-5}$ (Fig.~\ref{fig:comparisons}: upper-right). \textit{L}\textsubscript{IR} includes emissions from dust heated by a central AGN. The CO--IR ratio for quasars and Hot DOGs could therefore be larger than SMGs when considering purely the star-forming component. Nevertheless, in mid-\textit{J} CO, Hot DOGs are as luminous as expected given the superlative IR luminosity of their hosts.
\subsection{CO Luminosity Ratios}
The mid-\textit{J} CO in Hot DOGs is on average a factor of 450 more luminous than CO(1--0) in the same sources (Table \ref{table:lumin_mass}), whereas in SMGs the factor is approximately 20 (Fig.~\ref{fig:comparisons}: upper-left). Given that the mid-\textit{J} CO is as luminous as expected, CO(1--0) is significantly underluminous in Hot DOGs. The widths of CO(1--0) lines in Hot DOGs (75--342\,km\,s$^{-1}$) are also narrow compared to SMGs and high-redshift optically-selected quasars (Fig.~\ref{fig:comparisons}: lower-left).\\
\indent The excitation potential of CO(1--0) is 5.5\,K, while the temperature of the cosmic microwave background (CMB) at a redshift \textit{z}~$\geq$~3 is at least 11\,K [\textit{T}\textsubscript{z}~=~\textit{T}\textsubscript{CMB}~(1~+~\textit{z})]. A hotter CMB contributes to increased line excitation, but also produces a brighter observational background \citep{carilli2013cool}. The effects of the CMB at high redshift are therefore most pronounced when examining the lowest excitation CO gas. It is plausible that the CMB has thermalised the coldest molecular gas in these systems. This would explain the narrow CO(1--0) lines observed in Hot DOGs, as a thermalised component would not be turbulent. Mid-\textit{J} CO emission lines observed in this work, due to hotter excitation temperatures (\textit{T}\textsubscript{ex}~$\geq$~55\,K), are more easily observed in Hot DOGs, and would not be strongly affected by the CMB. Low-\textit{J} CO may be unreliable in high-redshift (\textit{z}~$\geq$~3) Hot DOGs for inferring properties of the molecular gas.\\
\indent The $\frac{L^{\prime}\textsubscript{CO(4--3)}}{L^{\prime}\textsubscript{CO(1--0)}}$ luminosity ratios in Hot DOGs range from 4.7--11.5. These luminosity ratios are close to unity for thermalised, optically thick regions of gas \citep{carilli2013cool}. If current CO(1--0) detections are underluminous, these ratios may not accurately reflect the ISM conditions in the host galaxies.\\
\indent The line ratios of W2246--0526 are closer to unity, though, notably, this source does not have a CO(1--0) detection. The $\frac{L^{\prime}\textsubscript{CO(7--6)}}{L^{\prime}\textsubscript{CO(5--4)}}$ ratio in this source is 1.1~$\pm$~0.1, meaning that molecular gas could be thermalised at energies around mid-\textit{J} CO.
\section{Summary} \label{sec:summary}
This paper analysed mid-\textit{J} CO ALMA observations of ten of the most luminous WISE-selected Hot DOGs and obtained the following results:
\begin{enumerate}
    \item The CO dispersions are consistently broad, with nine out of ten sources containing at least one broad ($\geq$~400\,km\,s$^{-1}$) spectral component. These large dispersions suggest that the molecular ISM of Hot DOGs is turbulent. The turbulence is plausibly caused by low-velocity shocks, caused by dynamical friction during galactic mergers. These mergers are likely the triggers of AGN activity in these sources.
    \item W0831+0140 plausibly comprises two galaxies in the process of merging, and the radial velocity map is consistent with this scenario or the presence of a rotating disc. As with [CII] observations of this source, we note the possibility of a third galaxy in this merger. W2246--7143 shows a similar possible merging scenario, though deeper observations are needed in both cases.
    \item A molecular halo is identified in W2305--0039, and is plausibly supplied by a 940\,km\,s$^{-1}$, momentum-driven, molecular outflow.
    \item All sources are detected in mid-\textit{J} CO with four detections at S/N~$\geq$~5, providing resolved detail in 40 per cent of the sources. Continuum emission is recovered in nine of the ten sources, and W0831+0140 and W2305--0039 likely either have greater quantities of dust, or hotter average dust than other extremely luminous Hot DOGs.
    \item CO(1--0) is significantly underluminous compared to the mid-\textit{J} CO presented in this work. We reason that the low-\textit{J} CO emission has been thermalised by the CMB at \textit{z}~$\geq$~3, as less distant dusty galaxies, such as SMGs, show more modest mid-\textit{J} to low-\textit{J} luminosity ratios.
\end{enumerate}
\indent To uncover whether additional structure exists in the fainter targets in this sample, we will require high-resolution and deeper observations to examine their profiles. While CO SLEDs of Hot DOGs contain few data points (due to limited numbers of observations), the CO luminosity of some targets (e.g. W2246--0526) is shown to increase up to at least CO(7--6) \citep{aranda2024benchmark}. Observations of high-\textit{J} CO, accompanied by JVLA CO(1--0) and mid-\textit{J} transitions in this work could begin to fill the gaps and acquire a better CO SLED coverage. This would allow us to model and characterise the physical properties of the molecular gas in this obscured quasar phase.
\section*{Acknowledgements}
We thank an anonymous reviewer for his/her constructive comments. LRM thanks the staff at the University of Leicester for their support, particularly Sergei Nayakshin and Martin Barstow for their helpful discussions. LRM was supported by the Science and Technologies Facilities Council (STFC) studentship. TDS acknowledges the research project was supported by the Hellenic Foundation for Research and Innovation (HFRI) under the "2nd Call for HFRI Research Projects to support Faculty Members \& Researchers" (Project Number: 3382). RJA was supported by FONDECYT grant number 1231718 and by the ANID BASAL project FB210003. HDJ was supported by the National Research Foundation of Korea (NRF) funded by the Ministry of Science and ICT (MSIT) of Korea (2022R1C1C2013543). RFA was supported by the European Research Council (ERC) under grant agreements
No. 771282. This paper makes use of the following ALMA data: ADS/JAO.ALMA\#2017.1.00358.S. ALMA is a partnership of ESO (representing its member states), NSF (USA) and NINS (Japan), together with NRC (Canada), MOST and ASIAA (Taiwan), and KASI (Republic of Korea), in cooperation with the Republic of Chile. The Joint ALMA Observatory is operated by ESO, AUI/NRAO and NAOJ. Portions of this research were carried out at the Jet Propulsion Laboratory, California Institute of Technology, under a contract with the National Aeronautics and Space Administration (NASA). 
\section*{Data Availability}
The data underlying this paper are available in the ALMA Science Archive, which can be accessed at https://almascience.nrao.edu/aq/.


\bibliographystyle{mnras}
\bibliography{references} 




\appendix
\section{Bayesian Information Criteria For Gaussian Fits}\label{sec:bic}
\begin{figure*}
\begin{tabular}{cc}
\noindent\subfloat{\includegraphics[width=8cm]{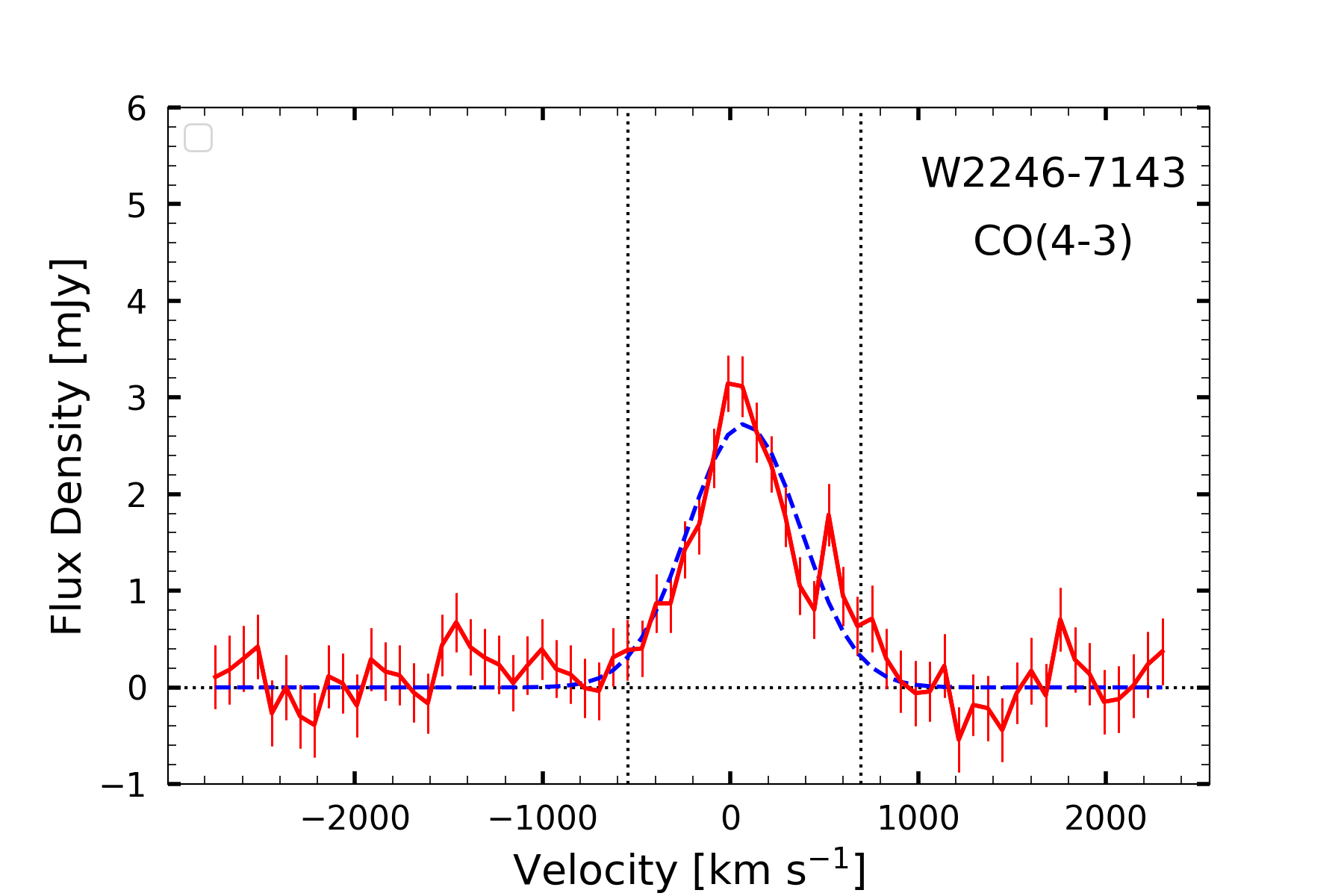}} \hspace{0.2cm}
\subfloat{\includegraphics[width=8cm]{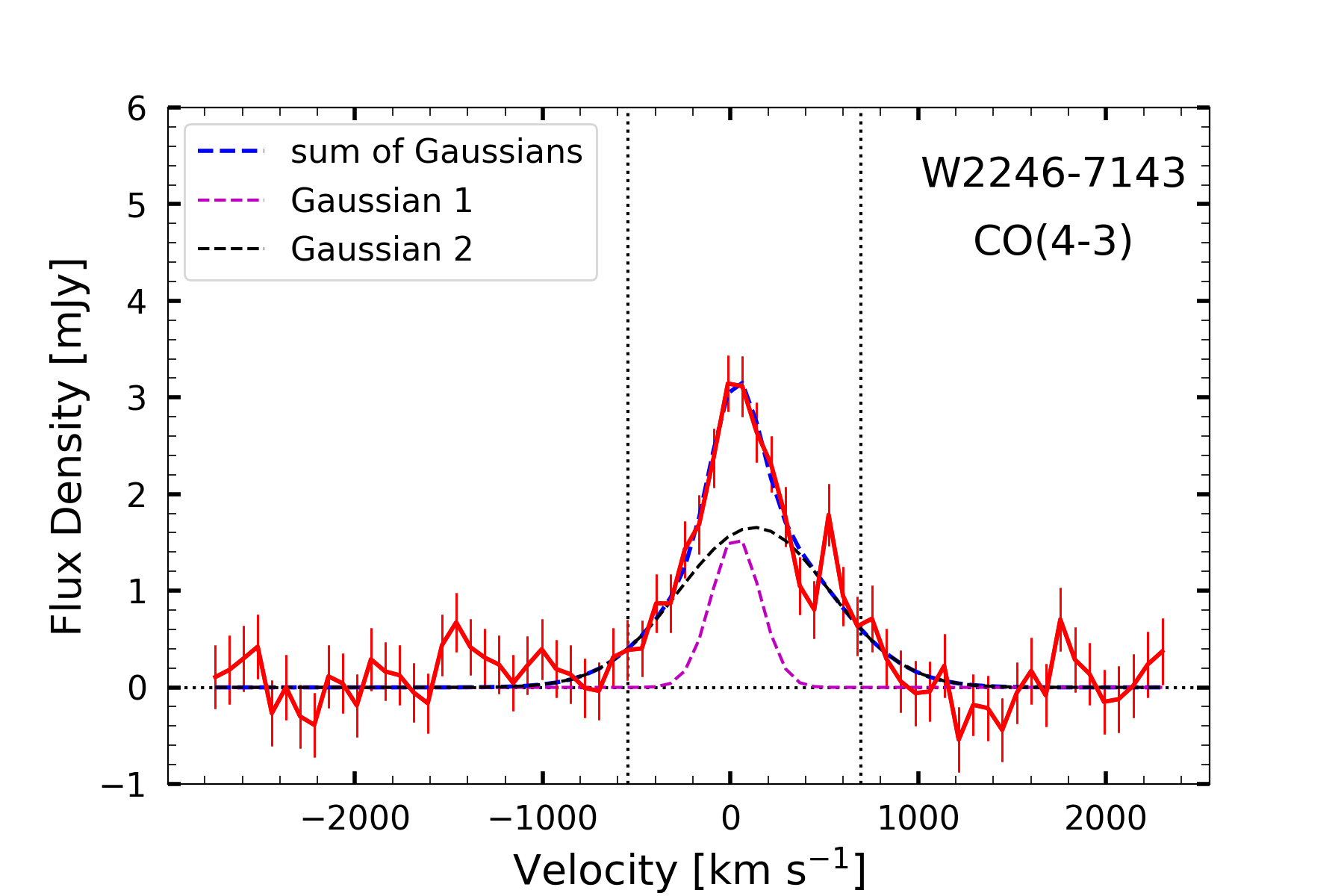}} \\
\subfloat{\includegraphics[width=8cm]{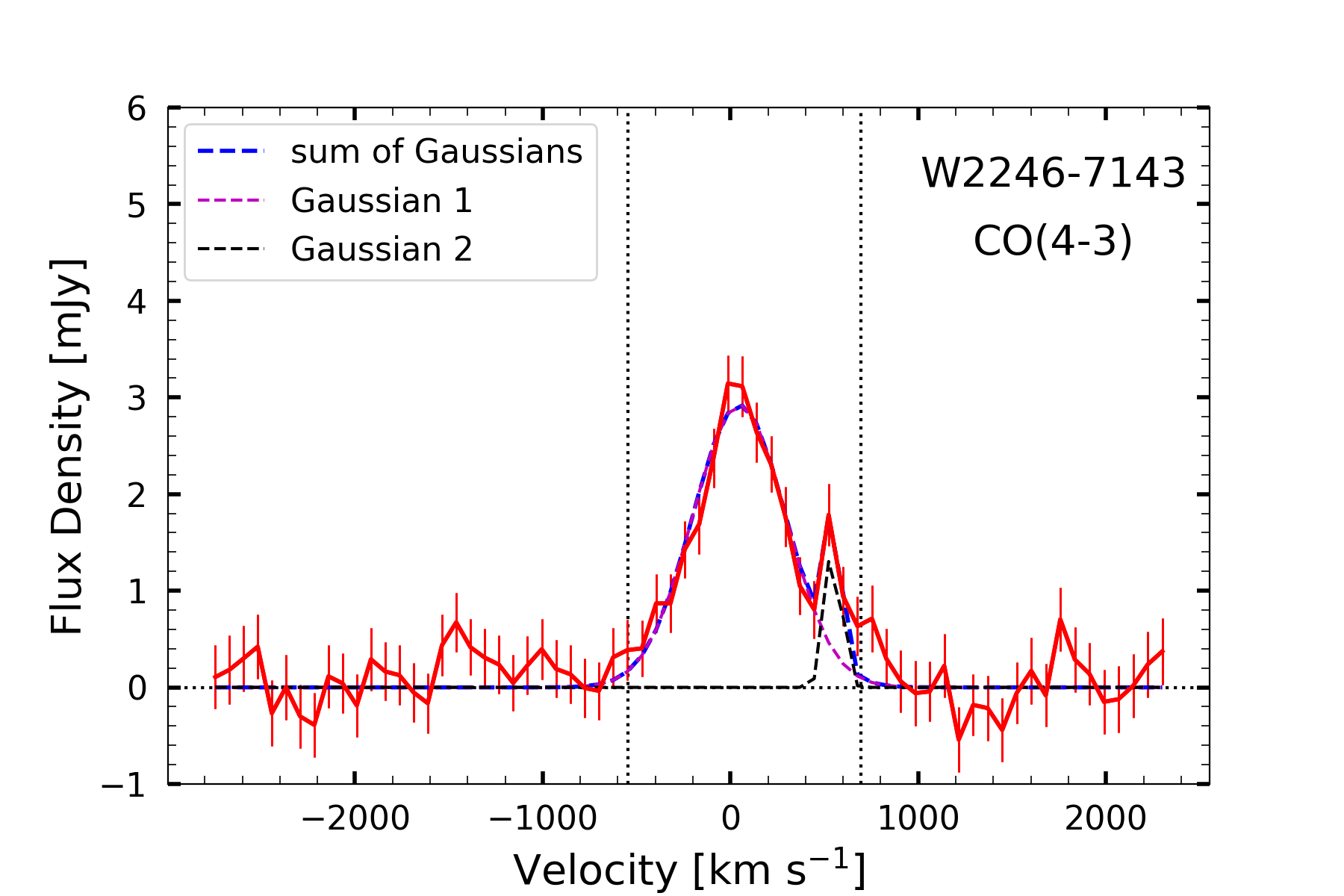}} \hspace{0.2cm}
\subfloat{\includegraphics[width=8cm]{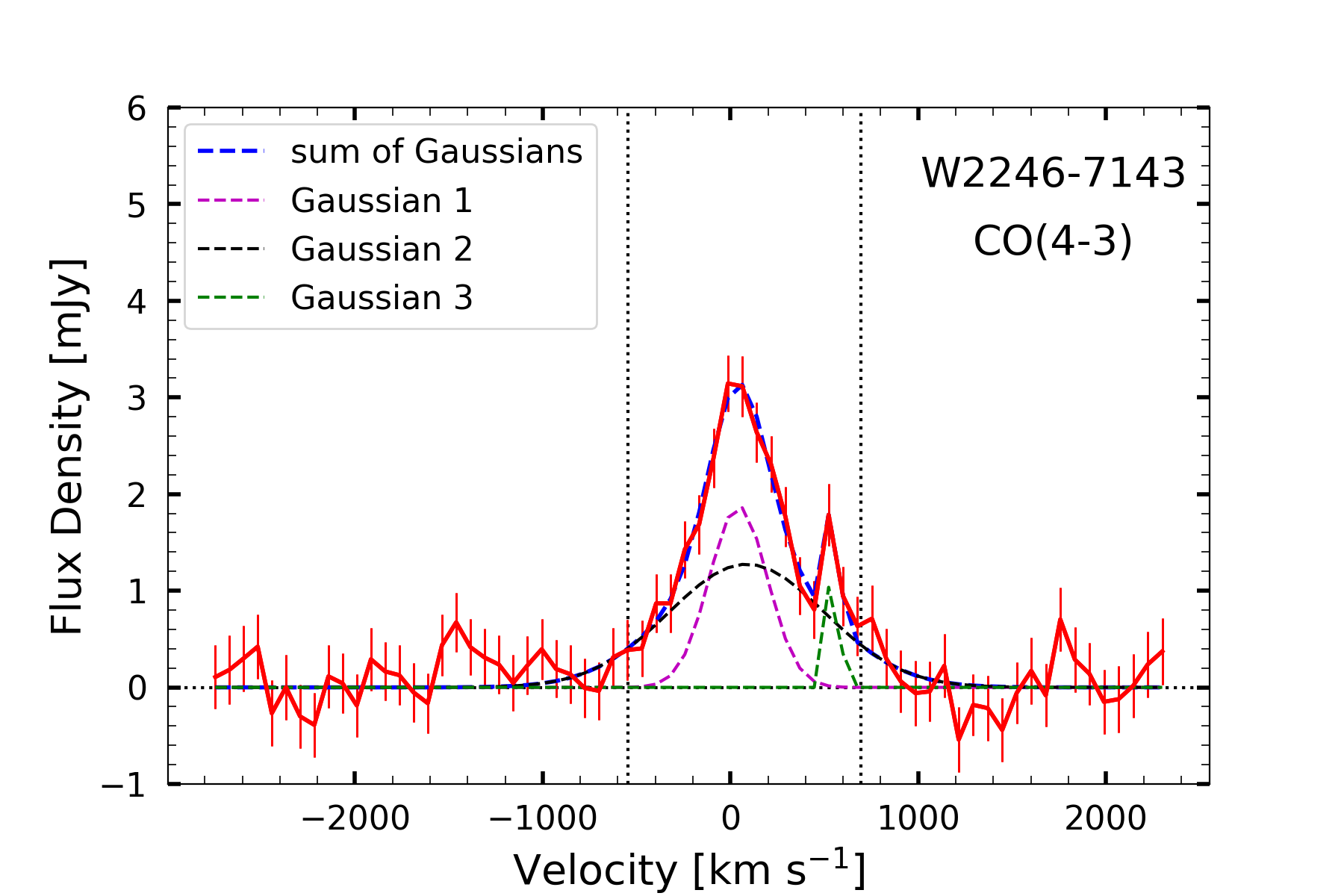}}\\
\subfloat{\includegraphics[width=8cm]{Spectra/W0831_double_gaussian.png}} \hspace{0.2cm}
\subfloat{\includegraphics[width=8cm]{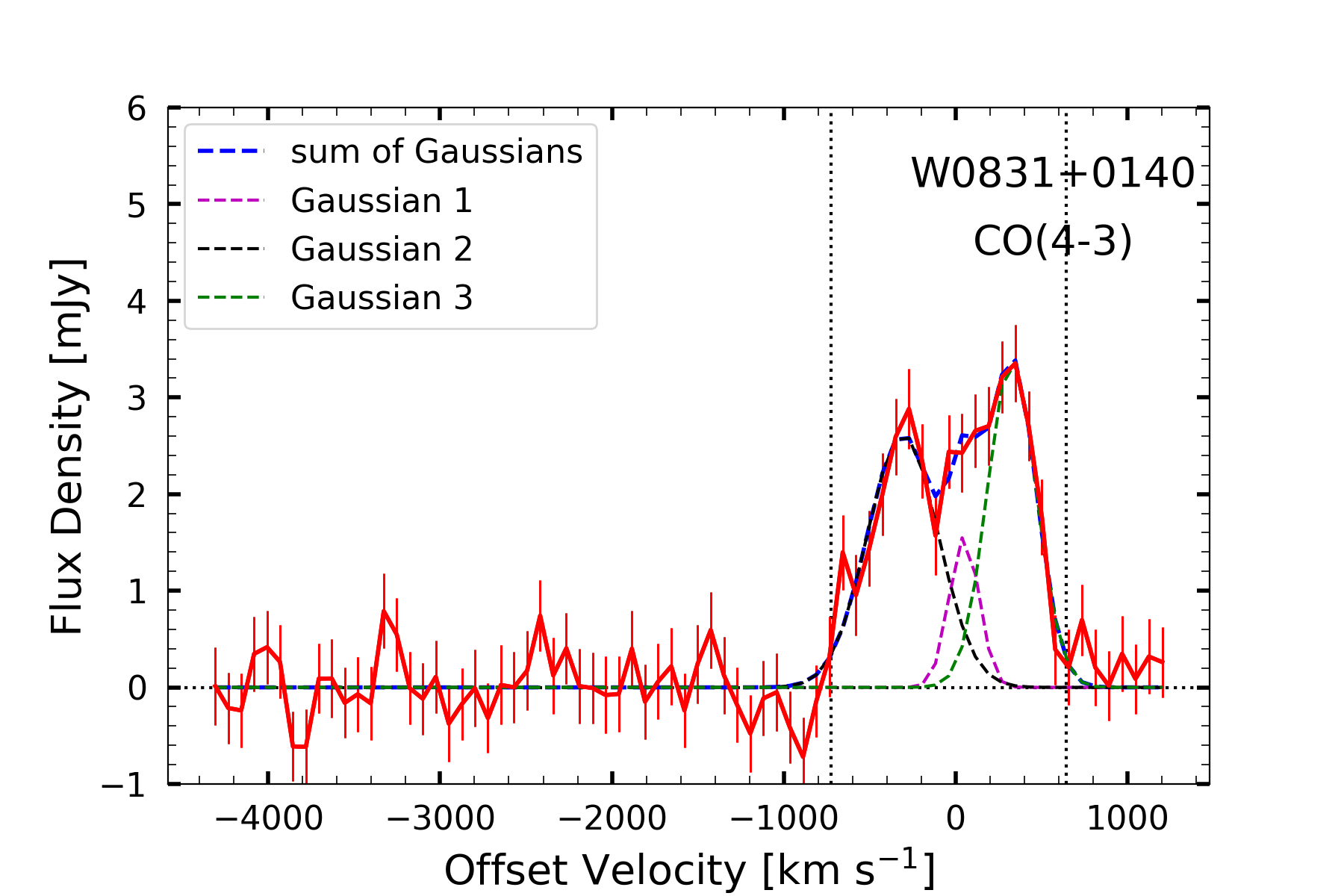}}\\
\end{tabular}
\caption{\textbf{Top two rows:} Four Gaussian fits to the CO(4--3) spectrum of W2246-7143. The double Gaussian (middle-left) is the most statistically significant fit. The triple Gaussian in W2246--7143 suffered from overfitting, and uncertainties could not be accurately estimated. \textbf{Bottom row:} Two of the W0831+0140 Gaussian fits. The double Gaussian fit (lower-left) was marginally more statistically significant ($\Delta$BIC$<$2) than the triple Gaussian fit (lower-right).}
\label{fig:W2246-7143_fits}
\end{figure*}

BIC statistics were used to measure the goodness of fit for single, double, and triple Gaussian fits to the CO spectra. The complete list of BIC values are shown in Table \ref{table:BIC}.
\begin{table*}
    \centering
    \caption{BIC statistics of Gaussian fits to spectra. (1) BIC value for a single Gaussian fit; (2) BIC value for a double Gaussian fit. For W2246--7143, two different double Gaussians were fit to the spectra; (3) BIC value for a triple Gaussian fit; (4) Comparison of the two lowest BIC statistics.}
    \label{table:BIC}
    \setlength\extrarowheight{2pt}
    \begin{tabular}{c c c c c}
        \hline
        Source & 1-G BIC & 2-G BIC & 3-G BIC & $\Delta$BIC\\
        & (1) & (2) & (3) & (4) \\
        \hline
        W0116--0505 & $-$43.7 & $-$43.9 & \dots & 0.2 \\
        W0134--2922 & $-$31.7 & \dots & \dots & \dots \\
        W0615--5716 & $-$25.6 & \dots & \dots & \dots \\
        W0831+0140 & $-$10.2 & $-$22.0 & $-$20.1 & 1.9 \\
        W1248--2154 & $-$38.1 & \dots & \dots & \dots \\
        W1322--0328 & $-$31.8 & \dots & \dots & \dots\\
        W2042--3245 & $-$22.2 & -40.3 & \dots & 18.1\\
        W2246--0526 & $-$23.5 & \dots & \dots & \dots \\
        W2246--7143 & $-$22.3 & $-$28.2/$-$34.8 & \dots & 6.6\\
        W2305--0039 & $-$26.1 & $-$
        41.9 & \dots & 15.8\\
        \hline
    \end{tabular}
    \vspace{0.2cm} \\
\end{table*}
In W2246-7143, we examined the goodness of fit between four Gaussian fits to the spectra. These included a single Gaussian, two double Gaussians, and a triple Gaussian. In W0831+0140, a triple Gaussian was also fit to the spectrum of the source. Each fit is shown in Fig.~\ref{fig:W2246-7143_fits}, and the widths of each Gaussian component are shown in Table \ref{table:stats}.
\begin{table*}
    \centering
    \caption{Fit parameters of each of the fits to the spectrum of W0831+0140 and W2246--7143. (1) The type of Gaussian fit applied to the spectrum; (2) The BIC value associated with the Gaussian fit; (3) The full-width at half-maximum of each of the Gaussian components.}
    \label{table:stats}
    \setlength\extrarowheight{2pt}
    \begin{tabular}{c c c}
        \hline 
         Gaussian Fit & BIC & FWHM \\
                      &          & [km~s$^{-1}$] \\
        (1) & (2) & (3) \\
        \hline
        W2246--7143 Single  & $-$22.3 & 704 $\pm$ 42 \\
        W2246--7143 Double(A) & $-$26.9 & [304 $\pm$ 71], [940 $\pm$ 114]\\
        W2246--7143 Double(B) & $-$34.2 & [548 $\pm$ 38], [100 $\pm$ 52]  \\
        W2246--7143 Triple  & \dots & [364], [980], [48] \\
        W0831+0140 Double & $-$22.0 & [528 $\pm$ 80], [434 $\pm$ 50] \\
        W0831+0140 Triple & $-$20.1 & [204 $\pm$ 95], [484 $\pm$ 81], [334 $\pm$ 57] \\
        \hline
    \end{tabular}
    \vspace{0.2cm} \\  
\end{table*}

\newpage
\section{Defining Extended Continuum and CO Emission}\label{sec:bem_subtractions}
\begin{figure*}
\begin{tabular}{cc}
\noindent\subfloat{\includegraphics[width=15cm]{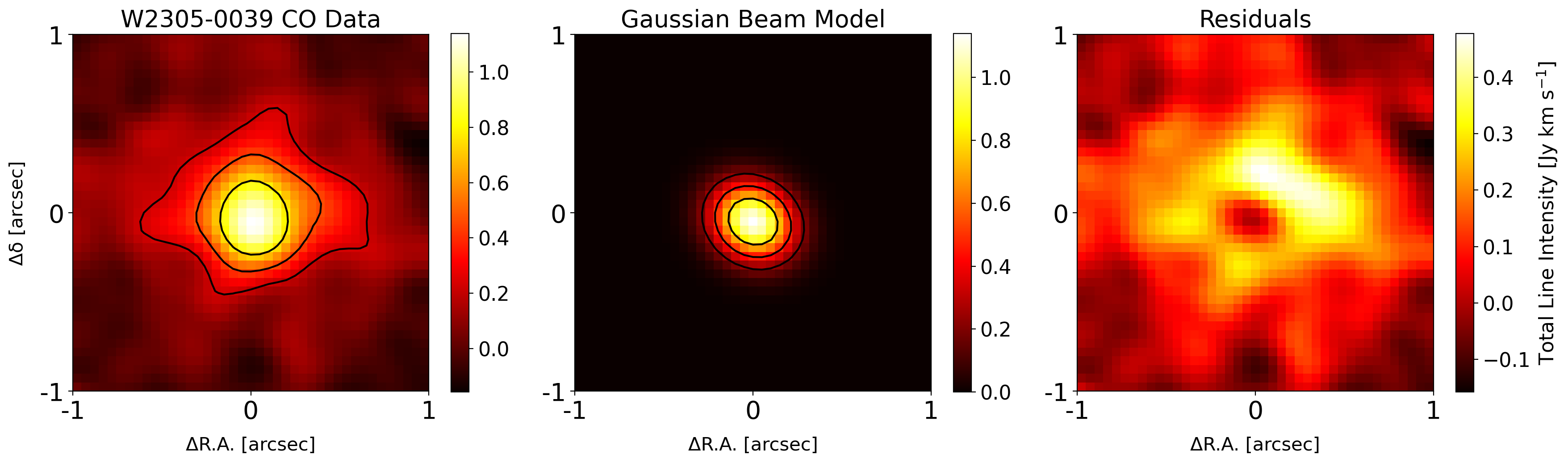}}
\end{tabular}
\caption{Example of a 2-D elliptical Gaussian subtraction performed on the continuum and CO flux maps. The left image shows the flux map with 3, 5, and 10$\sigma$ contours. The middle image shows the elliptical Gaussian (the beam), centred on the peak of emission and normalised to the peak flux. The right image shows the residuals after performing the 2-D subtraction. In this case, 60 per cent of the CO flux remains in the residuals within the 3$\sigma$ region after the subtraction.}
\label{fig:beam_subtraction_example}
\end{figure*}

To quantify the extension of both the continuum and CO in these observations, we performed 2-D elliptical Gaussian subtractions (representative of the clean beam) on each of the flux (moment-0) maps. The peak value of the Gaussian corresponds to the maximum flux of the data, meaning that we subtract the brightest possible unresolved point source. The estimations of extended emission are therefore plausibly conservative, and the level of extension may be greater than we calculated. An example of one of the subtractions is shown in Fig.~\ref{fig:beam_subtraction_example}.
The amount of extension is quantified as the fraction of flux remaining in the residuals within the 3$\sigma$ region of the original moment-0 map.

\bsp	
\label{lastpage}
\end{document}